\documentclass[12pt]{article}
\usepackage{pbox}
\usepackage{graphicx}
\usepackage{amssymb}
\usepackage{cite}
\usepackage{mathtools}
\usepackage{amsmath}
\usepackage{multirow}
\newcommand*{\Scale}[2][4]{\scalebox{#1}{\ensuremath{#2}}}%
\usepackage{xcolor,colortbl}
\usepackage[a4paper,left=2cm,right=2cm,top=2cm,bottom=2.5cm] {geometry}
\begin{document}
\title{\textbf{Multi-Qubit Dynamical Quantum Search Algorithm with
Dissipation}}
\author{A. H. Homid$^{1,3,}\footnote{E-mail: ahomid86@gmail.com}$, M. Abdel-Aty$^{1}$ and
A.-S. F. Obada$^{2}$\\
\small$^{1}$ University of Science and Technology, Zewail City for
Science and Technology, Giza, Egypt\\
\small$^{2}$ Faculty of Science, Al-Azhar University, Nasr city
11884, Cairo, Egypt\\
\small$^{3}$Faculty of Science, Al-Azhar University, Assiut 71524,
Egypt}
\date{}
\maketitle
\begin{abstract}
We invoke an efficient search algorithms as a key challenge in
multi-qubit quantum systems. An original algorithm called dynamical
quantum search algorithm from which Grover algorithm is obtained at
a specified time is presented. This algorithm is distinguished by
accuracy in obtaining high probability of finding any marked state
in a shorter time than Grover algorithm time. The algorithm
performance can be improved with respect to the different values of
the controlled phase. A new technique is used to generate the
dynamical quantum gates in the presence of dissipation effect that
helps in implementing the current algorithm.
\end{abstract}

\section{Introduction}
\indent
In the early 1996s, Grover search algorithm (GA) has attracted much
interest, since it could work in quantum computer.
The quantum search, from computational point of view, is proved to
get approximately $O(\sqrt{N})$ operations (in comparison with the
$O(N)$ classical operations), which indicates a quadratic speedup
\cite{gro96,gro97}.
Many efforts have been devoted to achievement of this algorithm
theoretically and experimentally by using superconducting qubits
\cite{dic09}, trapped ions \cite{yan08}, atom cavity QED systems
\cite{den05,yan07} and nuclear magnetic resonance \cite{chu98}.
Also, most previous studies for improving the quantum GA were
general and do not depend on a physical system.
Further, all previous implementations of quantum GA have been
developed depending on Grover visualizations, where the control
phase parameter for the controlled phase gates or oracle is switched
to $\pi $ or $\frac{\pi }{3}$ \cite{gro97,gro05}.
However, an interesting work by Long et al \cite{lon99} has replaced
the Grover's arbitrary phase by $\frac{\pi }{2}$.
The combination of strong resonant interactions with tight
subwavelength traps gives an excellent opportunity to implement much
faster quantum gates.
However, there is no way to achieve single site addressability in
those systems. Existing proposals circumventing this problem
\cite{gro97,gro05} do not consider an arbitrary controlled phase
parameter or dispassion effect, as the interaction need not to be
restricted by certain values of the phase control.
\\ \indent
In this article, we propose to add new factors allowing for
significant speedup.
By combining a multi-qubit system with dissipation taking into
account an arbitrary phase control, we can achieve a fast search in
a realistic setting.
We assume that the control phase is time-dependent to realize the
dynamical quantum controlled phase gates.
It is shown that the involved time is the crucial factor for
controlling and searching of any marked state regardless the number
of qubits.
Consequently, we define a new algorithm called a dynamical quantum
search algorithm (DQSA).
The proposed algorithm is a generalization of Grover algorithm and
(i) gives a higher probability of finding any marked state in
shorter time, (ii) there is no restriction on the controlled phase
parameter and (iii) the presence of qubits dissipation is allowed.
\section{Quantum Search Algorithm with Dissipation}
The effect of a thermal bath has been used to improve the
performance of a quantum adiabatic search algorithm \cite{ami08}.
Vega et al \cite{veg10} compared the effects of the field
dissipation on the algorithm performance with those of a structured
environment similar to the one encountered in systems coupled to an
electromagnetic field that exists within a photonic crystal. They
show that, the algorithm performance has been improved by tuning the
environment parameters to certain regimes \cite{mac10}. We consider
the problem with a different perspective, where the algorithm may be
implemented within a system taking into account the qubit
dissipation.
\begin{figure*}[tbp]
\centering\includegraphics[width=10cm,height=8cm]{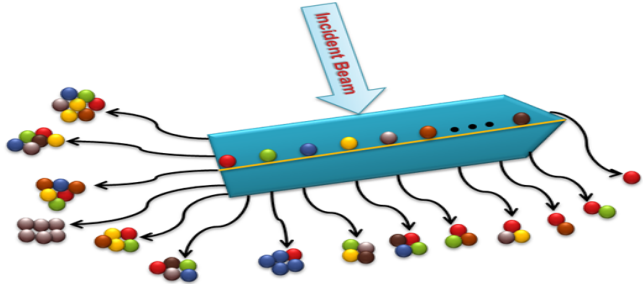}
\vspace{-0.3cm} \caption{Schematic diagram of quantum dots ensemble.
A single surface plasmon injected from outside is coherently
scattered by the dots. Our interest is to consider all the possible
interactions between any arbitrary pair or more inside the sample.
The dissipation effect resulted from the direct contact between the
qubits and the plasmone surface.}\label{ff-1}
\end{figure*}
Here, we consider a dense multi-qubit system sample interacting with
a single surface plasmon injected from outside. In this case, we
consider all the possible interactions between any arbitrary pair or
more inside the sample (see Fig.\ref{ff-1}). The direct contact
between the qubits and the plasmone surface yields dissipation
effects. As a consequence, we assume that the dipole-dipole
interaction potential is nonidentical for all qubits. Other
appropriate systems could be superconducting qubits
\cite{che07,xia09} or double quantum dot excitonic systems
\cite{che03}. Such a model can be written as follows
\begin{eqnarray}\label{z1}
\hat{H}&=&-\lambda\Big\{\sum_{s_{1}=1}^{N}J_{s_{1}}\,
\hat{\sigma}^{s_{1}}_{z}+\sum_{s_{1},s_{2}=1}^{N}J_{s_{1}s_{2}}\,
\hat{\sigma}^{s_{1}}_{z}\hat{\sigma}^{s_{2}}_{z}+\sum_{s_{1},s_{2},s_{3}=1}^{N}
J_{s_{1}s_{2}s_{3}}\,\hat{\sigma}^{s_{1}}_{z}\hat{\sigma}^{s_{2}}_{z}\hat{\sigma}^{s_{3}}_{z}\nonumber\\&&
+...+\sum_{s_{1},s_{2},s_{3},...,s_{n}=1}^{N}
J_{s_{1}s_{2}s_{3}...s_{n}}\,
\hat{\sigma}^{s_{1}}_{z}\hat{\sigma}^{s_{2}}_{z}\hat{\sigma}^{s_{3}}_{z}...\hat{\sigma}^{s_{n}}_{z}\Big\}
-i\sum_{v=1}^{N}\frac{\gamma_{v}}{2}\,\hat{\sigma}_{+}^{v}\hat{\sigma}_{-}^{v},
\end{eqnarray}
where $S=\{s_{1},s_{2},...,s_{n}\}$ refer to the labels of the
qubits, $N$ is the number of qubits, $\gamma _{u}$ are the
dissipation rates of different qubits, $J_{s_{1}}$,..., $J_{s_{n}}$
are the energies biases, $\lambda $ is the coupling constant and
$J_{s_{1}s_{2}}$, $J_{s_{1}s_{2}s_{3}}$, $J_{s_{1}s_{2}s_{3}s_{4}}$,
..., $ J_{s_{1}...s_{n}}$ are the interaction energies between two
qubits or more.
In the proposed model, the interaction energy between qubits is set
to provide the maximum ferromagnetic coupling.
Here, we consider the number of qubits is $N=9$ and
$S=\{s_{1},s_{2},...,s_{9}\}$.
To realize the dynamics of the controlled phase gates for different
qubits, we write, for simplicity, that the sub-indexes $s_{1}\equiv
s$, $s_{2}\equiv j$, $s_{3}\equiv k$ and $s_{4}\equiv l$,
$s_{5}\equiv m$, $s_{6}\equiv n$ and so on.
\\ \indent
First, we use the case of $N=2$ and $S=\{s_{1},s_{2}\}$.
Hence, the time evolution of system (\ref{z1}) in these cases is
given by $\hat{U}_{2}(t)$.
Therefore, we can realize a two-qubit dynamical controlled phase
gates  in the presence of qubits dissipation in the following
context.
\begin{itemize}
\item If $J_{1}=J_{2}=\pm(J_{11}+J_{22})$ and
$J_{12}=J_{21}=\pm\frac{1}{2}J_{1}$, we get
$\hat{U}_{2}(t)\rightarrow\hat{P}_{ee}(t;\gamma_{1},\gamma_{2})$ and
$\hat{U}_{2}(t)\rightarrow\hat{P}_{gg}(t;\gamma_{1},\gamma_{2})$.
\item If $J_{1}=\pm(J_{11}+J_{22})$, $J_{2}=-J_{1}$ and
$J_{12}=J_{21}=\pm\frac{1}{2}J_{2}$, we obtain
$\hat{U}_{2}(t)\rightarrow\hat{P}_{eg}(t;\gamma_{1},\gamma_{2})$ and
$\hat{U}_{2}(t)\rightarrow\hat{P}_{ge}(t;\gamma_{1},\gamma_{2})$.
\end{itemize}
\indent
Second, we use the case of  $N=3$ and $S=\{s_{1}, s_{2}, s_{3}\}$.
In this case, it is assumed that (i) $J_{kkj}=-J_{kjk}$,
$J_{jkk}=-\frac{1}{2}J_{jjj}$ and $J_{kj}=J_{jk}$, $\,\forall\,
j\neq k$; (ii) $F_{sjk}=J_{ss}+J_{jj}+J_{kk}$ and $J_{sjk}=w_{1}$,
$\,\forall\, s\neq j\neq k$.
Hence, after applying these steps, the time evolution of system
(\ref{z1}) is given by $\hat{U}_{3}(t)$.
So, we can realize three-qubit dynamical controlled phase gates in
the presence of qubits dissipation as follows.
\begin{itemize}
\item If $J_{s}=\pm F_{sjk}$, $J_{jk}=\pm\frac{1}{2}J_{s}$\,$(\forall\,
s\neq j\neq k)$ and $w_{1}=\pm\frac{1}{6}F_{sjk}$, we get
$\hat{U}_{3}(t)\rightarrow\hat{P}_{eee}(t;\gamma_{1},\gamma_{2},\gamma_{3})$
and
$\hat{U}_{3}(t)\rightarrow\hat{P}_{ggg}(t;\gamma_{1},\gamma_{2},\gamma_{3})$.
\item If $J_{1}=J_{2}=\pm F_{sjk}$, $J_{3}=\mp F_{sjk}$,
$J_{12}=\pm\frac{1}{2}J_{1}$, $J_{13}=J_{23}=\mp\frac{1}{2}J_{2}$
and $w_{1}=\mp\frac{1}{6}F_{sjk}$ we obtain
$\hat{U}_{3}(t)\rightarrow\hat{P}_{eeg}(t;\gamma_{1},\gamma_{2},\gamma_{3})$
and
$\hat{U}_{3}(t)\rightarrow\hat{P}_{gge}(t;\gamma_{1},\gamma_{2},\gamma_{3})$.
\item If $J_{1}=J_{3}=\pm F_{sjk}$, $J_{2}=\mp F_{sjk}$,
$J_{12}=J_{23}=\pm\frac{1}{2}J_{2}$, $J_{13}=\pm\frac{1}{2}J_{3}$
and $w_{1}=\mp\frac{1}{6}F_{sjk}$, then
$\hat{U}_{3}(t)\rightarrow\hat{P}_{ege}(t;\gamma_{1},\gamma_{2},\gamma_{3})$
and
$\hat{U}_{3}(t)\rightarrow\hat{P}_{geg}(t;\gamma_{1},\gamma_{2},\gamma_{3})$.
\item If $J_{1}=\pm F_{sjk}$, $J_{2}=J_{3}=\mp F_{sjk}$,
$J_{12}=J_{13}=\pm\frac{1}{2}J_{3}$, $J_{23}=\pm\frac{1}{2}J_{1}$
and $w_{1}=\pm\frac{1}{6}F_{sjk}$, one obtains
$\hat{U}_{3}(t)\rightarrow\hat{P}_{egg}(t;\gamma_{1},\gamma_{2},\gamma_{3})$
and
$\hat{U}_{3}(t)\rightarrow\hat{P}_{gee}(t;\gamma_{1},\gamma_{2},\gamma_{3})$.
\end{itemize}
\indent
Third, when $N=4$ and $S=\{s_{1}, s_{2}, s_{3}, s_{4}\}$, and assume
that (i) $J_{jsks}=-J_{sjsk}$ and
$J_{sjks}=-J_{jssk}$,\,$\forall\,s\neq j \neq k $;
(ii) $J_{kss}=-\frac{1}{3}J_{kkk}$, $J_{ssk}=-J_{sks}$ and
$J_{sksk}=-J_{skks}$,\,$\forall\,s\neq k$; (iii) $J_{jjkk}=0$,
$J_{jk}=J_{kj}$ and $J_{ssjk}=-J_{jkss}$,\,$\forall\, j\neq k$,
the time evolution of system (\ref{z1}) is given by
$\hat{U}_{4}(t)$.
After that, we write (i)
$E_{sjkl}=\big(J_{ss}+J_{ssss}+J_{jj}+J_{jjjj}+
J_{kk}+J_{kkkk}+J_{ll}+J_{llll}\big)$,\,$\forall\,s\neq j \neq k
\neq l$;
(ii) $w_{1}=J_{sjk} \,(s, j, k=1, 2, 3)$, $w_{2}=J_{sjk} \,(s, j,
k=1, 2, 4)$, $w_{3}=J_{sjk} \,(s, j, k=1, 3, 4)$ and $w_{4}=J_{sjk}
\,(s, j, k=2, 3, 4)$, $\,\forall\,s\neq j \neq k$;
(iii) $\frac{1}{6}E_{sjk4}=a_{1}$, $\frac{1}{6}E_{sjk3}=a_{2}$,
$\frac{1}{6}E_{sjk2}=a_{3}$ and $\frac{1}{6}E_{sjk1}=a_{4}$.
Now, one can realize the four-qubit dynamical controlled phase gates
in following.
\begin{itemize}
\item If $J_{sjkl}=\frac{1}{24}E_{sjkl}$, $J_{s}=\pm E_{sjkl}$,
$J_{jk}=\pm\frac{1}{2}J_{s}$\,$(\forall\,s\neq j\neq k)$, $w_{1}=\pm
a_{1}$, $w_{2}=\pm a_{2}$, $w_{3}=\pm a_{3}$ and $w_{4}=\pm a_{4}$,
one obtains
$\hat{U}_{4}(t)\rightarrow\hat{P}_{eeee}(t;\gamma_{1},...,\gamma_{4})$
and
$\hat{U}_{4}(t)\rightarrow\hat{P}_{gggg}(t;\gamma_{1},...,\gamma_{4})$.
\item If $J_{sjkl}=-\frac{1}{24}E_{sjkl}$, $J_{1}=J_{2}=J_{3}=\pm
E_{sjkl}$, $J_{4}=\mp E_{sjkl}$,
$J_{12}=J_{13}=J_{23}=\pm\frac{1}{2}J_{2}$,
$J_{14}=J_{24}=J_{34}=\mp\frac{1}{2}J_{3}$, $w_{1}=\pm a_{1}$,
$w_{2}=\mp a_{2}$, $w_{3}=\mp a_{3}$ and $w_{4}=\mp a_{4}$, then
$\hat{U}_{4}(t)\rightarrow\hat{P}_{eeeg}(t;\gamma_{1},...,\gamma_{4})$
and
$\hat{U}_{4}(t)\rightarrow\hat{P}_{ggge}(t;\gamma_{1},...,\gamma_{4})$.
\item If $J_{sjkl}=-\frac{1}{24}E_{sjkl}$, $J_{1}=J_{2}=J_{4}=\pm
E_{sjkl}$, $J_{3}=\mp E_{sjkl}$,
$J_{12}=J_{14}=J_{24}=\mp\frac{1}{2}J_{3}$,
$J_{13}=J_{23}=J_{34}=\pm\frac{1}{2}J_{3}$, $w_{1}=\mp a_{1}$,
$w_{2}=\pm a_{2}$, $w_{3}=\mp a_{3}$ and $w_{4}=\mp a_{4}$, obtain
$\hat{U}_{4}(t)\rightarrow\hat{P}_{eege}(t;\gamma_{1},...,\gamma_{4})$
and
$\hat{U}_{4}(t)\rightarrow\hat{P}_{ggeg}(t;\gamma_{1},...,\gamma_{4})$.
\item If $J_{sjkl}=\frac{1}{24}E_{sjkl}$, $J_{1}=J_{2}=\pm E_{sjkl}$,
$J_{3}=J_{4}=\mp E_{sjkl}$, $J_{12}=J_{34}=\pm\frac{1}{2}J_{1}$,
$J_{13}=J_{14}=J_{23}=J_{24}=\pm\frac{1}{2}J_{4}$, $w_{1}=\mp
a_{1}$, $w_{2}=\mp a_{2}$, $w_{3}=\pm a_{3}$ and $w_{4}=\pm a_{4}$,
then
$\hat{U}_{4}(t)\rightarrow\hat{P}_{eegg}(t;\gamma_{1},...,\gamma_{4})$
and $
\hat{U}_{4}(t)\rightarrow\hat{P}_{ggee}(t;\gamma_{1},...,\gamma_{4})$.
\item If $J_{sjkl}=-\frac{1}{24}E_{sjkl}$, $J_{1}=J_{3}=J_{4}=\pm
E_{sjkl}$, $J_{2}=\mp E_{sjkl}$,
$J_{12}=J_{23}=J_{24}=\mp\frac{1}{2}J_{4}$,
$J_{13}=J_{14}=J_{34}=\pm\frac{1}{2}J_{1}$, $w_{1}=\mp a_{1}$,
$w_{2}=\mp a_{2}$, $w_{3}=\pm a_{3}$ and $w_{4}=\mp a_{4}$, one
obtains
$\hat{U}_{4}(t)\rightarrow\hat{P}_{egee}(t;\gamma_{1},...,\gamma_{4})$
and
$\hat{U}_{4}(t)\rightarrow\hat{P}_{gegg}(t;\gamma_{1},...,\gamma_{4})$.
\item If $J_{sjkl}=\frac{1}{24}E_{sjkl}$, $J_{1}=J_{3}=\pm E_{sjkl}$,
$J_{2}=J_{4}=\mp E_{sjkl}$,
$J_{12}=J_{14}=J_{23}=J_{34}=\pm\frac{1}{2}J_{4}$,
$J_{13}=J_{24}=\mp\frac{1}{2}J_{4}$, $w_{1}=\mp a_{1}$, $w_{2}=\pm
a_{2}$, $w_{3}=\mp a_{3}$ and $w_{4}=\pm a_{4}$, we get
$\hat{U}_{4}(t)\rightarrow\hat{P}_{egeg}(t;\gamma_{1},...,\gamma_{4})$
and
$\hat{U}_{4}(t)\rightarrow\hat{P}_{gege}(t;\gamma_{1},...,\gamma_{4})$.
\item If $J_{sjkl}=\frac{1}{24}E_{sjkl}$, $J_{1}=J_{4}=\pm E_{sjkl}$,
$J_{2}=J_{3}=\mp E_{sjkl}$,
$J_{12}=J_{13}=J_{24}=J_{34}=\pm\frac{1}{2}J_{2}$,
$J_{14}=J_{23}=\pm\frac{1}{2}J_{4}$, $w_{1}=\pm a_{1}$, $w_{2}=\mp
a_{2}$, $w_{3}=\mp a_{3}$ and $w_{4}=\pm a_{4}$, then
$\hat{U}_{4}(t)\rightarrow\hat{P}_{egge}(t;\gamma_{1},...,\gamma_{4})$
and
$\hat{U}_{4}(t)\rightarrow\hat{P}_{geeg}(t;\gamma_{1},...,\gamma_{4})$.
\item If $J_{sjkl}=-\frac{1}{24}E_{sjkl}$, $J_{1}=\pm E_{sjkl}$,
$J_{2}=J_{3}=J_{4}=\mp E_{sjkl}$,
$J_{12}=J_{13}=J_{14}=\mp\frac{1}{2}J_{1}$,
$J_{23}=J_{24}=J_{34}=\mp\frac{1}{2}J_{2}$, $w_{1}=\pm a_{1}$,
$w_{2}=\pm a_{2}$, $w_{3}=\pm a_{3}$ and $w_{4}=\mp a_{4}$, one
obtains
$\hat{U}_{4}(t)\rightarrow\hat{P}_{eggg}(t;\gamma_{1},...,\gamma_{4})$
and
$\hat{U}_{4}(t)\rightarrow\hat{P}_{geee}(t;\gamma_{1},...,\gamma_{4})$.
\end{itemize}
In the above notations (First, Second and Third) the top sign is
used to realize the first part while the lower sign is used to
realize the second part.
\\ \indent
Using the same technique, one realizes the dynamical controlled
phase gates in the presence of the dissipation for the case of
$N=5,6,7,8$ and $9$.
The calculations of the dynamics gates when $N=5,6,7,8$ and $9$
lengthy and are not presented here.
But we present some results of application of these dynamic
controlled phase gates for different qubits in calculating the
probabilities of marked and un-marked states in the presence of
dissipation, see appendix.
If $\gamma_{i}=0, i=1,2,...,9$, one obtains the ideal dynamical
controlled phase gates.
The dynamical conditional phase gates for $N$ qubits is governed by
a dynamical control parameter phase
\begin{equation}
\beta _{N}(t)=\frac{2^{N}\theta _{N}}{\hbar}\lambda t.
\end{equation}
The parameter $\theta _{N}$ for different qubits is given by:
\begin{eqnarray*}
\begin{array}{ll}
\theta_{2}=\sum_{r=1}^{2}J_{rr}, \,\theta_{3}=\theta_{2}+J_{33},\,
\theta_{4}=\sum_{r=1}^{4}(J_{rr}+J_{rrrr}),\\
\theta_{5}=\theta_{4}+J_{55},\,
\theta_{6}=\sum_{r=1}^{6}(J_{rr}+J_{rrrr}+J_{rrrrrr}),\,
\theta_{7}=\theta_{6}+J_{77},\\
\theta_{8}=\sum_{r=1}^{8}(J_{rr}+J_{rrrr}+J_{rrrrrr}+J_{rrrrrrrr})
\,\,\textrm{and}\,\, \theta_{9}=\theta_{8}+J_{99}.
\end{array}
\end{eqnarray*}
\\ \indent
For an individual qubit and based on the available experimental data
as given in \cite{Joh1,Har}, we assume that the value of $\lambda
J_{s}=\theta _{N}$ $(s=1,2,3,...,N)$ to realize one qubit gate in
the presence of dissipation.
Now, one can realize a one-qubit gate $\hat{W}(\gamma _{s})$ in the
following context.
The time evolution of system under $\hat{H}_{s}=-\lambda J_{s}
\hat{\sigma}_{z}^{s}$ is given by $\hat{V}_{1}^{s}(t)$, and the time
evolution of system under $ \hat{H}^{s}=\exp (i\pi
\hat{\sigma}_{y}^{s}/4)\hat{H}_{s}\exp (-i\pi \hat{
\sigma}_{y}^{s}/4)-\frac{i\gamma
_{s}}{2}\,\hat{\sigma}_{+}^{s}\hat{\sigma}_{-}^{s}, $ is given by
$\hat{V}_{2}^{s}(t)$.
Then, it is concluded that
\begin{eqnarray}
\hat{W}(\gamma _{s})&=&e^{\frac{i\pi
}{2}}\hat{V}_{1}^{s}\Big(\frac{\pi \hbar}{4\theta
_{N}}\Big)\hat{V}_{2}^{s}\Big(\frac{\pi \hbar }{4\xi _{s}}
\Big)\hat{V}_{1}^{s}\Big(\frac{\pi \hbar }{4\theta _{N}}\Big) =
\frac{1}{\sqrt{2}}\exp (\frac{-\pi \gamma _{s}}{16\xi
_{s}})\Big\{(1+\frac{\gamma _{s}}{4\xi _{s}})|g_{s}\rangle \langle
g_{s}|\nonumber\\&&+\frac{\theta _{N} }{\xi _{s}}|g_{s}\rangle
\langle e_{s}|+\frac{\theta _{N}}{\xi _{s}} |e_{s}\rangle \langle
g_{s}|-(1-\frac{\gamma _{s}}{4\xi _{s}})|e_{s}\rangle \langle
e_{s}|\Big\},
\end{eqnarray}
with $4\xi _{s}=\sqrt{16\theta _{N}^{2}-\gamma _{s}^{2}}$.
\\ \indent
The current algorithm has two registers, the $N$ qubits in the first
register and one qubit in the second register.
Therefore, the proposed system contains $N+1$ qubits.
Initialize an $N+1$ qubits system to the state
$|g_{1},g_{2},...,g_{N}\rangle$ $|e_{N+1}\rangle $, where the state
in the second register refers to the excited state of an auxiliary
working qubit.
Now, we implement the dynamical quantum search algorithm in the
presence of dissipation as follows:
Performing $\hat{W}^{\otimes N}(\gamma _{1},\gamma _{2},...,\gamma
_{N})$ on the first register and performing $\hat{W}(\gamma _{N+1})$
on the second register.
Then, we apply the dynamical controlled phase gates
$\hat{P}_{x}(t;\gamma _{1},...,\gamma _{N})$ on the first register
\big( where $x\in\Lambda$ and
$\Lambda=\{g_{1}g_{2}...g_{N-1}g_{N}$,\thinspace\
$g_{1}g_{2}...g_{N-1}e_{N}$,\thinspace\
$g_{1}g_{2}...g_{N-2}e_{N-1}g_{N}$,\thinspace ...,\thinspace\ $
e_{1}e_{2}...e_{N-1}g_{N}$,\thinspace\
$e_{1}e_{2}...e_{N-1}e_{N}\}$\big) and apply $\hat{W}^{-1}(\gamma
_{N+1})$ on the second register.
Consequently, it is to observe the state of $\Lambda$ which has been
stored in the data qubits before applying the dynamical controlled
phase gates or oracles.
Also, we note that after applying this step the dynamical controlled
phase gates have  effects only on the states to be searched in the
first register and the state of the second register does not change.
So, we can omit the auxiliary working qubit or one can discard the
state of the second register at this point via normalization of this
state.
Finally, we apply a new dynamical diffusion transform matrix,
$\hat{D}(t;\gamma _{1},...,\gamma _{N})=\exp (i\beta
_{N}(t))\hat{W}^{\otimes N}(\gamma _{1},...,\gamma
_{N})\hat{P}_{g_{1}g_{2}...g_{N}}(t;\gamma _{1},...,\gamma
_{N})\hat{W}^{\otimes N}(\gamma _{1},...,\gamma _{N})$, on the first
register and measure the resulting state.
Consequently, the final formula of the dissipation DQSA is given by:
\begin{equation}
\mathop{\mathrm{DQSA}}\nolimits=\prod_{i=1}^{N-1}\Big\{\hat{D}(t;\gamma
_{1},...,\gamma _{N})\hat{P}_{x}(t;\gamma _{1},...,\gamma
_{N})\Big\}^{N-i}\hat{W}^{\otimes N}(\gamma _{1},...,\gamma
_{N})|g_{1}g_{2}...g_{N-1}g_{N}\rangle ,
\end{equation}
where $N-1$ is the number of iterations.
In the absence of dissipation, the matrix elements of $ \hat{D},$ at
any time, are $D_{ij}=D_{ji}$ for $i\neq j$ $ (i,j=1,2,...,2^{N}) $
while all diagonal elements $D_{ii}$ are equal.
Also, in the presence of qubits dissipation rates, whether similar
or different values, the matrix elements at any time are
$D_{ij}=D_{ji}$ for $i\neq j$ while all diagonal elements $D_{ii}$
are often close to each others.
\\ \indent
Now, we give the probabilities definitions of finding any marked
state $\rho_{r_{1}r_{2}...r_{N}}$ and finding unmarked states (or
remaining states) $\varrho_{j_{1}j_{2}...j_{N}}$.
The probabilities are exactly calculated in the presence of
different values of dissipation rates and different $N$, where both
of $r_{1}r_{2}...r_{N}$ and $j_{1}j_{2}...j_{N}$ are one subset of
the set $\Lambda$.
The probabilities $\rho_{r_{1}r_{2}...r_{N}}$ and
$\varrho_{j_{1}j_{2}...j_{N}}$ are give by:
\begin{eqnarray}\label{k1}
\rho_{r_{1}r_{2}...r_{N}}=\rho_{r_{1}r_{2}...r_{N}}(t;\gamma
_{1},...,\gamma _{N})= \Big|\langle
r_{1}r_{2}...r_{N}|I_{N-1}(t;\gamma _{1},...,\gamma _{N})
\hat{W}^{\otimes N}(\gamma _{1},...,\gamma
_{N})|g_{1}g_{2}...g_{N}\rangle\Big|^{2}.
\end{eqnarray}
\vspace{-1cm}
\begin{eqnarray}\label{k2}
\varrho_{j_{1}j_{2}...j_{N}}=\varrho_{j_{1}j_{2}...j_{N}}(t;\gamma
_{1},...,\gamma _{N})= \Big|\langle
j_{1}j_{2}...j_{N}|I_{N-1}(t;\gamma _{1},...,\gamma _{N})
\hat{W}^{\otimes N}(\gamma _{1},...,\gamma
_{N})|g_{1}g_{2}...g_{N}\rangle\Big|^{2},
\end{eqnarray}
where $I_{N-1}(t;\gamma _{1},...,\gamma
_{N})=\displaystyle\prod_{i=1}^{N-1}\Big\{\hat{D}(t;\gamma
_{1},...,\gamma _{N})\hat{P} _{r_{1}r_{2}...r_{N}}(t;\gamma
_{1},...,\gamma _{N})\Big\}^{N-i}$.
From Eq.\ref{k2}, in calculating the probabilities of unmarked
states, it should be noted that at least one of $j_{1}j_{2}...j_{N}$
is not equal to one of $r_{1}r_{2}...r_{N}$ for the same qubit.
Here, we write, for simplicity, that the sub-indexes in the marked
state $\rho_{e_{1}e_{2}}\equiv\rho_{ee}$ and un-marked states
($\varrho_{e_{1}g_{2}}$, $\varrho_{g_{1}e_{2}}$,
$\varrho_{g_{1}g_{2}}$ $\equiv$ Pr. r. s.); marked state
$\rho_{e_{1}g_{2}e_{3}}\equiv\rho_{ege}$ and unmarked states
($\varrho_{e_{1}e_{2}e_{3}}$, $\varrho_{e_{1}e_{2}g_{3}}$,
$\varrho_{e_{1}g_{2}g_{3}}$, $\varrho_{g_{1}e_{2}e_{3}}$,
$\varrho_{g_{1}e_{2}g_{3}}$, $\varrho_{g_{1}g_{2}e_{3}}$,
$\varrho_{g_{1}g_{2}g_{3}}$ $\equiv$ Pr. r. s.); and so on for all
states of the $N-$qubit.
Also, $\rho_{g_{1}e_{2}g_{3}e_{4}}\equiv\rho_{gege}$ and so on.
\section{Discussion}
\begin{figure*}[tbp]
\includegraphics[width=7cm,height=5cm]{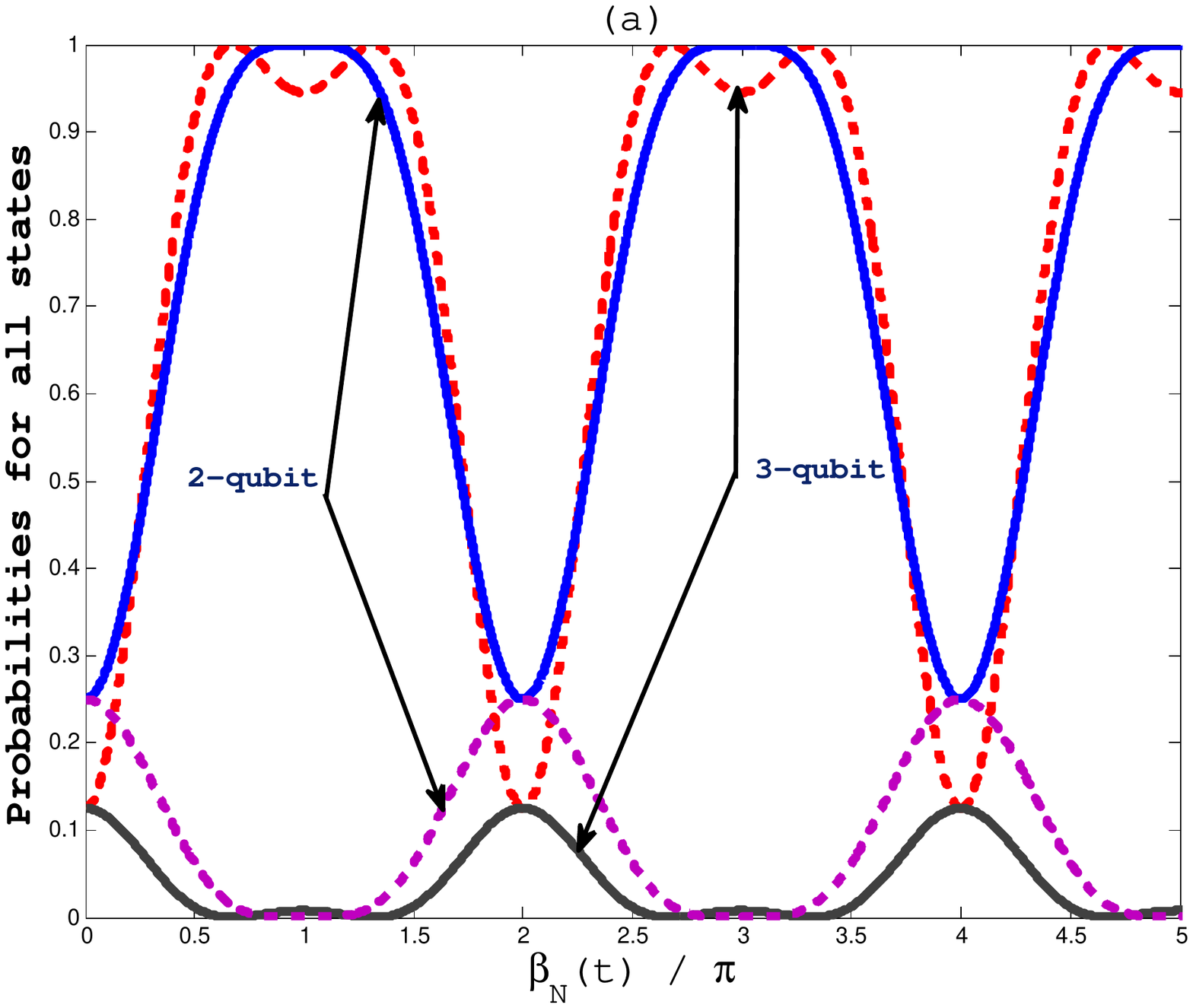}
\includegraphics[width=7cm,height=5cm]{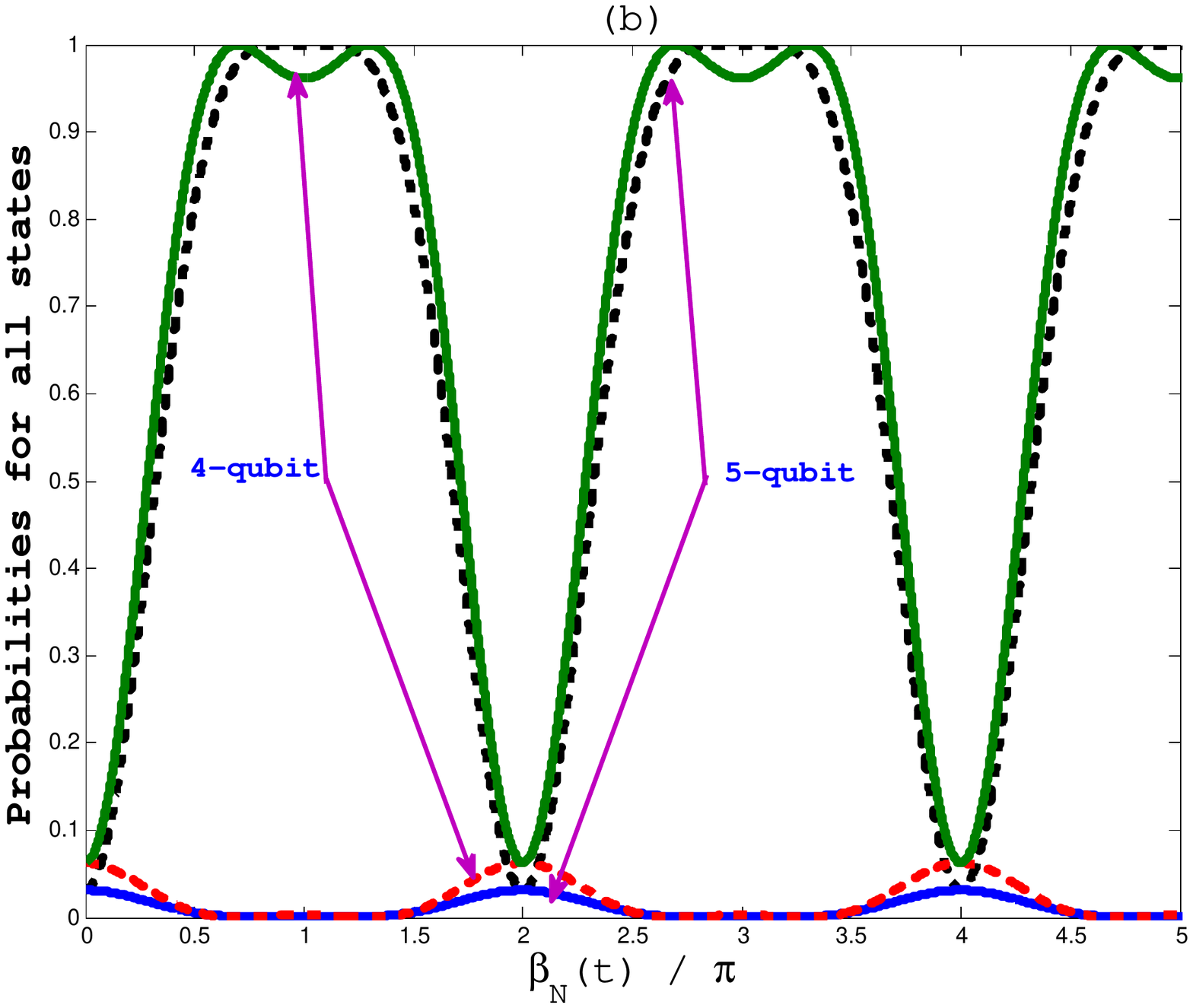}\\
\includegraphics[width=7cm,height=5cm]{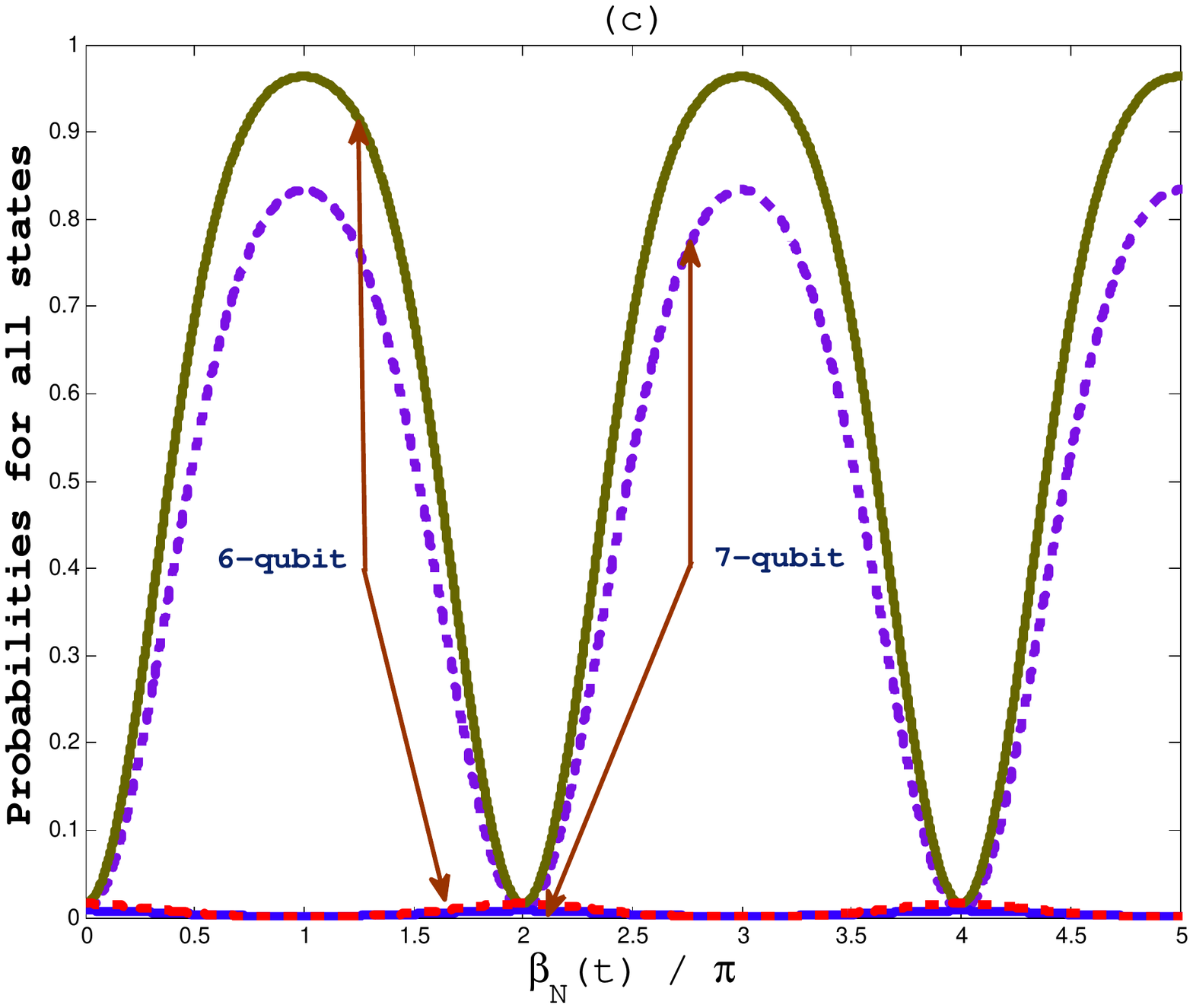}
\includegraphics[width=7cm,height=5cm]{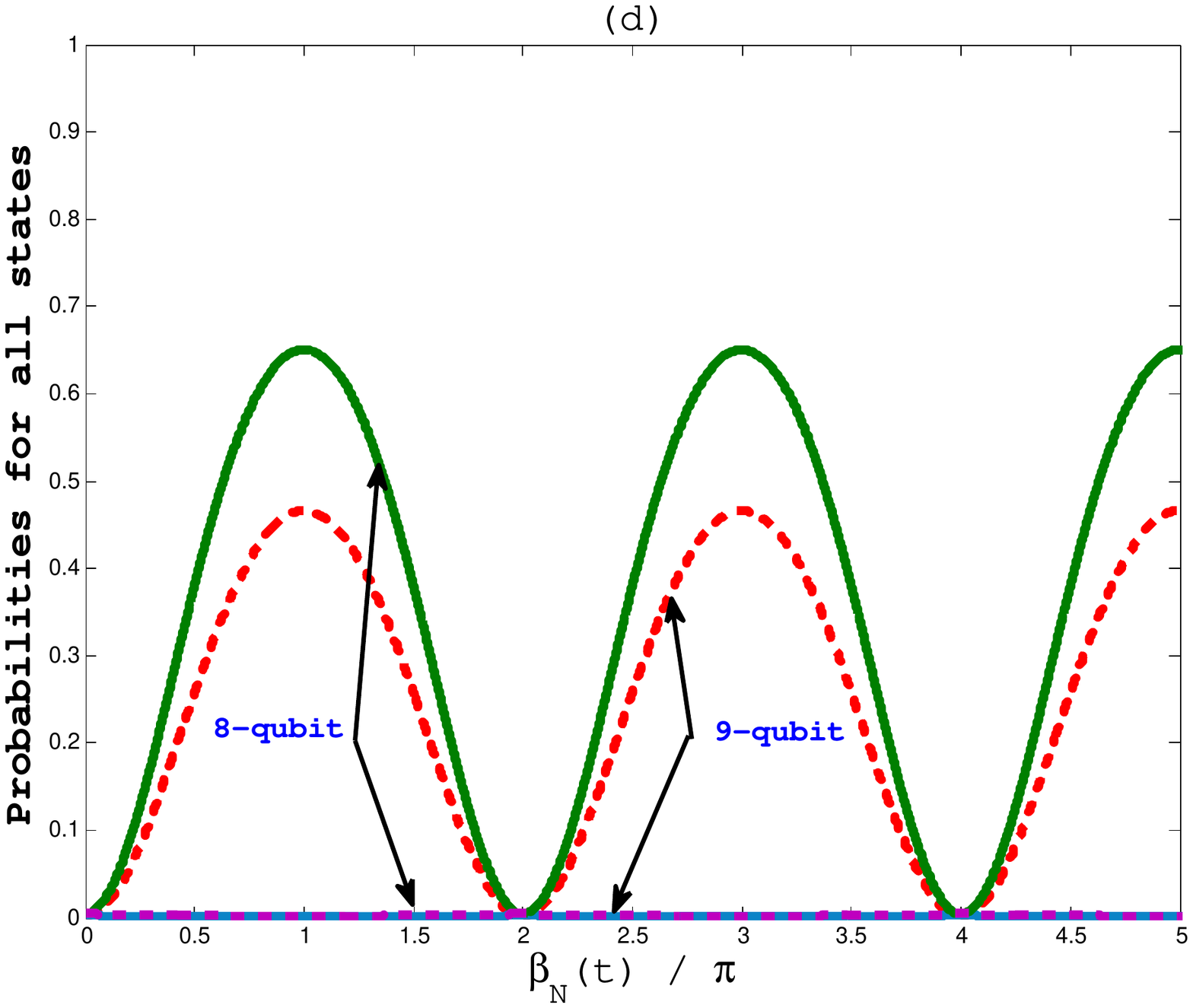} \vspace{-0.3cm}
\caption{Dynamical quantum search algorithm for the multi qubits in
the absence of dissipation for all marked states (solid lines for
$2-,4-,6-$ and $8-$qubit; dashed lines for $3-,5-,7-$ and $9-$qubit)
and all un-marked states (dashed lines for $2-,4-,6-$ and $8-$qubit;
solid lines for $3-,5-,7-$ and $9-$qubit), where $N=2,3$ in (a),
$N=4,5$ in (b), $N=6,7$ in (c) and $ N=8,9$ in (d).}\label{ff2}
\end{figure*}

\begin{table}[tbp] \centering{\renewcommand{\arraystretch}{1.4}
\renewcommand{\tabcolsep}{0.4cm} {\scriptsize
\begin{tabular}{|>{ }c|c|c|c|c|}
\hline No. & $\lambda t_{p}/\hbar$ & Present Pr. & $\lambda
t_{G}/\hbar$ & Grover Pr. \\ \hline
2 & $(0.9425\pi)/4\theta_{2}$ & 1 & $\pi/4\theta_{2}$ & 1
\\ \hline
3 & $(0.6723\pi)/8\theta_{3}$ & 1 & $\pi/8\theta_{3}$ & 0.9453
\\ \hline
4 & $(0.6933\pi)/16\theta_{4}$ & 1 & $\pi/16\theta_{4}$ & 0.9613
\\ \hline
5 & $(0.8661\pi)/32\theta_{5}$ & 1 & $\pi/32\theta_{5}$ & 0.9992
\\ \hline
6 & $(0.9899\pi)/64\theta_{6}$ & 0.9635 & $\pi/64\theta_{6}$ &
0.9635 \\ \hline
7 & $(0.9906\pi)/128\theta_{7}$ & 0.8335 & $\pi/128\theta_{7}$ &
0.8335 \\ \hline
8 & $(0.9906\pi)/256\theta_{8}$ & 0.6503 & $\pi/256\theta_{8}$ &
0.6503 \\ \hline
9 & $(0.995\pi)/512\theta_{9}$ & 0.4662 & $\pi/512\theta_{9}$ &
0.4662 \\ \hline
\end{tabular}
}}
\vspace{-0.1cm}\caption{A comparison of the time ($t_{p}$) of the
present algorithm and Grover time ($t_{G}$) for finding higher
probability (Pr.) of any marked state of different qubits in the
case of $\protect\gamma _{i}=0,\,i=1,2,...,9$.} \label{ff1}
\end{table}
 \indent
In table (\ref{ff1}), we compare our proposed algorithm (OA) results
with Grover algorithm's results.
It is shown that the probability reach $100\%$ for the two-qubit
case, in both: our case and GA case, but the time needed in our
case, $ t_{p},$ is shorter than the Grover time, $t_{G},$ (say, $
t_{p}=0.9425\pi \hbar /4\lambda \theta _{2}$ and $t_{G}=\pi \hbar
/4\lambda \theta _{2}$).
This means that the efficiency of our algorithm is higher than the
efficiency of GA. Once we increase the number of involved qubits, we
see that a significant difference between OA and GA, is seen.
The probability of finding any marked state using 3 qubits, 4 qubits
and 5 qubits in OA is $100\%$ while the corresponding case using GA
does not exceed $ 94.53\%$ for 3 qubits, $96.13\%$ for 4 qubits and
$99.92\%$ for 5 qubits.
Moreover, the times needed of finding any marked state are shorter
than Grover times.
This is an interesting advantage of the proposed algorithm since the
sensitivity is much higher than the Grover sensitivity to the number
of qubits.
Also, we can say that using OA, exact observation is clearly
obtained even whether the number of involved qubits gets higher (say
5) and one still have $100\%$ probability of finding any marked
states.
When we consider the number of qubits is greater than or equal 6, it
is shown that the same probability of finding any marked state is
obtained using OA and GA, but the time needed to obtain such
probability is shorter in our case, which is still gives OA an
advantage.
It is interesting to observe here that as the number of qubits is
increased the probability find any desired state still coincide
using OA while in Grover case there does not coincide, where from
$1-6$ qubits the results do not coincide with the fact that
increasing the number of qubits leads to decreasing the probability.
Moreover, as the number is increased further, GA shows the logic
behavior.
In these calculations, we expect that our results (times and
probabilities) are almost the same as GA for the large number of
qubits (say, $N>12$).
The reason of such agreement between our results and Grover results
is that when $N$ is increased the time for finding the highest
probability of any marked state is increased to reach almost
$\frac{\pi \hbar }{2^{13}\lambda\theta _{13}}$ for finding the
probability of any marked state when $N=13$.
After that, one obtains the highest probability for finding any
desired states for the different qubits at times $\frac{\pi
\hbar}{2^{r}\lambda\theta _{r}} \,(r=13,14,...,N)$ , this case is
compatible with Grover case.
This observation shows that using the dynamical quantum search
algorithm has some advantage, where GA is obtained as a special case
from the proposed general algorithm.
\\ \indent
In Fig.\ref{ff2} we plot the probabilities of finding any marked or
unmarked states as a function of dynamical phase $\beta _{N}(t)/\pi
,\,N=2,3,...,9$, in the absence of dissipation.
It is shown that for 2, 3, 4 and 5 qubits, see Fig.\ref{ff2}a and
Fig.\ref{ff2}b, the probabilities of marked states reach one at
several points rather than Grover observation, where GA shows the
maximum probability at $\beta _{N}(t)=\pi $ only.
In 3 qubits and 4 qubits cases, we find that the behavior of curves
peaks are between the two maxima there is a local minimum.
While 2 and 5 qubits cases the maximum stays longer as a function of
the phase, which means that the long lived maximum probability is
obtained for 2 and 5 qubits compared with 3 and 4 qubits cases where
the maximum stays for a short period.
This observation is no longer existing once the number of qubits is
increased (say $\geq 6$), where the maxima probability always less
than one, see Fig.\ref{ff2}c and Fig.\ref{ff2}d.
Also, the unmarked states are observed for the low number of qubits
and reach the maximum of $25\%$ for 2 qubits case, while this value
is decreased once the number of qubits is increased.
This phenomenon becomes more pronounced once the number of qubits
exceeds $6$ after this number of qubits the probabilities are very
close to zero.
From our further calculations (which are not presented here), the
probabilities of unmarked states vanish completely when the number
of qubits is greater than 20 qubits.
\begin{figure*}[tbp]
\includegraphics[width=7cm,height=5cm]{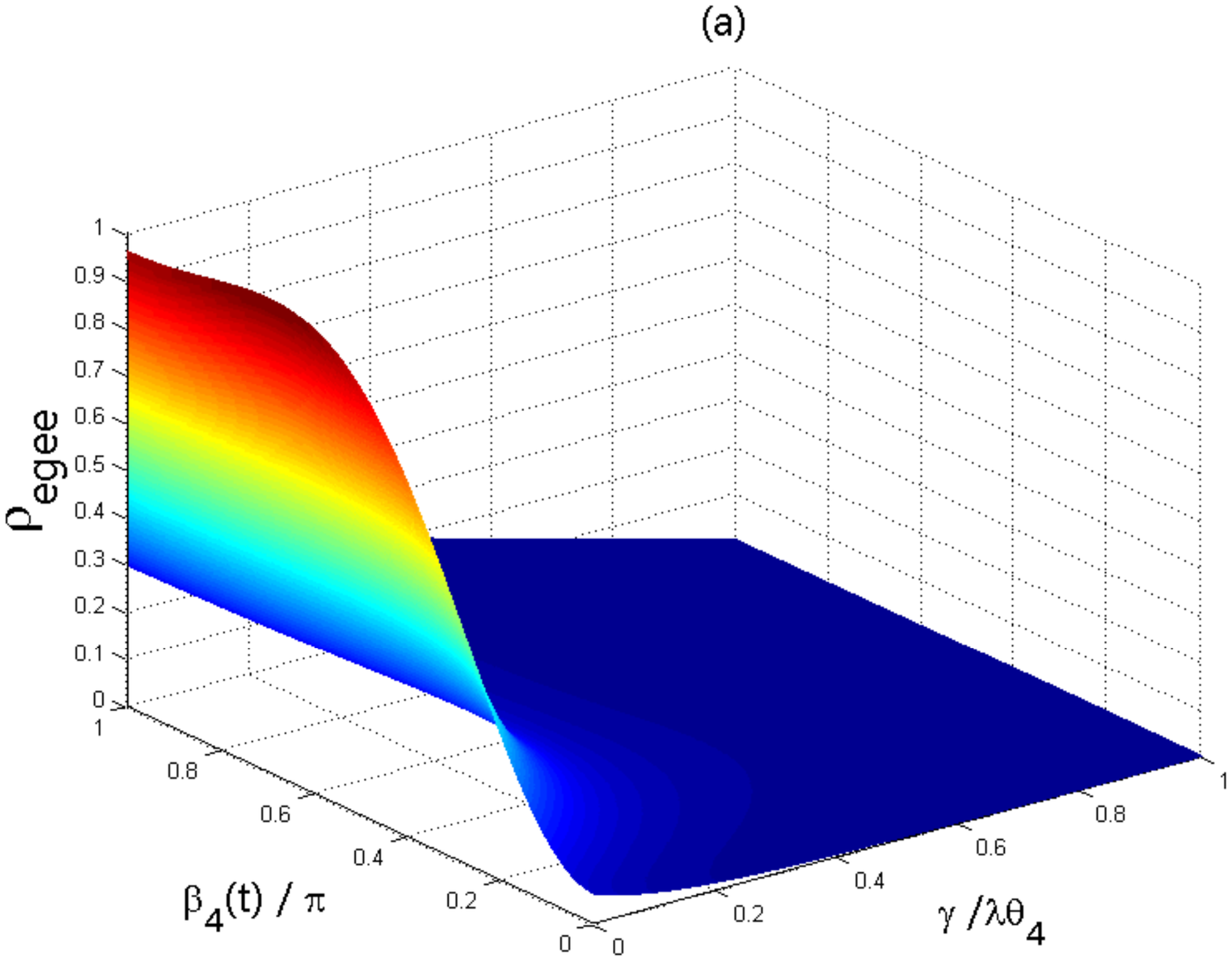}
\includegraphics[width=7cm,height=5cm]{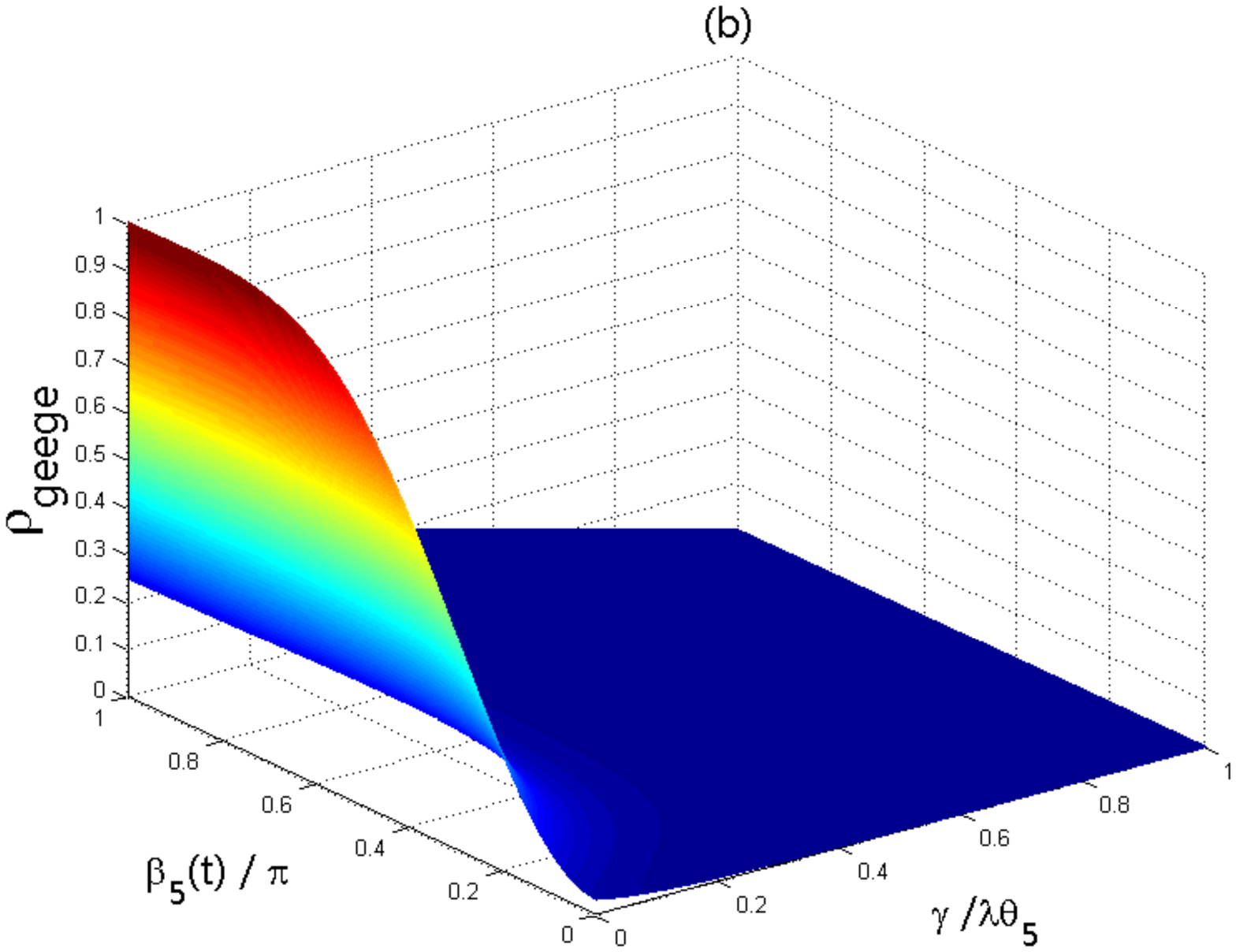}\\
\includegraphics[width=7cm,height=5cm]{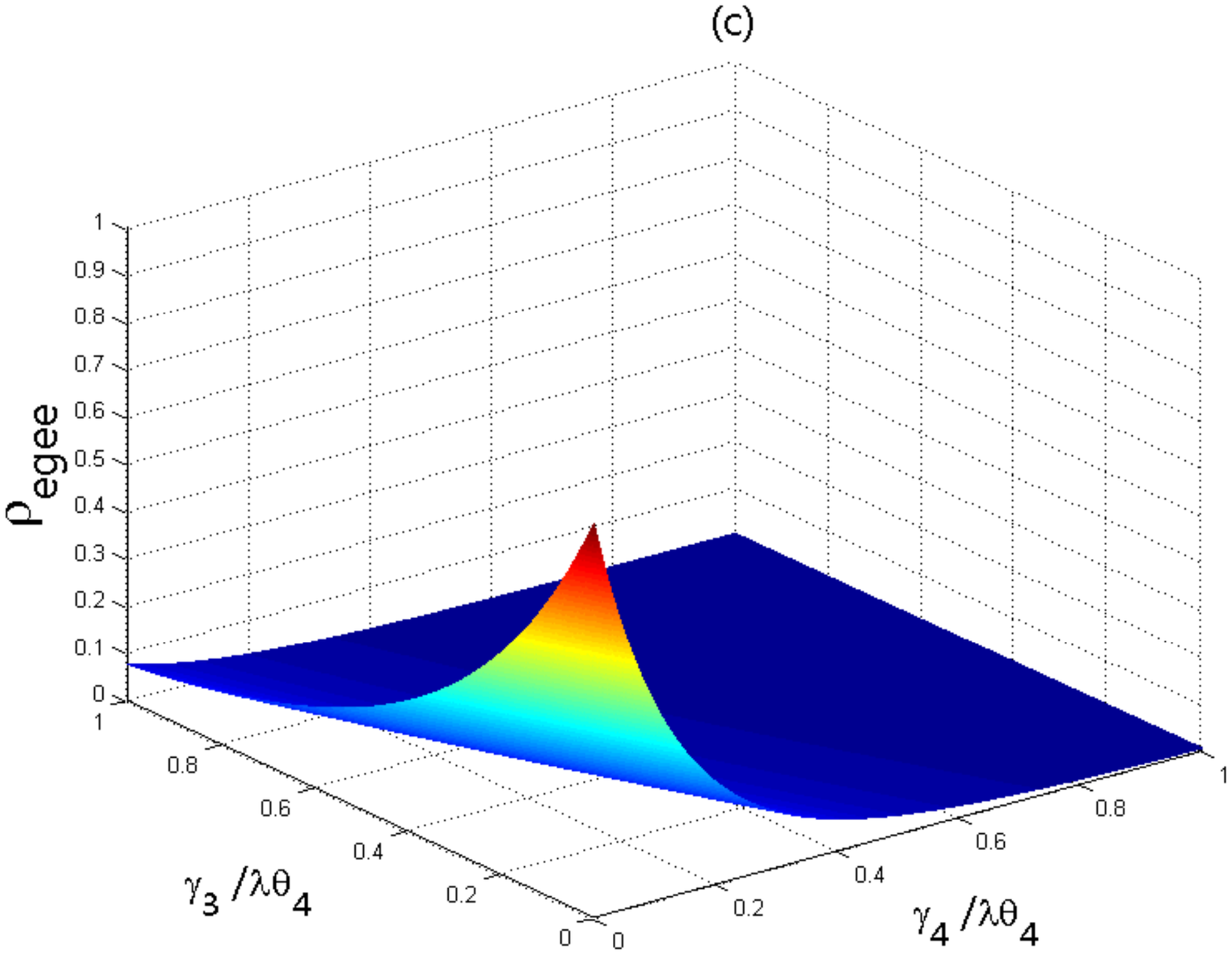}
\includegraphics[width=7cm,height=5cm]{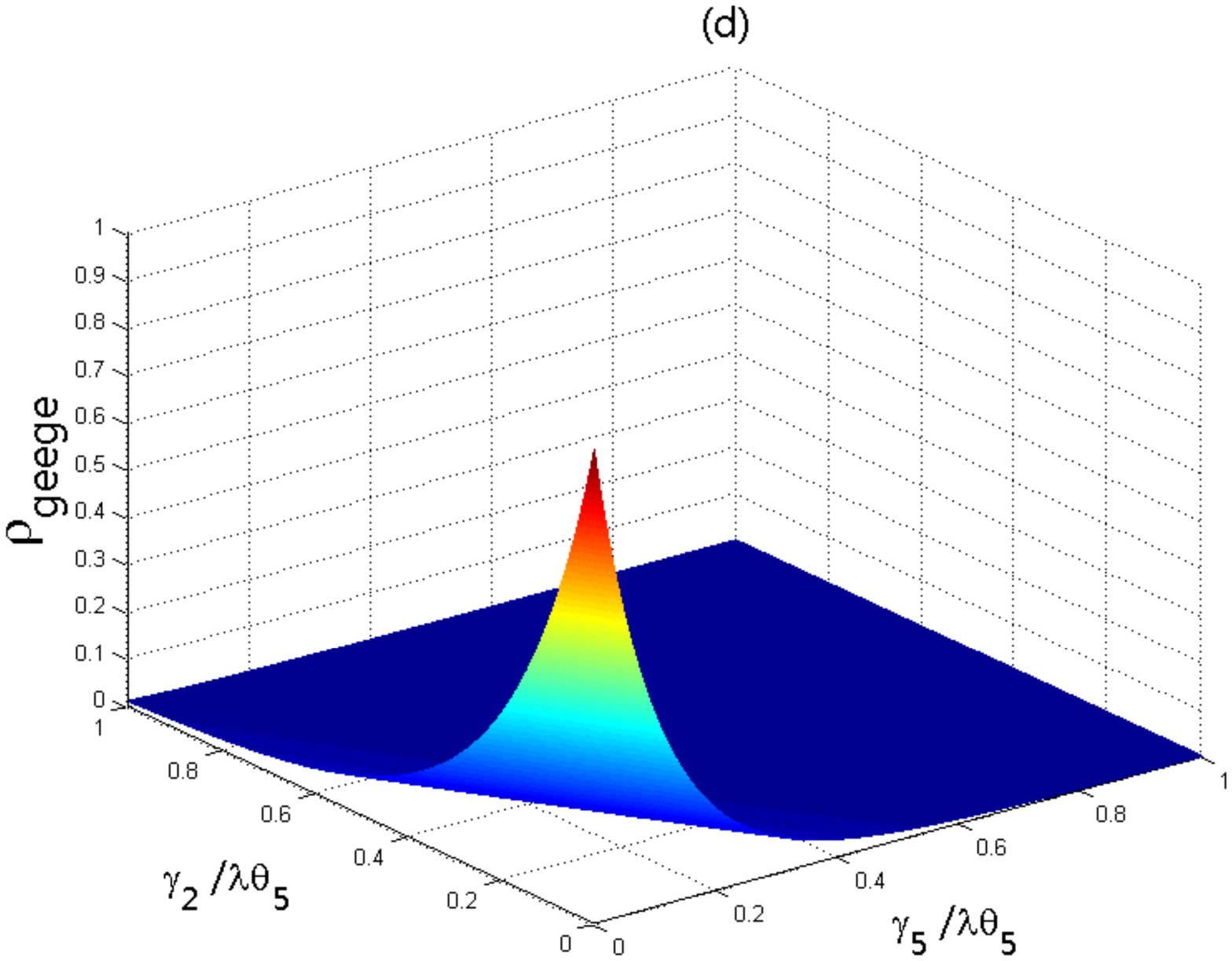} \vspace{-0.3cm}
\caption{Dynamical quantum search algorithm for the probabilities of
marked states $\rho_{egee}$ and $\rho_{geege} $ in the presence of
dissipation, where $\protect\gamma_{i}=\protect\gamma \,(i=1,...,4)$
in (a); $\protect\gamma_{j}=\protect\gamma \,(j=1,...,5)$ in (b);
$\protect \gamma _{1}=\protect\gamma_{2}=\protect\gamma _{4}$ and
$t_{p}=\frac{0.02813 \protect\pi \hbar}{\protect\lambda\theta _{4}}$
in (c); $\protect\gamma_{1}= \protect\gamma_{3}=\protect\gamma
_{5}$, $\protect\gamma_{2}=\protect \gamma_{4}$ and
$t_{p}=\frac{0.027065\protect\pi\hbar}{\protect\lambda\theta_{5}}$
in (d).}\label{ff3}
\end{figure*}
\begin{figure*}[tbp]
\includegraphics[width=9cm,height=6.5cm]{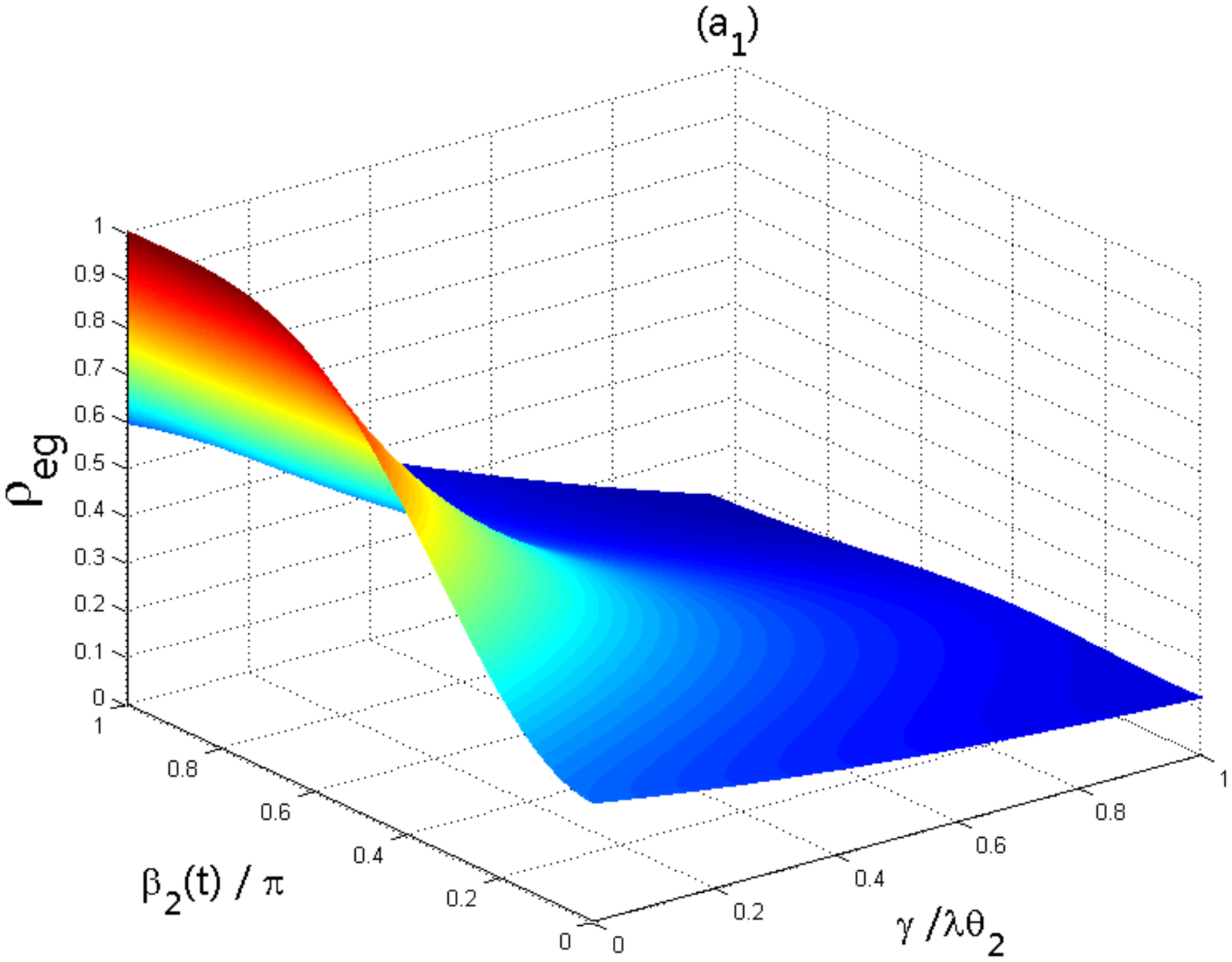}
\includegraphics[width=9cm,height=6.5cm]{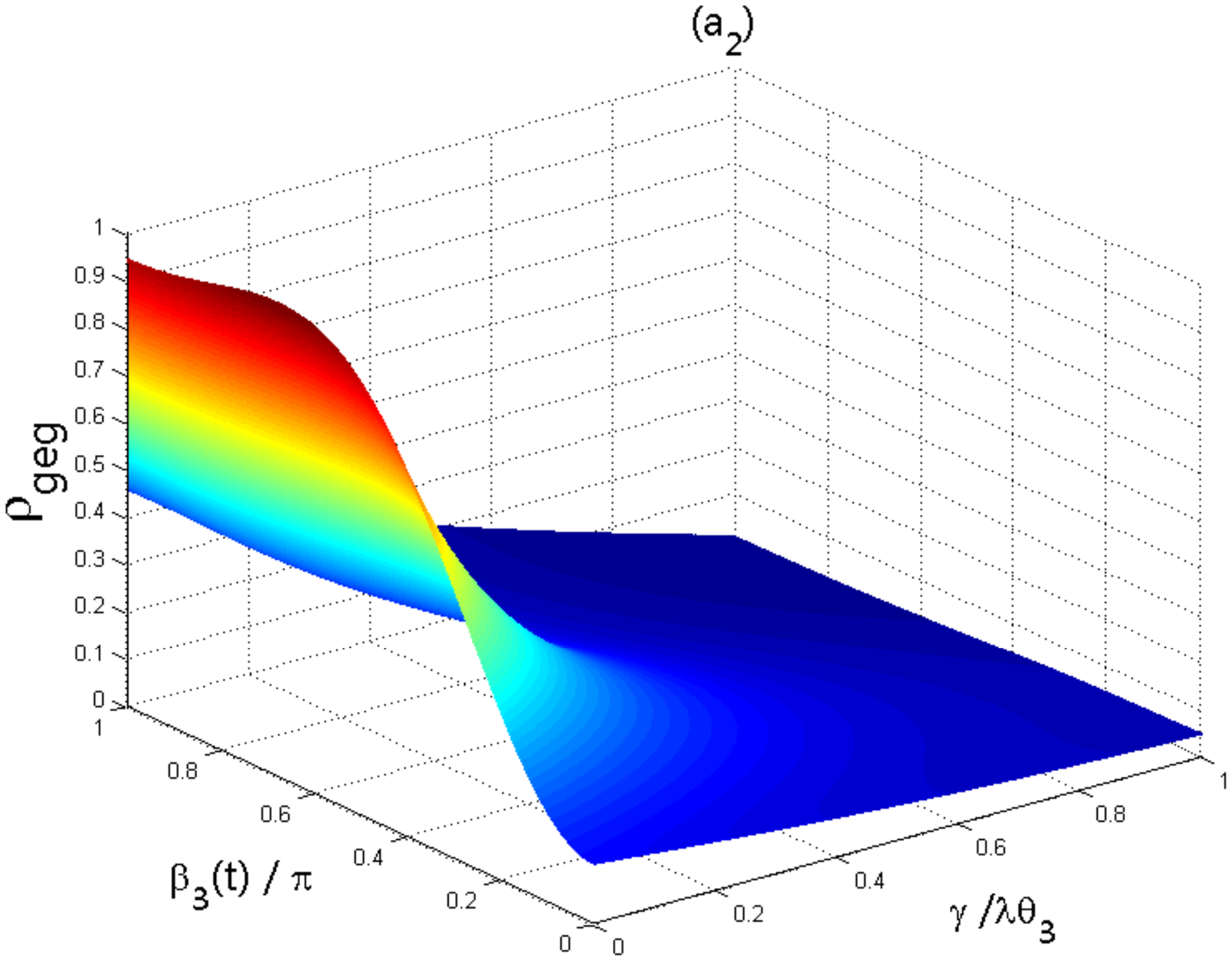}\\
\includegraphics[width=9cm,height=6.5cm]{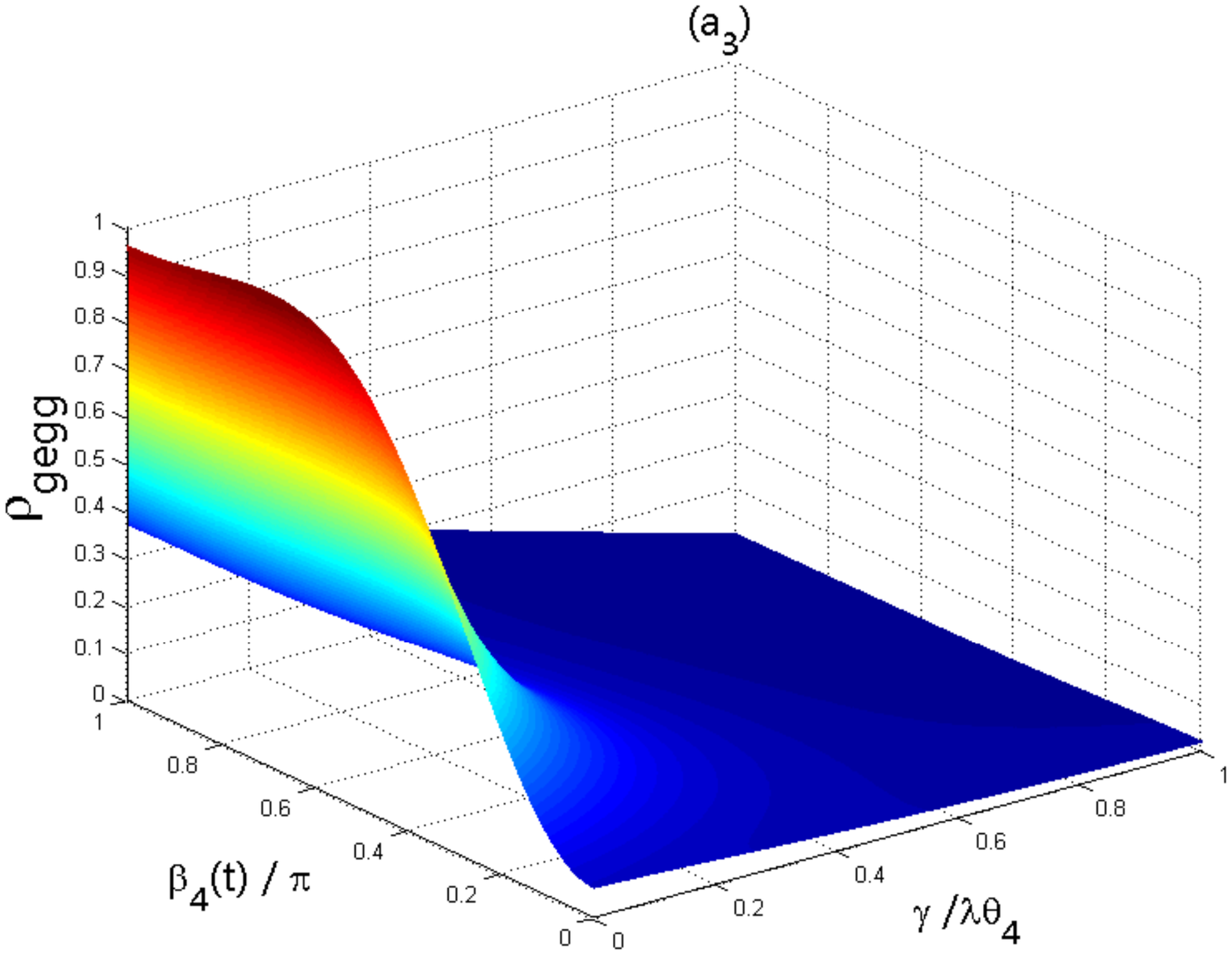}
\includegraphics[width=9cm,height=6.5cm]{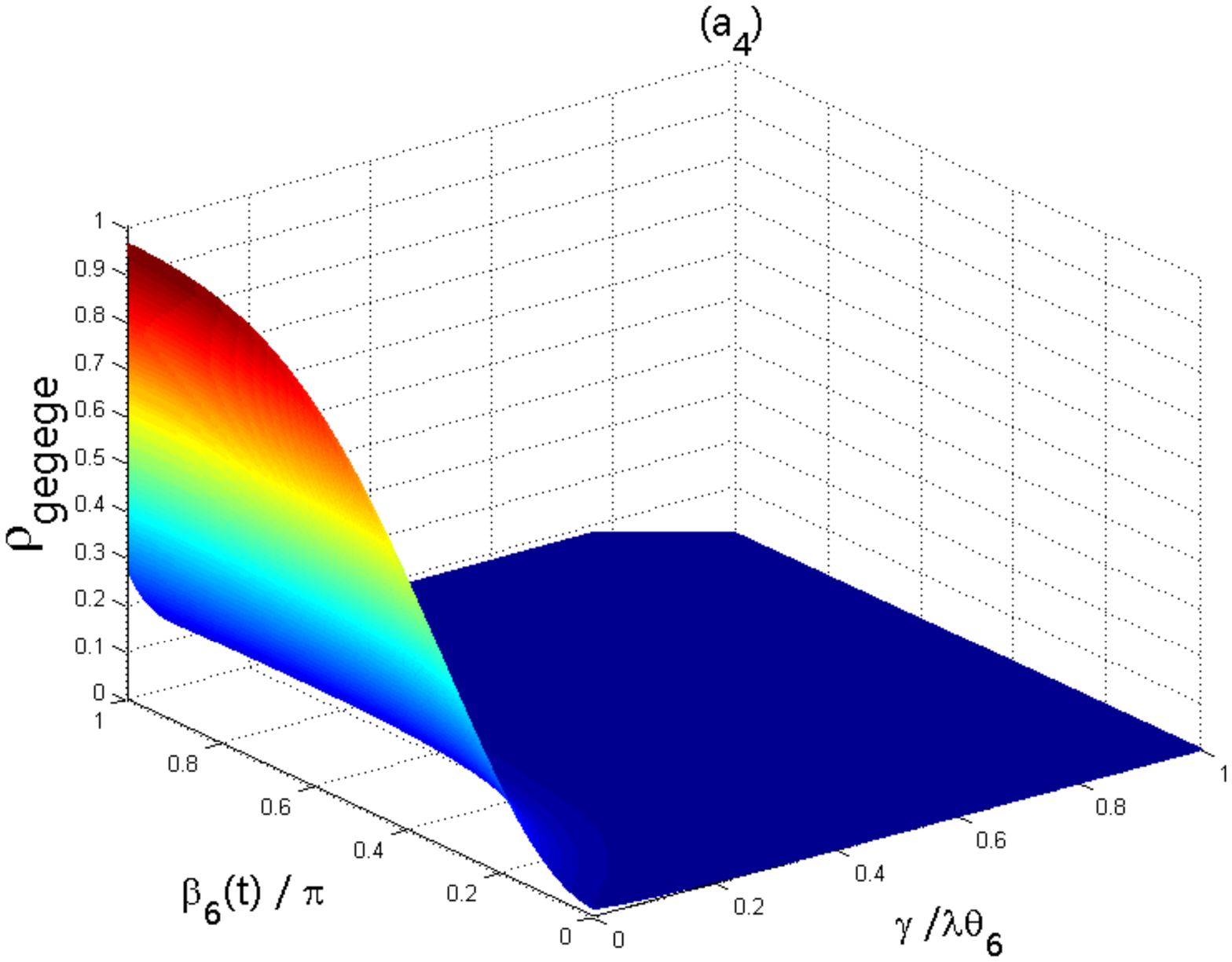}
\caption{Dynamical quantum search algorithm for the probabilities of
marked states $\rho_{eg}$, $\rho_{geg}$, $\rho_{gegg}$ and
$\rho_{gegege}$ in the presence of similar dissipation
rates.}\label{f1}
\end{figure*}
\begin{figure*}[tbp]
\includegraphics[width=9cm,height=6.5cm]{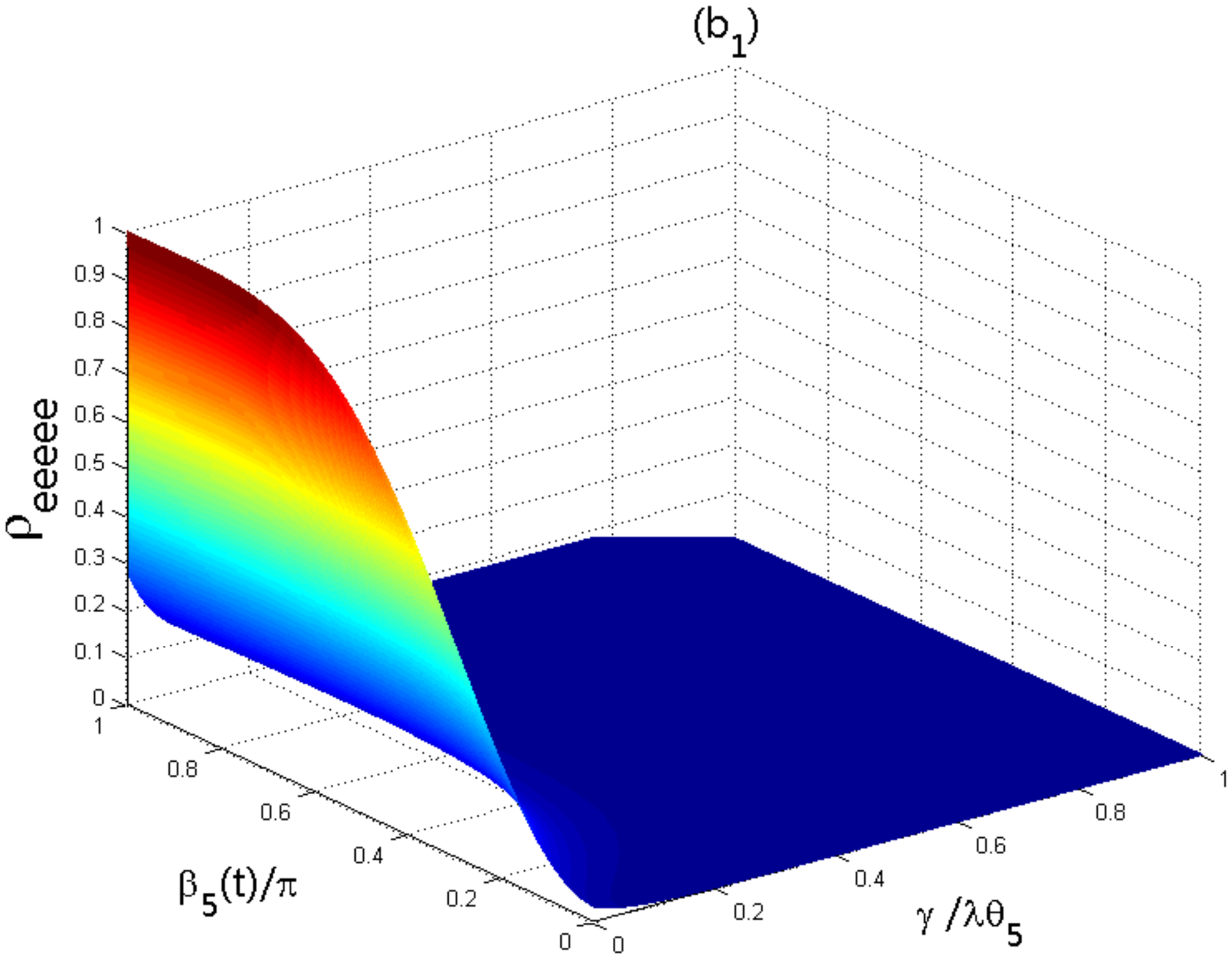}
\includegraphics[width=9cm,height=6.5cm]{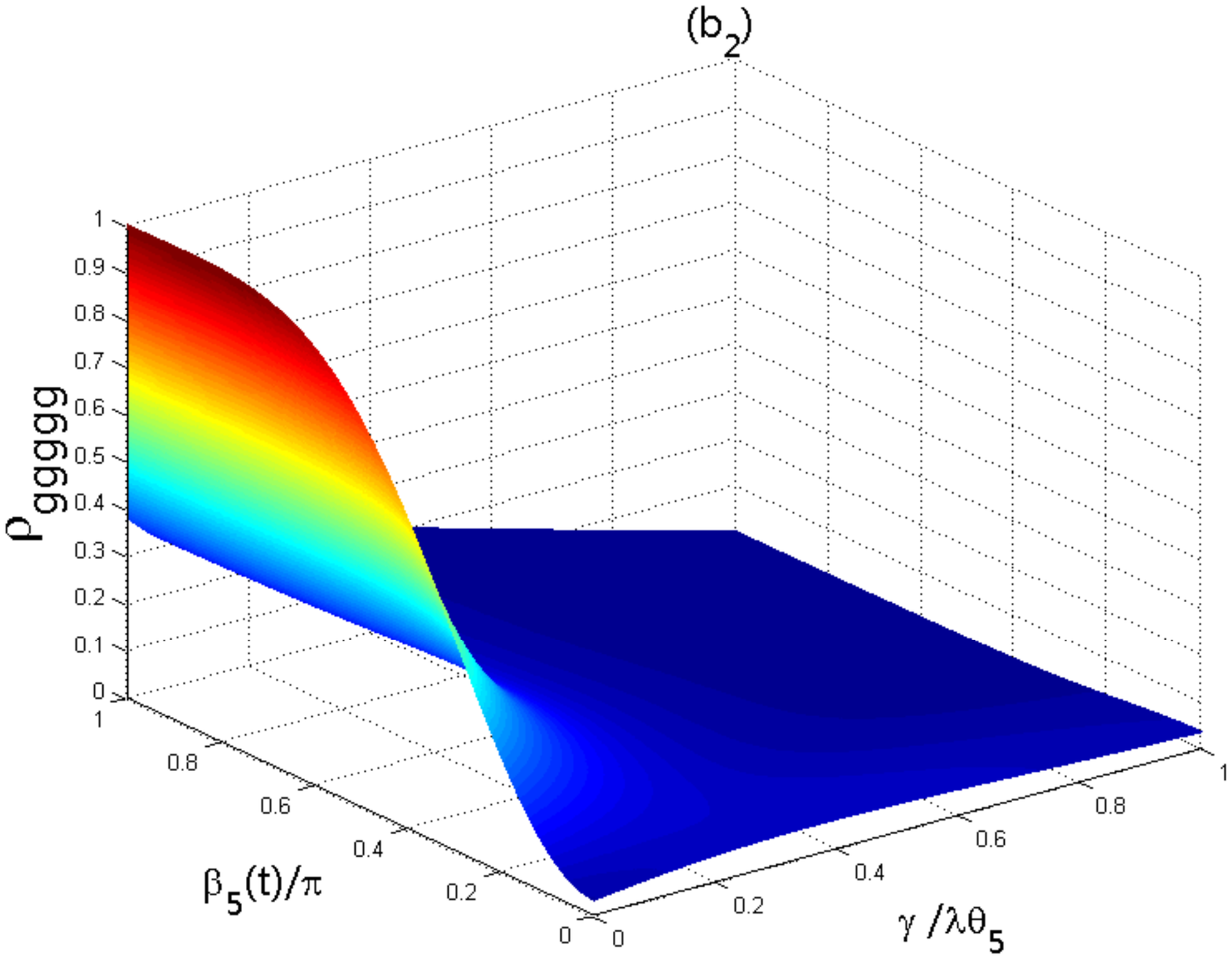}
\caption{Dynamical quantum search algorithm with dissipation for 5
qubits to compare the values of $\rho_{eeeee}$ and
$\rho_{ggggg}$.}\label{f2}
\end{figure*}
\\ \indent
In Fig.\ref{ff3}, we display the behavior of the dynamical search of
the probabilities of marked states $\rho_{egee}$ and $\rho_{geege}$
with the same dissipation rates.
In the absence of dissipation, see Fig.\ref{ff2}b and table
(\ref{ff1}), the probabilities of the current marked states or any
other marked states at times $t_{p}=0.04333\pi \hbar/\lambda\theta
_{4} $ for 4 qubits and $t_{p}=0.027065\pi \hbar/\lambda\theta _{5}
$ for 5 qubits are $100\%$ and the probabilities of un-marked states
are zero.
It is shown that when times are decreased from these values, the
probabilities values of finding any marked states are less than
$100\%$ and decrease more and more according to the decreasing
times.
With these time decreases, the probabilities of finding any marked
states are decreased and this decreasing is added to the un-marked
states probabilities.
This means that the revivals for the probabilities of un-marked
states when the time is decreased more and more to become larger
when the time is close to zero.
However, when the time reaches zero, the probabilities of any marked
or un-marked states are $ 6.25\%$ and $3.125\%$ for 4 and 5 qubits,
respectively.
This means that both marked and un-marked states have the same
chance to contribute.
In the presence of weak dissipation rate, one sees that the
probabilities of current marked states in Fig.\ref{ff3}a and
Fig.\ref{ff3}b or any other marked states at different values of the
time are gradually decreased.
The probabilities values of any desired state for any number of $N$
at a weak values of similar dissipation are often close to each
others.
If the dissipation value is taken to be
$\gamma_{i}=\lambda\theta_{N}$ $(i=1,2,..., N)$, the probabilities
values of any desired state are sometimes increased or decreased to
each other and one can not distinguish between the probability of
any marked and un-marked states at any time.
However, it is difficult to observe the probability of any marked
state for any qubit at large dissipation rates, where small values
of the probability of marked state is shown.
Which means that both marked and un-marked states have the same
chance to contribute.
\\ \indent
Also, we display the behavior of the probabilities of current marked
states in Fig.\ref{ff3}c and Fig.\ref{ff3}d for different values of
qubits dissipation rates with time $ t_{p}=0.02813\pi
\hbar/\lambda\theta _{4}$ for 4 qubits and $ t_{p}=0.027065\pi \hbar
/\lambda\theta _{5}$ for 5 qubits.
In the absence of dissipation the probabilities of any marked state
for 4 qubits and 5 qubits at these values of times start from 0.8332
and from 1, respectively.
When a weak value of qubits-dissipation is considered, one finds
that the probabilities of the current states in Fig.\ref{ff3}c and
Fig.\ref{ff3}d and any other marked states for 4 and 5 qubits are
gradually decreased.
The values of any desired states for 4 and 5 qubits in weak
dissipation are often close to each other.
It is shown that for strong dissipation, the probabilities of any
marked states for 4 and 5 qubits are decreased further, while one
can not distinguish between different probabilities of some
un-marked states.
It is interesting to mention here that the observation of any
desired state will be more difficult when dissipation rate is equal
to $\gamma_{3}=\gamma_{4}=\lambda\theta_{4}$  for 4 qubits and
$\gamma_{2}=\gamma_{5}=\lambda\theta_{5}$  for 5 qubits or more than
that, where there is some probabilities of un-marked states are
larger than marked state.
This means that both marked and un-marked states have the same
chance to contribute.
\\ \indent
However, the behavior of the probability of finding any remaining
marked state for any different number of $N$ in the dissipation
presence, is similar to the previous behavior, see Fig.\ref{f1}.
It is shown that, the time is a main factor in the appearance of
different probabilities of marked states using different number of
qubits.
Therefore, the probabilities for the remaining marked states (3
states for 2 qubits, 7 states for 3 qubits, 14 states for 4 qubits,
29 states for 5 qubits and 63 states for 6 qubits) have the same
behavior as in the figures displayed in the above.
In Fig.\ref{f2}, we plot DQSA for the probabilities
$\Scale[1]\rho_{eeeee}$ and $\Scale[1]\rho_{ggggg}$ when the
dissipation rates is considered.
It is shown that the probabilities $\Scale[1]\rho_{ggggg}$ and
$\Scale[1]\rho_{eeeee}$ are gradually decreased once dissipation
rate takes small values, and the values of $\Scale[1]\rho_{eeeee}$
are less than the values of $\Scale[1]\rho_{ggggg}$ at any time.
Also, one observes that the probability $\Scale[1]\rho_{ggggg}$ is
shown as  larger values compared with the probability
$\Scale[1]\rho_{eeeee}$ which shows small values once dissipation
rate $\gamma/\lambda\theta_{5}$ is increased.
Therefore, we can distinguish the probability
$\Scale[1]\rho_{gg...g}$ for multi qubits at any time once
dissipation rate is decreased or increased (or varied).
\\ \indent
Now, we give a short discussion to observe  un-marked states in the
presence of dissipation rates either similar or different.
The probabilities of un-desired states for weak dissipation are
distinguished by a very small values and non-zero compared with the
absence of dissipation.
While for strong dissipation, the probability of the desired state
is becoming less and is decreased due to the influence of strong
dissipation and some of this decrease is added to the un-desired
states.
On the other hand, the results of finding any marked and un-marked
states, in table (\ref{ff1}) are completely in agreement with the
results shown in Fig.\ref{ff3} and Fig.\ref{f1}, under the same
conditions.
\section{Conclusions}
We have demonstrated on-demand generation of dynamic multi-qubit
quantum search algorithm, which can be maintained for multi-qubit
system.
The algorithm we have demonstrated, gives favorable performance for
multi-qubit (up to 9 qubits) and would survive even in the presence
of the qubit dissipation.
We have focused our attention on the probabilities of the first 9
qubits and have obtained that their optimal values come either to
diagonalizing or to symmetrizing probabilities, that implies
analytical formula for its calculation.
Comparing these results with Grover algorithm analysis allows to
give a similar result when the number of involved qubits get much
larger, and is an example of a prefect observation which is suitable
for a particular purpose, e.g., for the time dependent interaction
and dissipation processes.
The long life time of this multi-qubit algorithm is due to a high
symmetry of the generated forms.
We have used different values of the controlled phase to improve the
suggested algorithm performance.
A new technique is used to generate the dynamical quantum gates that
helps in implementing the current algorithm.
The multi-qubit character of the interaction used for the creation
of dynamical algorithm allows for the potential applications and
extension of the present method to other models, possibly including,
superconducting materials or solid state models with long lived spin
states.
\section*{Appendix}
In this appendix, we took four optional values of times (say,
$\frac{0.331\pi\hbar}{2^{N}\lambda\theta_{N}}$,
$\frac{0.566\pi\hbar}{2^{N}\lambda\theta_{N}}$, $t_{Np}$ and
$\frac{\pi\hbar}{2^{N}\lambda\theta_{N}}$) to calculate the
probabilities of some marked and unmarked states for different
number of $N$.
The time $t_{Np}$ gives the highest probabilities for marked states
with different values of $N$.
\begin{table}[tbp] {\renewcommand{\arraystretch}{1.5}
\renewcommand{\tabcolsep}{0.1cm} {\tiny
\hspace{1.8cm}\begin{tabular}{|>{ }c|c|c|c|c|} \hline Times &
$t=\frac{0.331\pi\hbar}{4\lambda\theta_{2}}$ &
$t=\frac{0.566\pi\hbar}{4\lambda\theta_{2}}$
& $t_{2p}=\frac{0.9425\pi\hbar}{4\lambda\theta_{2}}$ & $t=\frac{\pi\hbar}{4\lambda\theta_{2}}$  \\
\noalign{\hrule height 2pt}
$\rho_{ee}$ &   0.5583  &  0.8537    &   0.9625   &   0.9618 \\
\hline
Pr. r. s. &
\pbox{5cm}{\vspace{0.1cm}0.1417, 0.1411, 0.1427}&
\pbox{5cm}{\vspace{0.1cm}0.0407, 0.0405, 0.0397}&
\pbox{5cm}{\vspace{0.1cm}  1.5605e, 1.5524e, 1.7800d}&
\pbox{5cm}{\vspace{0.1cm}7.7440d, 7.5690d, 0}
\\ \noalign{\hrule height 2pt}
$\rho_{ge}$ & 0.5634  &   0.8595  &  0.9668  &  0.9659 \\
\hline
Pr. r. s. &
\pbox{5cm}{\vspace{0.1cm}0.1389, 0.1405, 0.1428}&
\pbox{5cm}{\vspace{0.1cm} 0.0399, 0.0390, 0.0399}&
\pbox{5cm}{\vspace{0.1cm} 1.5556e, 1.6690d, 2.5570d}&
\pbox{5cm}{\vspace{0.1cm} 7.5690d, 0, 1.0000c}
\\ \noalign{\hrule height 2pt}
$\rho_{gg}$ & 0.5697  &  0.8667  &  0.9722   & 0.9712  \\
\hline
Pr. r. s. &
\pbox{5cm}{\vspace{0.1cm} 0.1378, 0.1404, 0.1401}&
\pbox{5cm}{\vspace{0.1cm} 0.0382, 0.0388, 0.0390}&
\pbox{5cm}{\vspace{0.1cm} 1.6490d, 9.6400c, 2.5700d}&
\pbox{5cm}{\vspace{0.1cm} 0, 1.0000c, 1.0000c}
\\ \noalign{\hrule height 2pt}
\end{tabular}}}
\caption{\footnotesize Outputs of dynamical quantum search algorithm
for some probabilities of marked  states $\rho_{r_{1}r_{2}}$\,
($r_{1},r_{2}=e,g$) and unmarked states or remaining states (Pr. r.
s.) of 2-qubit for different values of time, where
$\frac{\gamma_{1}}{\lambda}=\frac{\theta_{2}}{113}$ and
$\frac{\gamma_{2}}{\lambda}=\frac{\theta_{2}}{90}$;
$c=\times10^{-6}$, $d=\times10^{-5}$ and $e=\times10^{-4}$.}
\label{s1}
\end{table}
 \begin{table}[tbp]
{\renewcommand{\arraystretch}{1.5} \renewcommand{\tabcolsep}{0.1cm}
{\tiny
\hspace{1.6cm}\begin{tabular}{|>{ }c|c|c|c|c|} \hline Times &
$t=\frac{0.331\pi\hbar}{4\lambda\theta_{2}}$ &
$t=\frac{0.566\pi\hbar}{4\lambda\theta_{2}}$
& $t_{2p}=\frac{0.9425\pi\hbar}{4\lambda\theta_{2}}$ & $t=\frac{\pi\hbar}{4\lambda\theta_{2}}$  \\
\noalign{\hrule height 2pt}
$\rho_{ee}$ &  0.0961 &   0.1224  &  0.1142  &  0.1108 \\
\hline
Pr. r. s. &
\pbox{5cm}{\vspace{0.1cm}0.0921, 0.0933, 0.1792}&
\pbox{5cm}{\vspace{0.1cm}0.0625, 0.0632, 0.0680}&
\pbox{5cm}{\vspace{0.1cm} 0.0428, 0.0434, 0.0049}&
\pbox{5cm}{\vspace{0.1cm}0.0426, 0.0432, 0.0043 }
\\ \noalign{\hrule height 2pt}
$\rho_{ge}$ & 0.1683  &   0.1700 &  0.1148  &  0.1089 \\
\hline
Pr. r. s. &
\pbox{5cm}{\vspace{0.1cm} 0.0528, 0.0699, 0.1660}&
\pbox{5cm}{\vspace{0.1cm} 0.0491, 0.0248, 0.0486}&
\pbox{5cm}{\vspace{0.1cm} 0.0333, 0.0041, 5.7600e}&
\pbox{5cm}{\vspace{0.1cm} 0.0329, 0.0040, 3.1329e}
\\ \noalign{\hrule height 2pt}
$\rho_{gg}$ &  0.3244  &   0.2640  & 0.1098   & 0.0997  \\
\hline
Pr. r. s. &
\pbox{5cm}{\vspace{0.1cm}0.0308, 0.0713, 0.0712 }&
\pbox{5cm}{\vspace{0.1cm}0.0127, 0.0251, 0.0239 }&
\pbox{5cm}{\vspace{0.1cm} 0.0021, 3.2962e, 3.4306e }&
\pbox{5cm}{\vspace{0.1cm}  0.0021, 1.7640d, 9.2160d}
\\ \noalign{\hrule height 2pt}
\end{tabular}}}
\caption{\footnotesize The same as Table (\ref{s1}) but
$\frac{\gamma_{1}}{\lambda}=\frac{4\theta_{2}}{5}$ and
$\frac{\gamma_{2}}{\lambda}=\frac{7\theta_{2}}{9}$.} \label{s2}
\end{table}
\begin{table}[tbp]
{\renewcommand{\arraystretch}{1.5} \renewcommand{\tabcolsep}{0.1cm}
{\tiny
\begin{tabular}{|>{ }c|c|c|c|c|} \hline Times &
$t=\frac{0.331\pi\hbar}{8\lambda\theta_{3}}$ &
$t=\frac{0.566\pi\hbar}{8\lambda\theta_{3}}$
& $t_{3p}=\frac{0.6723\pi\hbar}{8\lambda\theta_{3}}$ & $t=\frac{\pi\hbar}{8\lambda\theta_{3}}$  \\
\noalign{\hrule height 2pt}
$\rho_{eee}$ & 0.5989 &  0.9085  & 0.9348  &  0.8776\\
\hline
Pr. r. s. &
\pbox{5cm}{\vspace{0.1cm} 0.0520, 0.0524, 0.0516, 0.0522,\\ 0.0516,
0.0517, 0.0518}&
\pbox{5cm}{\vspace{0.1cm}0.0052, 0.0053, 0.0046, 0.0052,\\ 0.0047,
0.0046, 0.0044 }&
\pbox{5cm}{\vspace{0.1cm} 1.0345e, 1.1197e, 4.1220d, 1.0625e,\\
4.5460d, 3.7480d,  2.5210d}&
\pbox{5cm}{\vspace{0.1cm} 0.0064, 0.0063, 0.0064, 0.0064,\\ 0.0065,
0.0064, 0.0063}
\\ \noalign{\hrule height 2pt}
$\rho_{eeg}$ & 0.6051  &  0.9170 &  0.9435  &  0.8855 \\
\hline
Pr. r. s. &
\pbox{5cm}{\vspace{0.1cm}  0.0515, 0.0512, 0.0522, 0.0511,\\ 0.0520,
0.0513, 0.0515}&
\pbox{5cm}{\vspace{0.1cm}0.0052, 0.0046, 0.0049, 0.0047,\\ 0.0048,
0.0044, 0.0042}&
\pbox{5cm}{\vspace{0.1cm} 9.5290d, 3.6000d, 5.9300d, 4.1780d,\\
5.6250d, 2.1130d, 1.3220d }&
\pbox{5cm}{\vspace{0.1cm}  0.0068, 0.0068, 0.0065, 0.0069,\\ 0.0066,
0.0067, 0.0065 }
\\ \noalign{\hrule height 2pt}
$\rho_{ege}$ & 0.6085   &  0.9218  &  0.9483  &  0.8898 \\
\hline
Pr. r. s. &
\pbox{5cm}{\vspace{0.1cm} 0.0517, 0.0509, 0.0517, 0.0509,\\ 0.0511,
0.0519, 0.0513}&
\pbox{5cm}{\vspace{0.1cm}  0.0053, 0.0046, 0.0046, 0.0046,\\ 0.0044,
0.0046, 0.0041 }&
\pbox{5cm}{\vspace{0.1cm} 9.8960d, 3.4340d, 3.4490d, 3.0920d,\\
1.9240d, 3.6170d, 8.8400c }&
\pbox{5cm}{\vspace{0.1cm} 0.0069, 0.0070, 0.0067, 0.0070,\\ 0.0069,
0.0067, 0.0067}
\\ \noalign{\hrule height 2pt}
$\rho_{gee}$ & 0.6065 &  0.9191 &  0.9454 &  0.8874\\
\hline
Pr. r. s. &
\pbox{5cm}{\vspace{0.1cm}  0.0516, 0.0510, 0.0511, 0.0512,\\ 0.0518,
0.0522, 0.0514 }&
\pbox{5cm}{\vspace{0.1cm}  0.0052, 0.0047, 0.0046, 0.0044,\\ 0.0047,
0.0048, 0.0041}&
\pbox{5cm}{\vspace{0.1cm}9.6260d, 4.0000d, 3.2500d, 2.0500d,\\
4.4680d, 5.0500d, 1.1250d }&
\pbox{5cm}{\vspace{0.1cm}  0.0069, 0.0069, 0.0069, 0.0068,\\ 0.0067,
0.0065, 0.0066}
\\ \noalign{\hrule height 2pt}
$\rho_{gge}$ & 0.6162 &  0.9325 &  0.9590 & 0.8998 \\
\hline
Pr. r. s. &
\pbox{5cm}{\vspace{0.1cm}0.0504, 0.0505, 0.0513, 0.0507,\\ 0.0515,
0.0507, 0.0515}&
\pbox{5cm}{\vspace{0.1cm} 0.0046, 0.0044, 0.0046, 0.0041,\\ 0.0048,
0.0041, 0.0041 }&
\pbox{5cm}{\vspace{0.1cm} 2.4820d, 1.5170d, 3.0440d, 6.6100c,\\
4.1810d, 7.7600c, 9.9700c }&
\pbox{5cm}{\vspace{0.1cm}0.0075, 0.0074, 0.0071, 0.0072,\\ 0.0071,
0.0072, 0.0069}
\\ \noalign{\hrule height 2pt}
\end{tabular}}}
\caption{\footnotesize Outputs of dynamical quantum search algorithm
for some probabilities of marked  states
$\rho_{r_{1}r_{2}r_{3}}$\,$(r_{1},r_{2},r_{3}=e,g)$ and unmarked
states of 3-qubit for different values of time, where
$\frac{\gamma_{1}}{\lambda}=\frac{\theta_{3}}{113}$,
$\frac{\gamma_{2}}{\lambda}=\frac{\theta_{3}}{90}$ and
$\frac{\gamma_{3}}{\lambda}=\frac{\theta_{3}}{140}$.} \label{s3}
\end{table}
 \begin{table}[tbp]
{\renewcommand{\arraystretch}{1.5} \renewcommand{\tabcolsep}{0.1cm}
{\tiny
\begin{tabular}{|>{ }c|c|c|c|c|} \hline Times &
$t=\frac{0.331\pi\hbar}{8\lambda\theta_{3}}$ &
$t=\frac{0.566\pi\hbar}{8\lambda\theta_{3}}$
& $t_{3p}=\frac{0.6723\pi\hbar}{8\lambda\theta_{3}}$ & $t=\frac{\pi\hbar}{8\lambda\theta_{3}}$  \\
\noalign{\hrule height 2pt}
$\rho_{eee}$ &  0.0217 &  0.0234  &  0.0201 & 0.0113 \\
\hline
Pr. r. s. &
\pbox{5cm}{\vspace{0.1cm} 0.0183, 0.0201, 0.0257, 0.0204,\\ 0.0261,
0.0289, 0.0596 }&
\pbox{5cm}{\vspace{0.1cm}0.0129, 0.0140, 0.0088, 0.0142,\\ 0.0089,
0.0094, 0.0044}&
\pbox{5cm}{\vspace{0.1cm}0.0093, 0.0100, 0.0054, 0.0101,\\ 0.0054,
0.0055, 4.8420e  }&
\pbox{5cm}{\vspace{0.1cm} 0.0033, 0.0034, 0.0026, 0.0034,\\ 0.0026,
0.0023, 0.0051}
\\ \noalign{\hrule height 2pt}
$\rho_{eeg}$ & 0.0317  & 0.0297  &  0.0247  &  0.0135 \\
\hline
Pr. r. s. &
\pbox{5cm}{\vspace{0.1cm} 0.0145, 0.0160, 0.0256, 0.0162,\\ 0.0261,
0.0238, 0.0539}&
\pbox{5cm}{\vspace{0.1cm} 0.0109, 0.0085, 0.0085, 0.0086,\\ 0.0087,
0.0050, 0.0027}&
\pbox{5cm}{\vspace{0.1cm} 0.0072, 0.0051, 0.0051, 0.0051,\\ 0.0052,
0.0024, 2.6240d}&
\pbox{5cm}{\vspace{0.1cm} 0.0018, 7.9524e, 0.0023, 7.7284e,\\
0.0023, 9.2416e, 0.0030 }
\\ \noalign{\hrule height 2pt}
$\rho_{ege}$ &  0.0356  &  0.0327  &  0.0268  &  0.0140 \\
\hline
Pr. r. s. &
\pbox{5cm}{\vspace{0.1cm} 0.0152, 0.0146, 0.0241, 0.0161,\\ 0.0207,
0.0282, 0.0520 }&
\pbox{5cm}{\vspace{0.1cm} 0.0111, 0.0075, 0.0071, 0.0081,\\ 0.0042,
0.0084, 0.0022}&
\pbox{5cm}{\vspace{0.1cm}0.0072, 0.0043, 0.0040, 0.0045,\\ 0.0019,
0.0047, 2.3050d}&
\pbox{5cm}{\vspace{0.1cm}0.0017, 5.8564e, 0.0016, 4.7089e,\\
7.2900e, 0.0016, 0.0026 }
\\ \noalign{\hrule height 2pt}
$\rho_{gee}$ & 0.0362 &  0.0332  & 0.0271  &  0.0141  \\
\hline
Pr. r. s. &
\pbox{5cm}{\vspace{0.1cm} 0.0153, 0.0145, 0.0159, 0.0203,\\ 0.0243,
0.0279, 0.0518 }&
\pbox{5cm}{\vspace{0.1cm} 0.0112, 0.0074,0.0079, 0.0040,\\ 0.0070,
0.0082, 0.0022}&
\pbox{5cm}{\vspace{0.1cm} 0.0073, 0.0042, 0.0044, 0.0018,\\ 0.0039,
0.0045, 3.1760d}&
\pbox{5cm}{\vspace{0.1cm} 0.0016, 5.4289e, 4.4521e, 7.0225e,\\
0.0015, 0.0015, 0.0025}
\\ \noalign{\hrule height 2pt}
$\rho_{gge}$ & 0.0737 & 0.0571  &  0.0448 &  0.0213 \\
\hline
Pr. r. s. &
\pbox{5cm}{\vspace{0.1cm}  0.0105, 0.0097, 0.0193, 0.0152,\\ 0.0192,
0.0153, 0.0420}&
\pbox{5cm}{\vspace{0.1cm}0.0045, 0.0026, 0.0087, 0.0010,\\ 0.0085,
9.8881e, 1.7730e}&
\pbox{5cm}{\vspace{0.1cm}0.0017, 7.9322e, 0.0041, 2.4912e,\\ 0.0039,
2.3850e, 9.4625e  }&
\pbox{5cm}{\vspace{0.1cm}2.7040d, 5.1840d, 1.4161e,\\ 4.8400c,
1.2100e, 5.7600c, 0.0014}
\\ \noalign{\hrule height 2pt}
\end{tabular}}}
\caption{\footnotesize The same as Table (\ref{s3}) but
$\frac{\gamma_{1}}{\lambda}=\frac{4\theta_{3}}{5}$,
$\frac{\gamma_{2}}{\lambda}=\frac{7\theta_{3}}{9}$ and
$\frac{\gamma_{3}}{\lambda}=\frac{5\theta_{3}}{8}$.} \label{s4}
\end{table}
 \begin{table}[tbp]
{\renewcommand{\arraystretch}{1.5} \renewcommand{\tabcolsep}{0.1cm}
{\tiny
\begin{tabular}{|>{ }c|c|c|c|c|} \hline Times &
$t=\frac{0.331\pi\hbar}{16\lambda\theta_{4}}$ &
$t=\frac{0.566\pi\hbar}{16\lambda\theta_{4}}$
& $t_{4p}=\frac{0.6933\pi\hbar}{16\lambda\theta_{4}}$ & $t=\frac{\pi\hbar}{16\lambda\theta_{4}}$  \\
\noalign{\hrule height 2pt}
$\rho_{eeee}$ & 0.5433  &  0.8570  & 0.8900  & 0.8499 \\
\hline
Pr. r. s. &
\pbox{5cm}{\vspace{0.1cm} 0.0269, 0.0267, 0.0259, 0.0269,\\ 0.0258,
0.0259, 0.0256, 0.0268,\\ 0.0259, 0.0259, 0.0256, 0.0258,\\ 0.0255,
0.0256, 0.0253}&
\pbox{5cm}{\vspace{0.1cm} 0.0032, 0.0032, 0.0030, 0.0033,\\ 0.0030,
0.0030, 0.0030, 0.0032,\\ 0.0030, 0.0030, 0.0030, 0.0030,\\ 0.0030,
0.0030, 0.0029 }&
\pbox{5cm}{\vspace{0.1cm} 6.5530d, 5.7800d, 5.3210d, 6.9200d,\\
5.7250d, 5.4500d,  6.3610d, 6.3370d,\\ 5.5210d, 5.2520d, 6.1450d,
5.5210d,\\ 6.5810d,  6.2820d, 7.2650d}&
\pbox{5cm}{\vspace{0.1cm} 0.0016, 0.0016, 0.0018, 0.0016,\\ 0.0018,
0.0018, 0.0019, 0.0016,\\ 0.0018, 0.0018, 0.0019, 0.0018,\\ 0.0019,
0.0019, 0.0020 }
\\ \noalign{\hrule height 2pt}
$\rho_{eegg}$ &  0.5611 & 0.8840  &   0.9178  &  0.8767 \\
\hline
Pr. r. s. &
\pbox{5cm}{\vspace{0.1cm} 0.0260, 0.0262, 0.0265, 0.0257,\\ 0.0254,
0.0255, 0.0260, 0.0257,\\ 0.0254, 0.0255, 0.0259, 0.0254,\\ 0.0251,
0.0252, 0.0248}&
\pbox{5cm}{\vspace{0.1cm} 0.0028, 0.0029, 0.0030, 0.0028,\\ 0.0027,
0.0028, 0.0029, 0.0028,\\ 0.0027, 0.0028, 0.0029, 0.0027,\\ 0.0027,
0.0027, 0.0027}&
\pbox{5cm}{\vspace{0.1cm} 1.3220d, 2.5780d, 2.8090d, 1.8010d,\\
2.3840d, 2.1730d, 4.4690d, 1.6850d,\\ 2.3050d, 2.0450d, 3.9290d,
2.3050d,\\ 3.0250d, 2.7250d, 3.5050d  }&
\pbox{5cm}{\vspace{0.1cm} 0.0020, 0.0019, 0.0018, 0.0020,\\ 0.0021,
0.0021, 0.0020, 0.0020,\\ 0.0021, 0.0021, 0.0020, 0.0021,\\ 0.0022,
0.0022, 0.0023}
\\ \noalign{\hrule height 2pt}
$\rho_{eggg}$ &  0.5729 &  0.9019  &  0.9364  & 0.8947   \\
\hline
Pr. r. s. &
\pbox{5cm}{\vspace{0.1cm} 0.0257, 0.0254, 0.0255, 0.0260,\\ 0.0254,
0.0256, 0.0259, 0.0254,\\ 0.0251, 0.0252, 0.0249, 0.0251,\\ 0.0248,
0.0249, 0.0253 }&
\pbox{5cm}{\vspace{0.1cm}  0.0027, 0.0026, 0.0026, 0.0028,\\ 0.0026,
0.0027, 0.0028, 0.0026,\\ 0.0026, 0.0026, 0.0026, 0.0026,\\ 0.0026,
0.0026, 0.0027 }&
\pbox{5cm}{\vspace{0.1cm} 3.9200c, 6.4800c, 5.4500c, 1.8180d,\\
6.8500c, 1.6290d, 1.8180d, 6.1300c,\\ 9.2500c, 8.4100c, 1.2010d,
1.0130d,\\ 1.4050d, 1.3010d, 2.8250d}&
\pbox{5cm}{\vspace{0.1cm}0.0022, 0.0022, 0.0022, 0.0021,\\ 0.0023,
0.0022, 0.0021, 0.0022,\\ 0.0023, 0.0023, 0.0024, 0.0023,\\ 0.0024,
0.0024, 0.0023}
\\ \noalign{\hrule height 2pt}
$\rho_{geeg}$ & 0.5629  & 0.8867  &  0.9206  & 0.8795 \\
\hline
Pr. r. s. &
\pbox{5cm}{\vspace{0.1cm}  0.0259, 0.0263, 0.0257, 0.0254,\\ 0.0256,
0.0253, 0.0254, 0.0251,\\ 0.0265, 0.0254, 0.0257, 0.0254,\\ 0.0259,
0.0251,  0.0248 }&
\pbox{5cm}{\vspace{0.1cm} 0.0028, 0.0029, 0.0028, 0.0027,\\ 0.0028,
0.0027, 0.0027, 0.0027,\\ 0.0030, 0.0027, 0.0028, 0.0027,\\ 0.0029,
0.0027, 0.0027 }&
\pbox{5cm}{\vspace{0.1cm} 1.1450d, 2.5480d, 1.4450d, 1.9010d,\\
1.6180d, 2.1530d, 1.9720d, 2.5610d,\\ 2.6330d, 1.8500d, 3.3650d,
2.0250d,\\ 4.2530d, 2.5000d, 3.1610d }&
\pbox{5cm}{\vspace{0.1cm} 0.0020, 0.0019, 0.0020, 0.0021,\\ 0.0021,
0.0022, 0.0021, 0.0022,\\ 0.0019, 0.0021, 0.0021, 0.0022,\\ 0.0021,
0.0022, 0.0023}
\\ \noalign{\hrule height 2pt}
\end{tabular}}}
\caption{\footnotesize Outputs of dynamical quantum search algorithm
for some probabilities of marked  states
$\rho_{r_{1}r_{2}r_{3}r_{4}}$\, ($r_{1},r_{2},r_{3},r_{4}=e,g$) and
unmarked states of 4-qubit for different values of time, where
$\frac{\gamma_{1}}{\lambda}=\frac{\theta_{4}}{113}$,
$\frac{\gamma_{2}}{\lambda}=\frac{\theta_{4}}{90}$,
$\frac{\gamma_{3}}{\lambda}=\frac{\theta_{4}}{140}$ and
$\frac{\gamma_{4}}{\lambda}=\frac{\theta_{4}}{100}$.} \label{s5}
\end{table}
 \begin{table}[tbp]
{\renewcommand{\arraystretch}{1.5} \renewcommand{\tabcolsep}{0.1cm}
{\tiny
\begin{tabular}{|>{ }c|c|c|c|c|} \hline Times &
$t=\frac{0.331\pi\hbar}{16\lambda\theta_{4}}$ &
$t=\frac{0.566\pi\hbar}{16\lambda\theta_{4}}$
& $t_{4p}=\frac{0.6933\pi\hbar}{16\lambda\theta_{4}}$ & $t=\frac{\pi\hbar}{16\lambda\theta_{4}}$  \\
\noalign{\hrule height 2pt}
$\rho_{eeee}$ & 0.0034  &  0.0025  & 0.0018  & 0.0017 \\
\hline
Pr. r. s. &
\pbox{5cm}{\vspace{0.1cm} 0.0038, 0.0035,  0.0041, 0.0038,\\ 0.0045,
0.0042, 0.0055, 0.0038,\\\ 0.0046,  0.0042, 0.0056, 0.0047,\\
0.0064, 0.0057, 0.0153 }&
\pbox{5cm}{\vspace{0.1cm} 0.0017, 0.0016, 0.0015, 0.0017,\\ 0.0016,
0.0015, 0.0011, 0.0018,\\ 0.0016, 0.0015, 0.0011, 0.0016,\\ 0.0010,
0.0011, 4.9361e }&
\pbox{5cm}{\vspace{0.1cm} 9.1801e, 8.9317e, 8.3357e, 9.2909e,\\
8.0002e, 8.2450e, 0.0011, 9.3466e,\\ 7.9141e, 8.2273e, 0.0011,
7.8829e,\\ 9.7501e, 0.0010, 0.0025}&
\pbox{5cm}{\vspace{0.1cm} 0.0010, 9.6100e, 7.7284e,  0.0010,\\
8.0089e, 7.7841e, 4.2025e, 0.0010,\\ 8.0089e, 7.7841e, 4.2436e,
8.0656e,\\ 4.4944e, 4.3264e, 1.0000$\times10^{-8}$  }
\\ \noalign{\hrule height 2pt}
$\rho_{eegg}$ & 0.0071 &  0.0048 &  0.0037  &  0.0026 \\
\hline
Pr. r. s. &
\pbox{5cm}{\vspace{0.1cm}0.0024, 0.0035, 0.0035, 0.0029,\\ 0.0037,
0.0035, 0.0049, 0.0030,\\ 0.0037, 0.0035, 0.0050, 0.0034,\\ 0.0047,
0.0043, 0.0127  }&
\pbox{5cm}{\vspace{0.1cm} 7.7860e, 0.0012, 0.0012, 8.0725e,\\
8.3394e, 8.4361e, 0.0010, 8.0770e,\\  8.3449e, 8.4817e, 0.0010,
7.5393e,\\ 4.4753e,  4.8322e, 6.1274e  }&
\pbox{5cm}{\vspace{0.1cm} 3.1058e, 4.9130e, 5.4666e, 2.7961e,\\
3.2020e, 3.4613e, 9.2273e, 2.7505e,\\ 3.1720e, 3.4289e, 9.1705e,
3.0266e,\\ 4.9689e, 5.3170e, 0.0018 }&
\pbox{5cm}{\vspace{0.1cm}4.0000e, 5.6169e, 5.8081e, 4.0000e,\\
3.4225e, 3.3489e, 2.5281e, 4.0000e,\\ 3.4596e, 3.3856e, 2.6244e,
3.2761e,\\ 1.1881e, 1.1449e, 8.4640d }
\\ \noalign{\hrule height 2pt}
$\rho_{eggg}$ & 0.0148 &   0.0100 &   0.0079 &   0.0044 \\
\hline
Pr. r. s. &
\pbox{5cm}{\vspace{0.1cm} 0.0019, 0.0026, 0.0025, 0.0045,\\ 0.0026,
0.0043, 0.0044, 0.0023,\\ 0.0025, 0.0024, 0.0033, 0.0025,\\ 0.0036,
0.0034, 0.0095 }&
\pbox{5cm}{\vspace{0.1cm} 4.1785e, 4.9360e, 5.0265e, 0.0011,\\
4.9034e, 9.2489e, 0.0010, 4.3810e,\\ 3.8849e, 3.9641e, 3.4597e,
3.8628e,\\ 3.0980e, 3.3857e, 0.0012 }&
\pbox{5cm}{\vspace{0.1cm} 1.3136e, 1.2770e, 1.3549e, 4.2365e,\\
1.2456e,3.3808e, 4.0637e, 1.3130e,\\ 1.4257e, 1.4788e, 3.0025e,
1.3925e,\\ 2.9314e, 2.9786e, 0.0019 }&
\pbox{5cm}{\vspace{0.1cm}1.2100e, 1.2996e, 1.2996e, 2.9584e,\\
1.2996e, 2.4649e, 2.8900e, 1.0816e,\\ 6.5610d, 6.5610d, 4.0000c,
6.5610d,\\ 2.8900c, 4.0000c, 4.9000d }
\\ \noalign{\hrule height 2pt}
$\rho_{geeg}$ & 0.0084  & 0.0054  &  0.0041  & 0.0029 \\
\hline
Pr. r. s. &
\pbox{5cm}{\vspace{0.1cm}0.0024, 0.0037, 0.0026, 0.0032,\\ 0.0029,
0.0037, 0.0029, 0.0038,\\ 0.0038, 0.0033, 0.0046, 0.0038,\\ 0.0056,
0.0040, 0.0122 }&
\pbox{5cm}{\vspace{0.1cm} 7.0922e, 0.0012, 6.8965e, 7.2320e,\\
7.0381e, 7.4378e, 6.4129e, 4.2530e,\\ 0.0012, 7.2041e, 8.5426e,
7.4321e,\\ 9.0625e, 4.1474e, 7.2585e  }&
\pbox{5cm}{\vspace{0.1cm} 2.6185e, 4.9985e, 2.3620e, 2.8793e,\\
2.1853e, 2.5348e, 2.7625e, 4.9369e,\\ 4.7890e, 2.7716e, 7.7681e,
2.4500e,\\ 7.5024e, 4.7545e, 0.0018 }&
\pbox{5cm}{\vspace{0.1cm} 3.4596e, 5.5696e, 3.2761e, 2.7225e,\\
3.4225e, 2.9584e, 2.5921e, 7.5690d,\\ 5.4756e, 2.7556e, 1.8496e,
2.9929e,\\ 2.4025e, 7.9210d, 1.2769e}
\\ \noalign{\hrule height 2pt}
$\rho_{ggeg}$ & 0.0195  &  0.0125 &  0.0095  & 0.0054 \\
\hline
Pr. r. s. &
\pbox{5cm}{\vspace{0.1cm}  0.0018, 0.0026, 0.0019, 0.0020,\\ 0.0027,
0.0051, 0.0021, 0.0027,\\ 0.0027, 0.0051, 0.0021, 0.0027,\\ 0.0050,
0.0028, 0.0083}&
\pbox{5cm}{\vspace{0.1cm} 3.2786e, 3.9785e, 3.4632e, 3.2057e,\\
3.9469e, 0.0010, 3.2050e, 3.3730e,\\ 3.9469e, 0.0010, 3.1688e,
3.3377e,\\ 9.7604e, 3.2677e, 0.0014}&
\pbox{5cm}{\vspace{0.1cm}  9.8410d, 8.8650d, 1.1365e, 1.2874e,\\
8.6210d, 3.5257e, 1.2789e, 2.8346e,\\ 8.5520d, 3.3962e, 1.2580e,
2.8157e,\\ 3.2185e,  2.7693e, 0.0019 }&
\pbox{5cm}{\vspace{0.1cm} 7.9210d, 9.0250d, 6.2410d, 2.9160d,\\
9.0250d, 2.7556e, 2.9160d, 1.4400c,\\ 9.0250d, 2.6569e, 2.9160d,
1.4400c,\\ 2.5921e,  1.4400c, 1.5625e }
\\ \noalign{\hrule height 2pt}
\end{tabular}}}
\caption{\footnotesize The same as Table (\ref{s5}) but
$\frac{\gamma_{1}}{\lambda}=\frac{4\theta_{4}}{5}$,
$\frac{\gamma_{2}}{\lambda}=\frac{7\theta_{4}}{9}$,
$\frac{\gamma_{3}}{\lambda}=\frac{5\theta_{4}}{8}$ and
$\frac{\gamma_{4}}{\lambda}=\frac{3\theta_{4}}{4}$.} \label{s6}
\end{table}
\begin{table}[tbp]
{\renewcommand{\arraystretch}{1.5} \renewcommand{\tabcolsep}{0.1cm}
{\tiny
\begin{tabular}{|>{ }c|c|c|c|c|} \hline Times &
$t=\frac{0.331\pi\hbar}{32\lambda\theta_{5}}$ &
$t=\frac{0.566\pi\hbar}{32\lambda\theta_{5}}$
& $t_{5p}=\frac{0.8661\pi\hbar}{32\lambda\theta_{5}}$ & $t=\frac{\pi\hbar}{32\lambda\theta_{5}}$  \\
\noalign{\hrule height 2pt}
$\rho_{eeeeg}$ &   0.4478   &  0.7707  &  0.8562  &    0.8536\\
\hline
Pr. r. s.&
\pbox{5cm}{\vspace{0.1cm}0.0159, 0.0153, 0.0156, 0.0153,\\ 0.0155,
0.0151,
0.0149, 0.0152,\\ 0.0156, 0.0150,  0.0148, 0.0151,\\
0.0149, 0.0148, 0.0146, 0.0153,\\ 0.0156, 0.0150, 0.0149, 0.0151,\\
0.0149, 0.0149, 0.0147,  0.0150,\\ 0.0148, 0.0148, 0.0146, 0.0149,\\
0.0146, 0.0146, 0.0144}&
\pbox{5cm}{\vspace{0.1cm}0.0035, 0.0035, 0.0037,  0.0035,\\ 0.0036,
0.0035, 0.0035, 0.0035,\\ 0.0037, 0.0035, 0.0036, 0.0035,\\ 0.0036,
0.0036, 0.0036, 0.0035,\\ 0.0036, 0.0035, 0.0036, 0.0035,\\ 0.0035,
0.0036, 0.0036, 0.0035,\\ 0.0036, 0.0036, 0.0036, 0.0036,\\ 0.0036,
0.0036, 0.0037}&
\pbox{5cm}{\vspace{0.1cm}\{5.9130, 3.9610,  5.5840,  4.1210,\\
5.1610, 3.7060,  3.4850,  3.9240,\\  5.8760,  3.6500,  3.4640,
3.7060,\\ 3.4850, 3.4640,  3.2850,  3.9610,\\  5.4410,  3.7060,
3.4850, 3.7370,\\ 3.5080,  3.4850,  3.2980,  3.6770,\\ 3.4640,
3.4450, 3.2740, 3.4640,\\ 3.2850,  3.2740, 3.1370\}d}&
\pbox{5cm}{\vspace{0.1cm} 1.0000a, 1.6000b, 8.1000b, 2.5000b,\\
2.5000b, 4.0000a, 1.0000a, 1.6000b,\\ 1.2100c, 0, 9.0000a, 1.0000a,\\
4.0000a, 9.0000a, 3.6000b, 2.5000b,\\ 6.4000b, 1.0000a, 4.0000a,
4.0000a,\\ 1.0000a, 4.0000a, 2.5000b, 1.0000a,\\ 4.0000a, 9.0000a,
4.9000b, 4.0000a,\\  2.5000b, 3.6000b, 1.0000c}
\\ \noalign{\hrule height 2pt}
$\rho_{eegeg}$ &   0.4553  &  0.7835  &   0.8707  &   0.8683  \\
\hline
Pr. r. s. &
\pbox{5cm}{\vspace{0.1cm}0.0154, 0.0156, 0.0152, 0.0150,\\ 0.0157,
0.0150, 0.0153, 0.0152,\\ 0.0150, 0.0149, 0.0147, 0.0150,\\ 0.0154,
0.0147, 0.0145,0.0152,\\ 0.0150, 0.0150, 0.0148, 0.0150,\\ 0.0153,
0.0148, 0.0146, 0.0149,\\ 0.0147, 0.0147, 0.0145, 0.0148,\\ 0.0146,
0.0145, 0.0143}&
\pbox{5cm}{\vspace{0.1cm}0.0033, 0.0035, 0.0034, 0.0034,\\ 0.0035,
0.0034, 0.0036,  0.0034,\\ 0.0034, 0.0035, 0.0035, 0.0034,\\ 0.0036,
0.0035, 0.0035, 0.0034,\\ 0.0034, 0.0034, 0.0035, 0.0034,\\ 0.0036,
0.0035, 0.0035, 0.0034,\\ 0.0035, 0.0035, 0.0035, 0.0035,\\ 0.0035,
0.0035, 0.0036}&
\pbox{5cm}{\vspace{0.1cm}\{ 3.1040, 4.1210, 2.857, 2.6440,\\ 4.2850,
2.6690, 4.0400, 2.8260,\\ 2.6440, 2.5220, 2.3530, 2.6440,\\ 4.1650,
2.3530, 2.2180, 2.8570,\\ 2.6690, 2.6440, 2.4650, 2.6690,\\ 3.9170,
2.4650, 2.3200, 2.6210,\\ 2.4500, 2.3400, 2.2130, 2.4650,\\ 2.3200,
2.2130, 2.1200\}d}&
\pbox{5cm}{\vspace{0.1cm} 3.2400c, 4.9000b, 1.9600c, 1.4400c,\\
4.9000b, 1.4400c, 4.0000a, 1.9600c,\\ 1.2100c, 1.0000c, 4.9000b,
1.2100c,\\ 1.6000b, 4.9000b, 1.6000b, 2.2500c,\\ 1.4400c, 1.2100c,
6.4000b, 1.4400c,\\ 0, 6.4000b, 2.5000b, 1.2100c,\\ 6.4000b,
4.9000b, 9.0000a,  6.4000b,\\ 2.5000b, 1.6000b, 0}
\\ \noalign{\hrule height 2pt}
$\rho_{egeee}$ &  0.4510   &  0.7761  & 0.8623  &  0.8597 \\
\hline
Pr. r. s. &
\pbox{5cm}{\vspace{0.1cm} 0.0161, 0.0153, 0.0152, 0.0150,\\ 0.0153,
0.0151, 0.0150, 0.0149,\\ 0.0154, 0.0155, 0.0148, 0.0154,\\ 0.0148,
0.0148, 0.0146, 0.0153,\\ 0.0151, 0.0150, 0.0148, 0.0151,\\ 0.0149,
0.0148, 0.0146, 0.0155,\\ 0.0148, 0.0147, 0.0145, 0.0148,\\ 0.0146,
0.0145,0.0143,}&
\pbox{5cm}{\vspace{0.1cm} 0.0036, 0.0034, 0.0034, 0.0035,\\ 0.0034,
0.0035, 0.0035, 0.0035,\\ 0.0036, 0.0036, 0.0035, 0.0036,\\ 0.0035,
0.0035, 0.0036, 0.0034,\\ 0.0035, 0.0035, 0.0035, 0.0035,\\ 0.0035,
0.0035, 0.0036, 0.0036,\\ 0.0035, 0.0035, 0.0036, 0.0035,\\ 0.0036,
0.0036, 0.0036}&
\pbox{5cm}{\vspace{0.1cm}\{ 5.9680, 3.5730, 3.4600, 3.2210,\\
3.6100, 3.3610, 3.2210, 3.0160,\\ 4.6450, 4.9130, 2.9800, 4.5140,\\
3.0160, 2.9970, 2.8340, 3.4600,\\ 3.2210, 3.1940, 2.9970, 3.3610,\\
3.0370, 3.0160, 2.8450, 4.7780,\\ 2.9970, 2.9800, 2.8250, 2.9970,\\
2.8340, 2.8250, 2.7040\}$d$}&
\pbox{5cm}{\vspace{0.1cm} 0, 8.1000b, 8.1000b, 3.6000b,\\
1.0000c, 4.9000b, 3.6000b, 9.0000a,\\ 9.0000a, 3.6000b, 1.0000a,
4.0000a,\\4.0000a, 4.0000a, 4.0000a, 8.1000b,
\\ 3.6000b, 2.5000b, 4.0000a, 3.6000b, \\ 9.0000a, 9.0000a, 1.0000a, 2.5000b,\\
4.0000a, 1.0000a, 4.0000a, 4.0000a,\\ 1.0000a, 4.0000a, 3.6000b}
\\ \noalign{\hrule height 2pt}
$\rho_{egegg}$ &  0.4704   &  0.8088  &   0.8987  &    0.8962 \\
\hline
Pr. r. s. &
\pbox{5cm}{\vspace{0.1cm} 0.0152, 0.0151, 0.0150, 0.0154,\\ 0.0151,
0.0149, 0.0148, 0.0146,\\ 0.0150, 0.0153, 0.0152, 0.0148,\\ 0.0146,
0.0145, 0.0147, 0.0150,\\ 0.0148, 0.0148, 0.0146, 0.0149,\\ 0.0147,
0.0146, 0.0144, 0.0148,\\ 0.0146, 0.0145, 0.0148, 0.0146,\\ 0.0144,
0.0143, 0.0141 }&
\pbox{5cm}{\vspace{0.1cm}  0.0032, 0.0032, 0.0032, 0.0034,\\ 0.0032,
0.0033, 0.0033, 0.0033,\\ 0.0032, 0.0034, 0.0034, 0.0033,\\ 0.0033,
0.0033, 0.0035, 0.0032,\\ 0.0033, 0.0033, 0.0033, 0.0033,\\ 0.0033,
0.0033, 0.0034, 0.0033,\\ 0.0033, 0.0033, 0.0035, 0.0033,\\ 0.0034,
0.0034, 0.0034 }&
\pbox{5cm}{\vspace{0.1cm} 1.3810d, 1.2100d, 1.2100d, 2.3780d,\\
1.3000d, 1.1530d, 1.0730d, 9.7000c,\\ 1.1890d, 2.1460d, 1.8450d,
1.0730d,\\ 9.7000c, 9.6500c, 1.5220d, 1.2100d,\\ 1.0730d, 1.0600d,
9.6500c, 1.0880d,\\ 9.7700c, 9.7000c, 9.0100c, 1.0600d,\\ 9.6500c,
9.6200c, 1.6820d, 9.6500c,\\ 9.0400c, 9.0900c, 8.3300c}&
\pbox{5cm}{\vspace{0.1cm}  1.3690d, 1.1560d, 1.0890d, 3.2400c,\\
1.1560d, 9.6100c, 9.6100c, 7.8400c,\\ 1.0890d, 3.6100c, 4.0000c,
9.0000c,\\ 7.2900c, 6.7600c, 2.2500c, 1.1560d,\\ 9.6100c, 9.0000c,
7.2900c, 9.6100c,\\ 7.8400c, 7.2900c, 5.7600c, 9.0000c,\\ 6.7600c,
6.7600c, 1.6900c, 7.2900c,\\ 5.7600c, 5.2900c, 3.6100c }
\\ \noalign{\hrule height 2pt}
$\rho_{geeee}$ &  0.4486   &  0.7721  & 0.8578  &   0.8553 \\
\hline
Pr. r. s. &
\pbox{5cm}{\vspace{0.1cm}  0.0160, 0.0153, 0.0153, 0.0151,\\ 0.0153,
0.0151, 0.0151, 0.0149,\\ 0.0152, 0.0150, 0.0150, 0.0148,\\ 0.0151,
0.0149, 0.0148, 0.0146,\\ 0.0155, 0.0156, 0.0149, 0.0155,\\ 0.0149,
0.0149,  0.0147, 0.0156,\\ 0.0148, 0.0148, 0.0146, 0.0148,\\ 0.0146,
0.0146, 0.0144 }&
\pbox{5cm}{\vspace{0.1cm}  0.0035, 0.0034, 0.0035, 0.0035,\\
0.0034,0.0035, 0.0035, 0.0035,\\ 0.0035, 0.0035, 0.0035, 0.0036,\\
0.0035, 0.0035, 0.0036, 0.0036,\\ 0.0036, 0.0037, 0.0035, 0.0036,\\
0.0035, 0.0035, 0.0036, 0.0037,\\ 0.0036, 0.0036, 0.0036, 0.0035,\\
0.0036, 0.0036, 0.0037 }&
\pbox{5cm}{\vspace{0.1cm}\{ 5.9130, 3.8420, 3.8050, 3.5890,\\
3.8810, 3.6200, 3.5890, 3.3700,\\ 3.8050, 3.5600, 3.5330, 3.3300,\\
3.5600, 3.3700, 3.3490, 3.1610,\\ 5.1220, 5.4020, 3.3490, 4.9850,\\
3.3930, 3.3700, 3.1850, 5.5450,\\ 3.3490, 3.3300, 3.0500, 3.3490,\\
3.1720, 3.1610, 3.0260 \}d }&
\pbox{5cm}{\vspace{0.1cm}1.0000a, 3.6000b, 3.6000b, 9.0000a,\\
4.9000b, 1.6000b, 9.0000a, 0,\\ 2.5000b, 4.0000a, 1.0000a,
4.0000a,\\
4.0000a, 1.0000a, 1.0000a, 2.5000b,\\ 3.6000b, 8.1000b, 1.0000a,
1.6000b,\\ 0, 0, 1.6000b, 1.0000c,\\ 1.0000a, 4.0000a, 2.5000b,
1.0000a,\\ 1.6000b, 2.5000b, 8.1000b
 }
\\ \noalign{\hrule height 2pt}
$\rho_{gegeg}$ &  0.4647 &  0.7993  &     0.8881  &   0.8857 \\
\hline
Pr. r. s. &
\pbox{5cm}{\vspace{0.1cm} 0.0153, 0.0151, 0.0151, 0.0149,\\ 0.0151,
0.0154, 0.0149, 0.0147,\\ 0.0150, 0.0148, 0.0148, 0.0146,\\ 0.0149,
0.0147, 0.0146, 0.0144,\\ 0.0151, 0.0153, 0.0149, 0.0147,\\ 0.0154,
0.0147, 0.0150, 0.0148,\\ 0.0146, 0.0146, 0.0144, 0.0146,\\ 0.0150,
0.0144, 0.0142}&
\pbox{5cm}{\vspace{0.1cm}  0.0033, 0.0033, 0.0033, 0.0033,\\ 0.0033,
0.0035, 0.0033, 0.0034,\\ 0.0033, 0.0034, 0.0033, 0.0034,\\ 0.0033,
0.0034, 0.0034, 0.0034,\\ 0.0033, 0.0034, 0.0033, 0.0034,\\ 0.0034,
0.0034, 0.0036, 0.0034,\\ 0.0034, 0.0034, 0.0035, 0.0034,\\ 0.0036,
0.0035, 0.0035 }&
\pbox{5cm}{\vspace{0.1cm}\{ 1.9700,  1.7690,  1.7440, 1.6020,\\
1.7960, 2.8260, 1.6020, 1.4690,\\ 1.7440, 1.5850, 1.4930, 1.3780,\\
1.6020, 1.4690, 1.3850, 1.2970,\\ 1.7690, 2.4730, 1.5850, 1.4600,\\
2.6960, 1.4690, 2.5360, 1.5080,\\ 1.3850, 1.3780, 1.2970, 1.4600,\\
2.6370, 1.2970, 1.2500\}d }&
\pbox{5cm}{\vspace{0.1cm}  9.0000c, 7.2900c, 6.7600c, 5.2900c,\\
7.2900c, 1.9600c, 5.7600c, 4.4100c,\\ 6.7600c, 5.2900c, 4.8400c,
3.6100c,\\ 5.2900c, 4.0000c, 3.6100c, 2.5600c,\\ 7.2900c, 2.5600c,
5.2900c, 4.0000c,\\ 2.2500c, 4.0000c, 3.6000b, 5.2900c,\\ 3.6100c,
3.6100c, 2.2500c, 4.0000c,\\ 1.6000b, 2.5600c, 1.4400c}
\\ \noalign{\hrule height 2pt}
$\rho_{ggege}$ &  0.4712 &  0.8104  &   0.9004   &   0.8979 \\
\hline
Pr. r. s. &
\pbox{5cm}{\vspace{0.1cm} 0.0152, 0.0150, 0.0150, 0.0148,\\ 0.0151,
0.0149, 0.0148, 0.0146,\\ 0.0150, 0.0148, 0.0152, 0.0145,\\ 0.0148,
0.0146, 0.0145, 0.0143,\\ 0.0150, 0.0148, 0.0154, 0.0146,\\ 0.0149,
0.0147, 0.0146, 0.0144,\\ 0.0153, 0.0145, 0.0147, 0.0146,\\ 0.0144,
0.0147, 0.0141 }&
\pbox{5cm}{\vspace{0.1cm} 0.0032, 0.0032, 0.0032, 0.0033,\\ 0.0032,
0.0033, 0.0033, 0.0033,\\ 0.0032, 0.0033, 0.0034, 0.0033,\\ 0.0033,
0.0033, 0.0033, 0.0034,\\ 0.0032, 0.0033, 0.0034, 0.0033,\\ 0.0033,
0.0033, 0.0033, 0.0034,\\ 0.0034, 0.0033, 0.0035, 0.0033,\\ 0.0034,
0.0035, 0.0034 }&
\pbox{5cm}{\vspace{0.1cm} 1.3140d, 1.1450d, 1.1240d, 9.9700c,\\
1.1450d, 1.0250d, 1.0100d, 9.0900c,\\ 1.1240d, 9.9700c, 1.9490d,
9.0100c,\\ 9.9700c, 9.0400c, 9.0100c, 8.5000c,\\ 1.1450d, 1.0100d,
2.2600d, 9.0400c,\\ 1.0100d, 9.0900c, 9.0400c, 8.4500c,\\ 2.0570d,
9.0100c, 1.5220d, 9.0400c,\\ 8.4500c, 1.4450d, 7.9300c}&
\pbox{5cm}{\vspace{0.1cm} 1.4440d, 1.2250d,  1.2250d, 1.0240d,\\
1.2250d, 1.0890d, 1.0240d, 8.4100c,\\ 1.1560d, 9.6100c, 4.4100c,
7.2900c,\\ 9.6100c, 7.8400c, 7.8400c, 5.7600c,\\ 1.2250d, 1.0240d,
3.6100c, 7.8400c,\\ 1.0240d, 8.4100c, 8.4100c, 6.2500c,\\ 4.0000c,
7.8400c, 2.2500c, 7.8400c,\\ 6.2500c, 2.5600c, 4.0000c}
\\ \noalign{\hrule height 2pt}
\end{tabular}}}
\caption{\small Outputs of dynamical quantum search algorithm for
some probabilities of marked  states
$\rho_{r_{1}r_{2}r_{3}r_{4}r_{5}}$\,
($r_{1},r_{2},r_{3},r_{4},r_{5}=e,g$) and unmarked states of 5-qubit
for different values of time, where
$\frac{\gamma_{1}}{\lambda}=\frac{\theta_{5}}{113}$,
$\frac{\gamma_{2}}{\lambda}=\frac{\theta_{5}}{90}$,
$\frac{\gamma_{3}}{\lambda}=\frac{\theta_{5}}{140}$,
$\frac{\gamma_{4}}{\lambda}=\frac{\theta_{5}}{100}$ and
$\frac{\gamma_{5}}{\lambda}=\frac{\theta_{5}}{125}$;
$a=\times10^{-8}$ and $b=\times10^{-7}$.} \label{s7}
\end{table}
\begin{table}[tbp]
{\renewcommand{\arraystretch}{1.5} \renewcommand{\tabcolsep}{0.1cm}
{\tiny
\hspace{-0.65cm}\begin{tabular}{|>{ }c|c|c|c|c|} \hline Times &
$t=\frac{0.331\pi\hbar}{32\lambda\theta_{5}}$ &
$t=\frac{0.566\pi\hbar}{32\lambda\theta_{5}}$
& $t_{5p}=\frac{0.8661\pi\hbar}{32\lambda\theta_{5}}$ & $t=\frac{\pi\hbar}{32\lambda\theta_{5}}$  \\
\noalign{\hrule height 2pt}
$\rho_{eeeeg}$ &  4.9268e   &  1.9396e & 1.9925e   & 1.4400e  \\
\hline
Pr. r. s. &
\pbox{5cm}{\vspace{0.1cm} 3.0069e,  3.9569e, 5.4196e, 3.6756e,\\
4.9844e, 4.5000e, 5.1885e, 3.9946e,\\ 5.5514e, 4.9860e, 6.0265e,
4.5785e,\\ 5.3434e, 4.7248e, 6.1245e, 4.0706e,\\ 5.6450e, 5.0705e,
6.1956e, 4.6580e,\\ 5.4445e, 4.8296e, 6.3040e, 5.1560e,\\ 6.3673e,
5.5953e, 8.0965e, 4.9360e,\\ 6.6002e, 5.5714e, 0.0032 }&
\pbox{5cm}{\vspace{0.1cm}  8.4500d, 7.7450d, 1.0345e, 7.7450d,\\
1.0100e, 8.4020d, 1.1545e, 7.7450d,\\ 1.0345e, 8.0900d, 1.0642e,
8.4020d,\\ 1.1338e, 1.1601e, 2.2978e, 7.7450d,\\ 1.0345e, 8.1170d,
1.0445e, 8.4250d,\\ 1.1133e, 1.1392e, 2.2394e, 7.9400d,\\ 1.0309e,
1.0280e, 1.9325e, 1.1234e,\\ 2.1818e, 2.4097e, 8.9730e }&
\pbox{5cm}{\vspace{0.1cm} 8.8820d,  8.4250d, 1.0413e, 8.3520d,\\
9.8810d, 8.4100d, 7.5560d, 8.6690d,\\ 1.0498e, 8.6530d,
7.8440d, 8.4100d,\\ 7.5560d, 7.2650d, 4.5050d, 8.6690d,\\
1.0498e, 8.6530d, 7.7620d,  8.4890d,\\ 7.6250d, 7.2650d, 4.4980d,
8.7380d,\\ 7.7620d, 7.6100d,  4.2410d, 7.3300d, \\ 4.4930d, 4.6600d,
1.7557e }&
\pbox{5cm}{\vspace{0.1cm}  5.3290d, 4.2250d, 5.4760d, 4.3560d,\\
5.4760d, 4.3560d, 5.4760d, 4.2250d,\\ 5.4760d, 4.0960d, 4.9000d,
4.3560d,\\ 5.3290d, 5.6250d, 1.1236e, 4.2250d,\\ 5.4760d, 4.0960d,
4.9000d, 4.3560d,\\ 5.3290d, 5.4760d, 1.1025e, 4.0960d,\\ 4.7610d,
4.9000d, 9.6040d, 5.4760d,\\ 1.0816e, 1.1881e, 4.5369e }
\\ \noalign{\hrule height 2pt}
$\rho_{eegeg}$ &  6.3684e   &  2.8265e  &  2.5492e  &  1.9321e \\
\hline
Pr. r. s. &
\pbox{5cm}{\vspace{0.1cm} 2.7266e, 4.1876e, 3.6496e, 5.0013e,\\
3.8408e, 4.3661e, 5.0960e, 3.7220e,\\ 5.1301e, 4.6045e, 5.4920e,
4.4420e,\\ 5.2370e, 4.3874e, 5.6180e, 3.7458e,\\ 5.2154e, 4.6472e,
5.5953e, 4.5256e,\\ 5.3800e, 4.4874e, 5.8154e, 4.7714e,\\ 5.7717e,
5.0338e, 7.2125e, 4.6309e,\\ 6.0842e, 4.9865e, 0.0031}&
\pbox{5cm}{\vspace{0.1cm} 5.9450d, 7.9570d, 6.1000d, 6.8050d,\\
8.9360d, 7.6050d, 1.3093e, 6.1000d,\\ 6.8050d, 6.5970d, 8.3210d,
7.4320d,\\ 1.3093e, 9.8930d, 2.1697e, 6.1000d,\\ 6.8240d, 6.5970d,
8.3210d, 7.4450d,\\ 1.3093e, 9.7330d, 2.1125e, 6.6100d,\\ 7.9690d,
8.7880d, 1.7890e, 9.5400d,\\ 2.0561e, 2.2045e, 8.7986e}&
\pbox{5cm}{\vspace{0.1cm} 6.2600d, 8.4500d, 6.9320d, 7.4500d,\\
8.5930d, 7.1540d, 7.9560d, 6.9320d,\\ 7.5290d, 7.2200d, 6.1970d,
7.2250d,\\ 8.1250d, 5.9680d, 3.7850d, 7.0010d,\\ 7.5290d, 7.2200d,
6.1970d, 7.2250d,\\ 8.1860d, 5.9680d, 3.7700d, 7.2970d,\\ 6.2660d,
6.0650d, 3.3680d, 5.8820d,\\ 3.7570d, 3.7210d, 1.8386e }&
\pbox{5cm}{\vspace{0.1cm} 3.2490d, 4.2250d, 3.1360d, 3.1360d,\\
4.7610d, 3.6000d, 6.2410d, 3.0250d,\\ 3.0250d, 3.1360d, 3.6000d,
3.4810d,\\ 6.0840d, 4.3560d, 9.8010d, 3.0250d,\\ 3.0250d, 3.0250d,
3.6000d, 3.4810d,\\ 6.0840d, 4.3560d, 9.6040d, 3.0250d,\\ 3.4810d,
3.8440d, 8.2810d, 4.2250d,\\ 9.4090d, 1.0201e, 4.2849e}
\\ \noalign{\hrule height 2pt}
$\rho_{egeee}$ &   4.7081e  &  1.9066e  &  1.9652e  &  1.3924e \\
\hline
Pr. r. s. &
\pbox{5cm}{\vspace{0.1cm} 3.0197e, 4.1732e, 3.9569e, 5.2925e,\\
3.6756e, 4.8690e, 4.5000e, 5.1209e,\\ 5.6052e, 5.1994e, 6.0740e,
4.7317e,\\ 5.3434e, 4.7912e, 6.1498e, 4.0706e,\\ 5.4689e, 5.0705e,
6.0930e, 4.6580e,\\ 5.3874e, 4.8296e, 6.2785e, 5.3737e,\\6.4160e,
5.7424e, 8.1252e, 5.0362e,\\ 6.6265e, 5.6186e, 0.0032 }&
\pbox{5cm}{\vspace{0.1cm} 8.4250d, 7.7480d, 7.7450d, 8.1460d,\\
7.7450d, 8.4500d, 8.5850d, 1.1285e,\\ 1.0548e, 1.0525e, 1.0900e,
1.0485e,\\ 1.1600e, 1.2125e, 2.3053e, 7.7480d,\\ 8.0000d, 8.2960d,
1.0000e, 8.6080d,\\ 1.0877e, 1.1650e, 2.2685e, 1.0525e,\\ 1.0565e,
1.1188e, 1.9669e, 1.1965e,\\ 2.2180e, 2.4473e, 8.9809e}&
\pbox{5cm}{\vspace{0.1cm}  8.7050d, 8.9170d, 8.6690d, 8.9050d,\\
8.3520d, 8.7370d, 8.5770d, 7.5560d,\\ 1.0874e, 1.0330e, 7.8440d,
9.9850d,\\ 7.7860d, 7.5850d, 4.6330d, 8.6690d,\\ 8.9920d, 8.8200d,
7.7620d, 8.6560d,\\ 7.6960d, 7.4890d, 4.6330d, 1.0600e,\\ 8.0000d,
7.9300d, 4.3720d, 7.6500d,\\ 4.4900d, 4.6600d, 1.7389e}&
\pbox{5cm}{\vspace{0.1cm} 5.1840d,  4.2250d, 4.3560d, 4.0960d,\\
4.3560d, 4.3560d, 4.4890d, 5.4760d,\\ 5.6250d, 5.6250d, 5.0410d,
5.6250d,\\ 5.4760d, 5.9290d, 1.1449e, 4.3560d,\\ 3.9690d, 4.2250d,
4.7610d, 4.4890d,\\ 5.1840d, 5.6250d, 1.1236e, 5.6250d,\\ 4.9000d,
5.3290d, 9.6040d, 5.7760d,\\ 1.0816e, 1.1881e, 4.5796e }
\\ \noalign{\hrule height 2pt}
$\rho_{egegg}$ & 0.0011    &  5.8981e  &  4.3525e  &  3.5721e \\
\hline
Pr. r. s. &
\pbox{5cm}{\vspace{0.1cm} 2.3722e, 3.6452e, 3.4866e, 5.6537e,\\
2.9377e, 3.8464e, 3.5989e, 4.1645e,\\ 3.5405e, 5.7330e, 5.5386e,
3.6650e,\\ 4.2577e, 3.9041e, 4.4434e, 3.2456e,\\ 4.3938e, 4.1296e,
5.2561e, 3.5620e,\\ 3.8020e, 3.3908e, 4.2761e, 4.2029e,\\ 5.3645e,
4.9045e, 6.5173e, 3.5066e,\\ 4.4629e, 3.7705e, 0.0027}&
\pbox{5cm}{\vspace{0.1cm} 4.3280d, 4.8200d, 4.9860d, 8.7300d,\\
4.6800d, 4.9640d, 5.2650d, 7.5850d,\\ 4.9860d, 8.3450d, 9.3160d,
5.1220d,\\ 7.4210d, 7.9370d, 2.4592e, 4.5140d,\\ 4.5250d, 4.6730d,
6.2900d, 5.2840d,\\ 7.3970d, 7.9210d, 1.8596e, 4.6730d,\\ 6.1330d,
6.7600d, 2.2160e, 7.7440d,\\ 1.8056e, 1.9700e, 8.7425e}&
\pbox{5cm}{\vspace{0.1cm} 4.0840d, 4.9370d, 4.8800d, 7.5560d,\\
4.4690d, 4.5620d, 4.5730d, 3.8050d,\\ 4.8800d, 7.5560d, 7.8130d,
4.5730d,\\ 3.8480d, 3.8420d, 4.1650d, 4.5730d,\\ 4.7450d, 4.6820d,
4.0930d, 4.5730d,\\ 3.6490d, 3.5730d, 2.4770d, 4.6820d,\\ 4.1480d,
4.1060d, 4.3720d, 3.5730d,\\ 2.4400d, 2.6930d, 2.0740e }&
\pbox{5cm}{\vspace{0.1cm}  1.7640d, 1.7640d, 1.8490d, 3.2490d,\\
1.7640d, 1.7640d, 1.8490d, 2.6010d,\\ 1.8490d, 3.1360d, 3.6000d,
1.8490d,\\ 2.5000d, 2.8090d, 9.8010d, 1.6810d,\\ 1.4440d, 1.6000d,
2.0250d, 1.8490d,\\ 2.5000d, 2.7040d, 7.2250d, 1.6000d,\\ 1.9360d,
2.2090d, 8.6490d, 2.7040d,\\ 6.8890d, 7.7440d, 3.8416e}
\\ \noalign{\hrule height 2pt}
$\rho_{geeee}$ &  4.7681e    & 1.9125e   &  1.9652e  & 1.4161e  \\
\hline
Pr. r. s. &
\pbox{5cm}{\vspace{0.1cm} 3.0197e, 4.1732e, 3.9569e, 5.2925e,\\
3.6756e, 4.8690e, 4.5000e, 5.1209e,\\ 4.0325e, 5.3802e, 4.9860e,
5.9714e,\\ 4.5785e, 5.2865e, 4.7248e, 6.0745e,\\ 5.6521e, 5.2445e,
6.1956e, 4.8177e,\\ 5.4445e, 4.8865e, 6.3730e, 5.3737e,\\ 6.4160e,
5.7424e, 8.1252e, 5.0362e,\\ 6.6265e, 5.6186e, 0.0032 }&
\pbox{5cm}{\vspace{0.1cm} 8.4500d, 7.7480d, 7.7450d, 8.1460d,\\
7.7450d, 8.2690d, 8.5850d, 1.1080e,\\ 7.7450d, 7.9690d, 8.2690d,
1.0000e,\\ 8.6080d, 1.0877e, 1.1601e, 2.3273e,\\ 1.0573e, 1.0525e,
1.0504e, 1.0282e,\\ 1.1393e, 1.1912e, 2.2469e, 1.0525e,\\ 1.0309e,
1.0985e, 1.9325e, 1.1912e,\\ 2.1893e, 2.4170e, 8.9730e}&
\pbox{5cm}{\vspace{0.1cm}  8.8820d, 8.7460d, 8.6690d, 8.9050d,\\
8.3520d, 8.7370d, 8.5770d, 7.5560d,\\ 8.6690d, 8.9920d, 8.8200d,
7.6850d,\\ 8.5770d, 7.6250d, 7.4240d, 4.6400d,\\ 1.0685e, 1.0413e,
7.9210d, 9.9850d,\\ 7.6250d, 7.5850d, 4.4980d, 1.0413e,\\ 8.0000d,
7.7690d, 4.3720d, 7.6500d,\\ 4.4900d, 4.6600d, 1.7389e }&
\pbox{5cm}{\vspace{0.1cm} 5.1840d, 4.2250d, 4.3560d, 4.0960d,\\
4.3560d, 4.2250d, 4.4890d, 5.3290d,\\ 4.3560d, 3.9690d, 4.2250d,
4.7610d,\\ 4.4890d, 5.3290d, 5.6250d, 1.1236e,\\ 5.6250d, 5.6250d,
4.9000d, 5.6250d,\\ 5.4760d, 5.7760d, 1.1025e, 5.6250d,\\ 4.9000d,
5.1840d, 9.6040d, 5.7760d,\\ 1.0816e, 1.1881e, 4.5796e }
\\ \noalign{\hrule height 2pt}
$\rho_{gegeg}$ &  9.8957e    &  5.4797e   &  3.9770e  & 3.3489e  \\
\hline
Pr. r. s. &
\pbox{5cm}{\vspace{0.1cm} 2.4250e, 3.6658e, 3.2125e, 4.2841e,\\
3.3181e, 5.2317e, 3.7402e, 4.4225e,\\ 3.2456e, 4.4005e, 3.9005e,
4.3552e,\\ 3.8081e, 4.5610e, 3.4697e, 4.3733e,\\ 3.5605e, 5.5009e,
4.1296e, 5.0866e,\\ 5.1280e, 4.1993e, 4.9613e, 4.2434e,\\ 5.2385e,
4.1282e, 5.7556e, 4.3749e,\\ 5.2496e, 4.0385e, 0.0027 }&
\pbox{5cm}{\vspace{0.1cm} 4.5370d, 5.0440d, 4.7140d, 4.9160d,\\
5.2370d, 9.0890d, 5.3780d, 7.7930d,\\ 4.7140d, 4.9160d, 5.3650d,
7.0810d,\\ 5.3650d, 7.6330d, 8.1040d, 1.8938e,\\ 5.0690d, 7.7800d,
5.0660d, 6.7880d,\\ 9.4900d, 8.1490d, 2.4973e, 5.0660d,\\ 6.6420d,
7.4120d, 1.6553e, 7.9700d,\\ 2.4697e, 1.9769e, 8.6333e }&
\pbox{5cm}{\vspace{0.1cm} 4.3730d, 5.0660d, 4.8250d, 4.9930d,\\
4.9540d, 7.6500d, 4.8800d, 4.2500d,\\ 4.8800d, 4.9930d, 4.9960d,
4.0570d,\\ 4.8100d, 4.2970d, 3.9610d, 2.6280d,\\ 5.1970d, 7.3090d,
4.9300d, 4.3290d,\\ 7.9130d, 4.2050d, 4.5890d, 4.9300d,\\ 4.3840d,
4.0100d, 2.3780d, 4.2050d,\\ 4.7050d, 2.8010d, 1.9720e }&
\pbox{5cm}{\vspace{0.1cm} 1.9360d, 1.9360d, 1.8490d, 1.7640d,\\
2.1160d, 3.7210d, 2.0250d, 2.8090d,\\ 1.8490d, 1.7640d, 1.9360d,
2.4010d,\\ 2.0250d, 2.7040d, 3.0250d, 7.5690d,\\ 1.9360d, 3.0250d,
1.8490d, 2.3040d,\\ 3.9690d, 2.9160d, 1.0201e, 1.7640d,\\ 2.2090d,
2.6010d, 6.4000d, 2.8090d,\\ 1.0000e, 7.9210d, 3.8416e}
\\ \noalign{\hrule height 2pt}
$\rho_{geggg}$ &  0.0022 &  0.0015 &  8.4776e  & 7.6729e   \\
\hline
Pr. r. s. &
\pbox{5cm}{\vspace{0.1cm} 2.1097e,  3.0185e, 2.9017e, 4.1992e,\\
2.7421e, 4.0482e, 3.9266e, 7.1018e,\\ 2.7508e, 3.4154e, 3.1912e,
3.3525e,\\ 2.9905e, 3.0625e, 2.8240e, 3.5197e,\\ 2.9597e, 4.3034e,
4.1353e, 6.9269e,\\ 3.9869e, 7.0690e, 7.1045e, 3.3125e,\\ 3.5033e,
3.1700e, 4.2562e, 2.9245e,\\ 3.7120e, 3.2840e, 0.0015}&
\pbox{5cm}{\vspace{0.1cm} 4.1410d, 4.2050d, 4.2640d, 4.4960d,\\
4.1490d, 4.7560d, 4.9300d, 1.2325e,\\ 4.5610d, 4.8010d, 4.9810d,
6.4660d,\\ 5.0320d, 6.6560d, 7.0130d, 1.6388e,\\ 4.2640d, 4.4960d,
4.6250d, 1.0169e,\\ 4.7970d, 1.1674e, 1.3252e, 4.9320d,\\ 6.2800d,
6.6250d, 1.5385e, 6.8170d,\\ 1.5885e, 1.6904e, 8.7125e }&
\pbox{5cm}{\vspace{0.1cm}2.2370d, 2.4650d, 2.4340d, 2.7040d,\\
2.4340d, 2.6650d, 2.7250d, 5.8130d,\\ 2.4050d, 2.3720d, 2.3120d,
1.7450d,\\ 2.2850d, 1.8000d, 1.7890d, 1.8320d,\\ 2.4650d, 2.6090d,
2.6650d, 4.8400d,\\ 2.6650d, 5.5250d, 6.1090d,  2.3410d,\\ 1.7450d,
1.7300d, 1.6490d, 1.7890d,\\ 1.7650d, 1.9400d, 1.5538e }&
\pbox{5cm}{\vspace{0.1cm} 8.4100c, 7.2900c, 7.8400c, 7.2900c,\\
7.8400c, 7.8400c, 8.4100c, 3.1360d,\\ 8.4100c, 7.8400c, 8.4100c,
1.1560d,\\ 9.0000c, 1.2960d, 1.4440d, 4.4890d,\\ 7.2900c, 6.7600c,
7.2900c, 2.3040d,\\ 8.4100c, 2.8090d, 3.3640d, 8.4100c,\\ 1.1560d,
1.2250d, 4.0960d, 1.3690d,\\ 4.3560d, 4.6240d, 3.3124e}
\\ \noalign{\hrule height 2pt}
$\rho_{ggege}$ &  0.0010    &  5.6410e   &  4.1425e  & 3.4225e  \\
\hline
Pr. r. s. &
\pbox{5cm}{\vspace{0.1cm} 2.3985e, 3.3800e, 3.5060e, 4.3181e,\\
2.9690e, 3.7700e, 3.6360e, 3.6682e,\\ 3.5405e, 4.3938e, 5.4401e,
5.2474e,\\ 3.7025e, 3.7953e, 3.9041e, 4.2400e,\\ 3.5752e, 4.4770e,
5.4721e, 5.3554e,\\ 3.7402e, 3.8792e, 3.9538e, 4.3893e,\\ 5.4920e,
5.4650e, 6.7010e, 4.0784e,\\ 4.5621e, 4.0673e, 0.0027 }&
\pbox{5cm}{\vspace{0.1cm} 4.4100d, 4.5810d, 5.0960d, 4.7860d,\\
4.7780d, 5.3930d, 5.3930d, 7.7480d,\\ 4.9570d, 4.6490d, 8.9170d,
6.6250d,\\ 5.3930d, 7.5730d, 8.3060d, 1.8913e,\\ 4.9570d, 4.6490d,
8.7300d, 6.4640d,\\ 5.3780d, 7.4050d, 8.1250d, 1.8640e,\\ 8.5280d,
6.3050d, 2.2525e, 7.9460d,\\ 1.8369e, 2.5856e, 8.7466e}&
\pbox{5cm}{\vspace{0.1cm} 4.2050d, 4.7530d, 5.0090d, 4.8680d,\\
4.6450d, 4.7530d, 4.6980d, 3.7640d,\\ 5.0090d, 4.8680d, 7.6500d,
4.2100d,\\ 4.7530d, 3.7640d, 4.0000d, 2.5700d,\\ 5.0660d, 4.9330d,
7.7170d, 4.2650d,\\ 4.7530d, 3.8050d, 4.0000d, 2.5700d,\\ 7.5560d,
4.3220d, 4.6280d, 4.0410d,\\ 2.5330d, 4.3520d, 2.0506e }&
\pbox{5cm}{\vspace{0.1cm} 1.8490d, 1.6810d, 1.9360d, 1.6000d,\\
1.8490d, 1.8490d, 1.9360d, 2.7040d,\\ 1.9360d, 1.6000d, 3.6000d,
2.2090d,\\ 1.9360d, 2.6010d, 2.9160d, 7.3960d,\\ 1.8490d, 1.6000d,
3.4810d, 2.1160d,\\ 1.9360d, 2.6010d, 2.8090d, 7.2250d,\\ 3.3640d,
2.1160d, 9.0250d, 2.8090d,\\ 7.0560d, 1.0609e, 3.8416e}
\\ \noalign{\hrule height 2pt}
$\rho_{gggeg}$ &  0.0023   &  0.0015  & 8.7754e  & 7.8961e  \\
\hline
Pr. r. s. &
\pbox{5cm}{\vspace{0.1cm} 2.1073e, 2.9890e, 2.6874e, 3.3256e,\\
2.7325e, 4.1000e, 2.9097e, 2.9250e,\\ 2.9017e, 4.2772e, 3.1714e,
3.2786e,\\ 3.9637e, 7.3225e, 2.7592e, 3.3898e,\\ 2.9306e, 4.3165e,
3.1912e, 3.3525e,\\ 4.0010e, 7.3053e, 2.7914e, 3.4697e,\\ 4.1738e,
7.1381e, 3.0985e, 4.1093e,\\ 7.3157e, 3.1322e, 0.0015 }&
\pbox{5cm}{\vspace{0.1cm} 4.1050d, 4.1490d, 4.5700d, 4.8500d,\\
4.0970d, 4.6660d, 5.0850d, 6.6890d,\\ 4.2100d, 4.4100d, 4.9010d,
6.4970d,\\ 4.8400d, 1.2356e, 7.0480d, 1.6400e,\\ 4.2100d, 4.4100d,
4.9010d, 6.3400d,\\ 4.7090d, 1.1920e, 6.8850d, 1.6145e,\\ 4.5370d,
9.9970d, 6.6890d, 1.5392e,\\ 1.3058e,  1.6916e, 8.8290e}&
\pbox{5cm}{\vspace{0.1cm} 2.1460d, 2.3720d, 2.3120d, 2.2500d,\\
2.3120d, 2.6650d, 2.1940d, 1.7060d,\\ 2.3410d, 2.6090d, 2.2210d,
1.6490d,\\ 2.6280d, 5.7130d, 1.6970d, 1.8180d,\\ 2.3410d, 2.6090d,
2.2210d, 1.6490d,\\ 2.6280d, 5.5700d, 1.6970d, 1.8180d,\\ 2.6090d,
4.7540d, 1.7170d, 1.6370d,\\ 6.1090d,   1.9300d, 1.6164e }&
\pbox{5cm}{\vspace{0.1cm} 8.4100c, 6.7600c, 7.8400c, 7.2900c,\\
7.2900c, 7.2900c, 9.0000c, 1.2960d,\\ 7.2900c, 6.2500c, 7.8400c,
1.1560d,\\ 7.8400c, 3.0250d, 1.3690d, 4.3560d,\\ 6.7600c, 6.2500c,
7.8400c, 1.0890d,\\ 7.8400c, 2.9160d, 1.3690d, 4.3560d,\\ 6.7600c,
2.2090d, 1.2250d, 3.9690d,\\ 3.2490d, 4.6240d, 3.3124e }
\\ \noalign{\hrule height 2pt}
\end{tabular}}}
\caption{\small The same as Table (\ref{s7}) but
$\frac{\gamma_{1}}{\lambda}=\frac{4\theta_{5}}{5}$,
$\frac{\gamma_{2}}{\lambda}=\frac{7\theta_{5}}{9}$,
$\frac{\gamma_{3}}{\lambda}=\frac{5\theta_{5}}{8}$,
$\frac{\gamma_{4}}{\lambda}=\frac{3\theta_{5}}{4}$ and
$\frac{\gamma_{5}}{\lambda}=\frac{6\theta_{5}}{7}$.} \label{s8}
\end{table}
\begin{table}[tbp]
{\renewcommand{\arraystretch}{1.7}
\renewcommand{\tabcolsep}{0.3cm} {\tiny
\hspace{4cm}\begin{tabular}{|>{ }c|c|c|c|c|} \hline Times &
$t=\frac{0.331\pi\hbar}{64\lambda\theta_{6}}$ &
$t=\frac{0.566\pi\hbar}{64\lambda\theta_{6}}$
& $t_{6p}=\frac{0.9899\pi\hbar}{64\lambda\theta_{6}}$ & $t=\frac{\pi\hbar}{64\lambda\theta_{6}}$  \\
\noalign{\hrule height 2pt}
$\rho_{eeeeee}$ &  0.3216  &  0.6082 & 0.7566   &   0.7566 \\
\hline
\noalign{\hrule height 2pt}
$\rho_{eeeegg}$ & 0.3365  & 0.6362  &  0.7914  &  0.7914 \\
\hline
\noalign{\hrule height 2pt}
$\rho_{eeegge}$ & 0.3381  &  0.6389 & 0.7949   & 0.7950  \\
\hline
\noalign{\hrule height 2pt}
$\rho_{eegeeg}$ &  0.3357  &  0.6346 &  0.7895  &  0.7894 \\
\hline
\noalign{\hrule height 2pt}
$\rho_{eeggee}$ & 0.3372  & 0.6374  &  0.7931  &  0.7930 \\
\hline
\noalign{\hrule height 2pt}
$\rho_{eegggg}$ &  0.3531 & 0.6670  &  0.8299  &  0.8299 \\
\hline
\noalign{\hrule height 2pt}
$\rho_{egeeeg}$ & 0.3394 &  0.6416 &  0.7982  & 0.7982  \\
\hline
\noalign{\hrule height 2pt}
$\rho_{egeegg}$ & 0.3471  &  0.6559  &  0.8162  &  0.8161 \\
\hline
\noalign{\hrule height 2pt}
$\rho_{egegge}$ & 0.3487  & 0.6588  & 0.8198   &  0.8197 \\
\hline
\noalign{\hrule height 2pt}
$\rho_{eggeeg}$ & 0.3463  & 0.6544  & 0.8141   &  0.8141 \\
\hline
\noalign{\hrule height 2pt}
$\rho_{egggee}$ & 0.3479  &  0.6573  &  0.8177  & 0.8178  \\
\hline
\noalign{\hrule height 2pt}
$\rho_{eggggg}$ &  0.3643 &  0.6879  &  0.8561  & 0.8562  \\
\hline
\noalign{\hrule height 2pt}
$\rho_{geeege}$ & 0.3370  & 0.6369  &  0.7923  &  0.7923 \\
\hline
\noalign{\hrule height 2pt}
$\rho_{geegeg}$ &  0.3469 &  0.6555 &  0.8155  &  0.8154 \\
\hline
\noalign{\hrule height 2pt}
$\rho_{geeggg}$ &  0.3548 & 0.6702  &  0.8339  & 0.8339  \\
\hline
\noalign{\hrule height 2pt}
$\rho_{gegegg}$ & 0.3519  &   0.6648 &  0.8272  & 0.8272  \\
\hline
\noalign{\hrule height 2pt}
$\rho_{geggeg}$ &  0.3539  &  0.6685 & 0.8318   & 0.8319  \\
\hline
\noalign{\hrule height 2pt}
$\rho_{gegggg}$ & 0.3619  &  0.6836  &  0.8506  & 0.8506  \\
\hline
\noalign{\hrule height 2pt}
$\rho_{ggeeeg}$ & 0.3480  & 0.6575  &  0.8181  &  0.8179 \\
\hline
\noalign{\hrule height 2pt}
$\rho_{ggegeg}$ &  0.3579 &  0.6760 &  0.8412  &  0.8411 \\
\hline
\noalign{\hrule height 2pt}
$\rho_{ggeggg}$ &  0.3661 & 0.6912  & 0.8601   &   0.8603 \\
\hline
\noalign{\hrule height 2pt}
$\rho_{gggege}$ & 0.3545  & 0.6699  &   0.8335  &  0.8336 \\
\hline
\noalign{\hrule height 2pt}
$\rho_{ggggeg}$ & 0.3652  &  0.6896 &  0.8582  &  0.8582  \\
\hline
\noalign{\hrule height 2pt}
$\rho_{gggggg}$ &  0.3736  &  0.7051 &  0.8777  & 0.8776  \\
\hline
\noalign{\hrule height 2pt}
\end{tabular}}}
\caption{\footnotesize Outputs of dynamical quantum search algorithm
for some probabilities of marked  states
$\rho_{r_{1}r_{2}r_{3}r_{4}r_{5}r_{6}}$\,
($r_{1},r_{2},r_{3},r_{4},r_{5}, r_{6}=e,g$) of 6-qubit for
different values of time, where
$\frac{\gamma_{1}}{\lambda}=\frac{\theta_{6}}{113}$,
$\frac{\gamma_{2}}{\lambda}=\frac{\theta_{6}}{90}$,
$\frac{\gamma_{3}}{\lambda}=\frac{\theta_{6}}{140}$,
$\frac{\gamma_{4}}{\lambda}=\frac{\theta_{6}}{100}$,
$\frac{\gamma_{5}}{\lambda}=\frac{\theta_{6}}{125}$ and
$\frac{\gamma_{6}}{\lambda}=\frac{\theta_{6}}{119}$.} \label{s9}
\end{table}
 \begin{table}[tbp]
{\renewcommand{\arraystretch}{1.5} \renewcommand{\tabcolsep}{0.1cm}
{\tiny
\hspace{-0.65cm}\begin{tabular}{|>{ }c|c|c|c|c|} \hline Times &
$t=\frac{0.331\pi\hbar}{64\lambda\theta_{6}}$ &
$t=\frac{0.566\pi\hbar}{64\lambda\theta_{6}}$
& $t_{6p}=\frac{0.9899\pi\hbar}{64\lambda\theta_{6}}$ & $t=\frac{\pi\hbar}{64\lambda\theta_{6}}$  \\
\noalign{\hrule height 2pt}
$\rho_{eeegge}$ & 3.7540d   &  1.6180d  &  1.4170d  & 1.4440d  \\
\hline
Pr. r. s. &
\pbox{5cm}{\vspace{0.1cm}  1.6250d,  2.4980d,  2.6090d, 3.8210d,\\
2.4660d, 3.4850d, 4.6250d, 2.1730d,\\ 3.1460d,  3.1690d, 3.9780d,
2.8730d,\\ 3.6010d, 3.4900d, 3.0880d, 2.4050d,\\ 3.5810d,  3.4850d,
4.6730d, 3.2800d,\\ 4.2500d, 4.0960d, 4.2120d, 2.9530d,\\ 3.6040d,
3.4820d, 3.2650d, 3.1360d,\\ 2.6120d, 2.4250d, 6.0020d, 2.4050d,\\
3.5810d, 3.6020d, 4.8100d, 3.2800d,\\ 4.2500d, 4.2250d, 4.3940d,
2.9530d,\\ 3.7300d, 3.6040d, 3.3620d, 3.2490d,\\ 2.7540d, 2.4740d,
6.2290d, 3.3700d,\\ 4.3920d, 4.2500d, 4.5460d, 3.8530d,\\ 3.8260d,
3.5240d, 8.4500d, 3.3650d,\\ 2.8090d, 2.6650d,  6.4660d, 2.0500d,\\
5.5810d, 5.3380d, 9.2909e }&
\pbox{5cm}{\vspace{0.1cm}6.5600c,  8.1800c, 8.6500c, 9.3200c,\\
8.6500c, 9.3200c, 1.1300d, 7.4000c,\\ 9.0100c, 9.0100c, 9.6200c,
8.5000c,\\ 9.4100c, 1.0210d, 1.1050d, 7.8500c,\\ 9.3200c, 9.3200c,
9.8500c, 9.0100c,\\ 9.6200c, 1.1050d, 1.1600d, 8.5000c,\\ 9.6200c,
9.4100c, 1.0980d, 9.4100c,\\ 1.1050d, 1.1140d, 3.6170d, 7.8500c,\\
9.3200c, 9.3200c, 9.8500c, 9.0100c,\\ 9.6200c, 1.1050d, 1.1570d,
8.5000c,\\ 9.6200c, 9.4100c, 1.0980d, 9.4100c,\\ 1.1050d, 1.1140d,
3.5300d, 9.3200c,\\ 9.8500c, 9.6200c, 1.0890d, 9.6200c,\\ 1.0930d,
1.1650d, 3.2080d,  9.4100c,\\ 1.1050d, 1.1050d, 3.5280d, 1.1140d,\\
3.7890d, 3.8800d,  2.7073e}&
\pbox{5cm}{\vspace{0.1cm} 5.7800c,  6.7400c,  7.4000c,  7.9300c,\\
7.4000c, 7.9300c, 8.4800c, 6.2500c,\\
7.2500c, 7.2500c, 7.4000c, 6.7400c,\\
6.7400c, 7.4000c, 3.6000c, 6.7400c,\\
7.9300c, 7.9300c, 7.9300c, 7.4000c,\\
7.4000c, 7.9300c, 4.4900c, 6.7400c,\\
6.7400c, 6.7400c, 3.6000c, 6.2500c,\\
3.2500c, 3.2500c, 7.3000b, 6.7400c,\\
7.9300c, 7.9300c, 7.9300c, 7.4000c,\\
7.4000c, 7.9300c, 4.4900c, 7.2500c,\\
6.7400c, 6.7400c, 4.1000c, 6.2500c,\\
3.2500c, 3.2500c, 7.3000b, 7.4000c,\\
7.9300c, 7.4000c, 4.9000c, 6.8900c,\\
4.1000c, 4.1000c, 2.5000b, 6.7400c,\\
3.6000c, 3.2500c, 5.8000b, 2.9200c,\\
8.5000b, 1.0400c, 6.9530d }&
\pbox{5cm}{\vspace{0.1cm} 5.2900c,  6.7600c,  7.2900c,  7.8400c,\\
7.2900c,  7.8400c,   8.4100c,  6.2500c,\\ 7.2900c,  7.2900c,
7.2900c, 6.7600c,\\ 6.7600c,  7.2900c,  3.6100c,  6.7600c,\\
7.8400c, 7.8400c, 7.8400c,  7.2900c,\\ 7.2900c,  7.8400c,  4.4100c,
7.2900c,\\ 6.7600c, 6.7600c,  3.6100c,  6.2500c,\\ 3.2400c, 3.2400c,
4.9000b, 6.7600c,\\ 7.8400c,  7.8400c,  7.8400c,  7.2900c,\\7.2900c,
8.4100c, 4.4100c, 7.2900c,\\ 6.7600c,  6.7600c,  4.0000c, 6.2500c,\\
3.2400c, 3.2400c, 4.9000b,  7.2900c,\\ 7.8400c,  7.2900c, 4.8400c,
7.2900c,\\ 4.0000c, 4.0000c,  9.0000a,  6.7600c,\\ 3.6100c, 3.2400c,
3.6000b, 2.8900c,\\ 6.4000b,  8.1000b,  6.8890d }
\\ \noalign{\hrule height 2pt}
$\rho_{eggeeg}$ &   4.5730d  &  2.1760d  &  1.7690d  &  1.7640d \\
\hline
Pr. r. s. &
\pbox{5cm}{\vspace{0.1cm} 1.5170d,  2.4400d,  2.3140d,  3.5810d,\\
2.1380d,  3.2570d,  3.1460d,  4.2500d,\\ 2.1250d,  3.3140d, 2.9530d,
3.8450d,\\ 2.7700d,  3.4810d,  3.2490d,  2.9000d,\\ 2.2600d,
3.5890d, 3.2570d,  4.5140d,\\ 3.0600d,  3.9780d,  3.7300d,
3.8260d,\\ 3.1330d, 3.4820d,  3.0420d,  3.1400d,\\ 2.4400d, 2.2100d,
5.3890d, 2.2250d,\\ 3.4850d,  3.2570d,  4.3920d,  3.0370d,\\
3.9850d, 3.7370d, 3.8970d, 2.8730d,\\ 3.7220d,  3.3650d,  3.2040d,
3.0260d,\\ 2.4740d, 2.1760d, 5.9240d,  3.1460d,\\ 4.2410d,  3.9850d,
4.2850d, 3.4850d,\\ 3.4900d, 3.0770d,  7.9010d,  3.2500d,\\
2.7620d, 2.4740d, 5.9130d, 1.8850d,\\ 5.0580d, 5.1210d,  9.0817e  }&
\pbox{5cm}{\vspace{0.1cm} 6.1700c,  7.7300c,  7.7300c,  9.1400c,\\
7.3000c,  8.8100c,  8.8100c,  9.5300c,\\ 7.7300c,  1.0180d, 8.3200c,
9.2800c,\\ 8.3200c,  9.0500c,  8.5000c,  1.0280d,\\ 7.7300c,
1.0530d, 8.8100c,  9.5300c,\\ 8.3200c,  9.2800c,  9.2800c,
1.0240d,\\ 1.0180d, 9.0500c, 1.2260d,  9.0500c,\\ 1.2290d,  1.0400d,
3.5300d, 7.7300c,\\ 9.1400c,  9.1400c,  9.5300c,  8.3200c,\\
9.5300c, 9.2800c, 1.0240d, 8.3200c,\\ 9.2800c,  9.2800c,  1.0280d,
8.5000c,\\ 1.0330d, 1.0330d, 3.4450d,  8.8100c,\\ 9.5300c,  9.5300c,
1.0250d, 9.2800c,\\ 1.0250d, 1.0250d,  3.1250d,  9.0500c,\\ 1.2290d,
1.0330d, 3.3620d, 1.0400d,\\ 3.6170d,  3.7060d,  2.6645e}&
\pbox{5cm}{\vspace{0.1cm}   4.7700c,  6.2500c,  6.2500c,  7.2500c,\\
5.7800c,  6.7400c,  6.7400c,  6.7400c,\\ 5.7800c,  7.9300c, 6.7400c,
6.7400c,\\ 6.2500c,  6.2500c,  6.2500c,  3.2500c,\\ 6.2500c,
7.9300c, 6.7400c,  7.2500c,\\ 6.7400c,  6.7400c,  6.7400c,
3.9700c,\\ 7.9300c, 6.2500c,  4.4900c,  5.7800c,\\ 3.7300c, 2.6100c,
8.5000b, 6.2500c,\\ 7.2500c,  7.2500c,  7.2500c,  6.7400c,\\
6.7400c, 6.7400c, 3.9700c, 6.2500c,\\ 6.2500c,  6.2500c,  3.2500c,
5.7800c,\\ 2.9200c, 2.9200c, 8.5000b,  6.7400c,\\ 6.7400c,  6.7400c,
3.9700c, 6.2500c,\\ 3.6000c, 3.6000c,  4.5000b,  5.7800c,\\ 4.1000c,
2.9200c, 6.8000b, 2.6100c,\\ 1.0400c,  1.2500c,  7.1370d}&
\pbox{5cm}{\vspace{0.1cm} 4.8400c,  6.2500c,  6.2500c,  7.2900c,\\
5.7600c,  6.7600c,  6.7600c,  6.7600c,\\ 5.7600c,  7.8400c, 6.7600c,
6.7600c,\\ 6.2500c,  6.2500c,  6.2500c,  3.2400c,\\ 6.2500c,
7.8400c, 6.7600c,  7.2900c,\\ 6.7600c,  6.7600c,  6.7600c,
4.0000c,\\ 7.8400c, 6.2500c,  4.4100c,  5.7600c,\\ 3.6100c, 2.5600c,
6.4000b, 6.2500c,\\ 7.2900c,  7.2900c,  7.2900c,  6.7600c,\\
6.7600c, 6.7600c, 4.0000c, 6.2500c,\\ 6.2500c,  6.2500c,  3.2400c,
5.7600c,\\ 2.8900c, 2.8900c, 6.4000b,  6.7600c,\\ 6.7600c,  6.7600c,
4.0000c, 6.2500c,\\ 3.6100c, 3.2400c,  2.5000b,  6.2500c,\\ 4.0000c,
2.8900c, 4.9000b, 2.5600c,\\ 1.0000c,  1.2100c,  7.0560d }
\\ \noalign{\hrule height 2pt}
$\rho_{geeggg}$ & 8.6440d  & 4.8050d   &  3.0980d  &   3.1360d \\
\hline
Pr. r. s. &
\pbox{5cm}{\vspace{0.1cm} 1.3000d,  2.1250d,  2.1250d,  3.4450d,\\
1.9610d,  3.2500d,  3.1410d,  5.2480d,\\ 1.7690d,  2.6260d, 2.5290d,
3.2530d,\\ 2.3410d,  2.9250d,  2.8250d,  2.4660d,\\ 1.8820d,
2.8730d, 2.7970d,  3.8480d,\\ 2.5970d,  3.4810d,  3.3640d,
3.7570d,\\ 2.3720d, 2.7200d,  2.6170d,  2.0410d,\\ 2.3400d, 1.5850d,
1.4120d, 4.5890d,\\ 2.0420d,  3.3610d,  3.2500d,  5.3780d,\\
3.0650d, 5.0000d, 4.8820d, 2.4340d,\\ 3.0340d,  2.9250d,  2.6930d,
2.6260d,\\ 2.2100d, 2.0530d, 3.1120d,  2.6960d,\\ 3.6010d,  3.4810d,
3.9860d, 3.1370d,\\ 3.3300d, 3.1250d,  4.7610d,  2.5250d,\\ 1.8050d,
1.6200d, 4.9330d, 1.2200d,\\ 4.3600d,  4.2010d,  8.1577e}&
\pbox{5cm}{\vspace{0.1cm} 6.1300c,  7.6100c,  7.6100c,  8.9000c,\\
7.2200c,  8.9000c,  8.5300c,  1.2850d,\\ 6.8500c,  8.0800c, 8.0800c,
8.7200c,\\ 8.0800c,  8.2100c,  7.9400c,  9.0000c,\\ 7.2200c,
8.4500c, 8.4500c,  9.3200c,\\ 8.0800c,  9.0100c,  9.0100c,
9.7000c,\\ 7.7300c, 8.4500c,  8.4500c,  9.0000c,\\ 7.9400c, 9.0100c,
9.0400c, 3.2020d,\\ 7.2200c,  8.9000c,  8.9000c,  1.2850d,\\
8.5300c, 1.2850d, 1.2850d, 8.0800c,\\ 8.2100c,  8.2100c,  9.6200c,
7.9400c,\\ 9.0000c, 9.0100c, 3.2180d,  8.0800c,\\ 9.0100c,  9.0100c,
9.7700c, 8.7200c,\\ 9.6500c, 9.6500c,  2.9860d,  7.9400c,\\ 9.0100c,
9.0100c, 3.0420d, 9.0400c,\\ 3.2850d,  3.3700d,  2.5801e }&
\pbox{5cm}{\vspace{0.1cm} 3.8600c,  5.2000c,  5.2000c,  6.2500c,\\
4.7700c,  5.6500c,  5.6500c,  8.4800c,\\ 4.7700c,  5.2000c, 5.2000c,
5.2000c,\\ 5.2000c,  4.7700c,  4.7700c,  2.2100c,\\ 4.7700c,
5.6500c, 5.6500c,  5.6500c,\\ 5.2000c,  5.2000c,  5.2000c,
3.2500c,\\ 5.2000c, 4.7700c,  4.7700c,  2.2100c,\\ 4.3600c, 1.9400c,
1.6900c, 1.7000c,\\ 4.7700c,  6.1200c,  6.1200c,  8.4800c,\\
5.6500c, 7.9300c, 7.9300c, 5.2000c,\\ 4.7700c,  4.7700c,  2.5000c,
4.7700c,\\ 2.2100c, 2.2100c, 5.8000b,  5.6500c,\\ 5.6500c,  5.6500c,
3.6000c, 5.2000c,\\ 2.8100c, 2.8100c,  1.7000b,  4.3600c,\\ 1.9400c,
1.9400c, 1.4500c, 1.6900c,\\ 1.9700c,  1.9700c,  7.6690d}&
\pbox{5cm}{\vspace{0.1cm} 4.0000c,  4.8400c,  4.8400c,  6.2500c,\\
4.8400c,  5.7600c,  5.7600c,  8.4100c,\\ 4.4100c,  5.2900c, 5.2900c,
5.2900c,\\ 5.2900c,  4.8400c,  4.8400c,  2.2500c,\\ 4.8400c,
5.7600c, 5.7600c,  5.7600c,\\ 5.2900c,  5.2900c,  5.2900c,
3.2400c,\\ 5.2900c, 4.8400c,  4.8400c,  2.2500c,\\ 4.4100c, 1.9600c,
1.6900c, 1.4400c,\\ 4.8400c,  6.2500c,  5.7600c,  8.4100c,\\
5.7600c, 8.4100c, 7.8400c, 5.2900c,\\ 4.8400c,  4.8400c,  2.5600c,
4.4100c,\\ 2.2500c, 1.9600c, 3.6000b,  5.7600c,\\ 5.7600c,  5.2900c,
3.2400c, 5.2900c,\\ 2.8900c, 2.8900c,  1.0000a,  4.4100c,\\ 1.9600c,
1.9600c, 1.2100c, 1.6900c,\\ 1.6900c,  1.9600c,  7.5690d }
\\ \noalign{\hrule height 2pt}
$\rho_{gegggg}$ &  1.5541e   &  1.1444e   &  5.7600d  &  5.6250d \\
\hline
Pr. r. s. &
\pbox{5cm}{\vspace{0.1cm} 1.3010d,  1.9720d,  1.9720d,  2.9840d,\\
1.8320d,  2.7880d,  2.7380d,  4.1960d,\\ 1.7650d,  2.6450d, 2.5540d,
3.8650d,\\ 2.3780d,  3.6500d,  3.5600d,  6.2450d,\\ 1.8850d,
2.6010d, 2.6010d,  3.1690d,\\ 2.4250d,  2.8730d,  2.6960d,
2.2850d,\\ 2.1760d, 2.5290d,  2.4340d,  1.8640d,\\ 2.2810d, 1.5700d,
1.4050d, 2.8250d,\\ 1.9010d,  2.8850d,  2.7880d,  4.3250d,\\
2.6450d, 3.9650d, 3.8420d, 6.4810d,\\ 2.4650d,  3.7690d,  3.6500d,
6.2570d,\\ 3.4450d, 6.0840d, 6.0880d,  2.4250d,\\ 2.9780d,  2.8730d,
2.5290d, 2.5970d,\\ 2.0800d, 1.8850d,  4.0180d,  2.3410d,\\ 1.6640d,
1.5700d, 3.1060d, 1.2500d,\\ 2.6690d,  2.5000d,  4.2250e  }&
\pbox{5cm}{\vspace{0.1cm} 5.4900c,  6.5000c,  6.5000c,  7.6100c,\\
6.5000c,  7.6100c,  7.6100c,  8.6500c,\\ 6.5000c,  7.6100c, 7.6100c,
8.3200c,\\ 7.6100c,  8.3200c,  8.3200c,  1.5130d,\\ 6.5000c,
7.6100c, 7.2400c,  8.0100c,\\ 7.2400c,  8.0100c,  7.7200c,
7.9300c,\\ 7.2400c, 7.2500c,  7.2500c,  7.8800c,\\ 7.2500c, 7.8500c,
7.8500c, 2.4580d,\\ 6.5000c,  7.6100c,  7.6100c,  8.6500c,\\
7.6100c, 8.3200c, 8.3200c, 1.4170d,\\ 7.6100c,  8.3200c,  8.3200c,
1.5130d,\\ 8.3200c, 1.5880d, 1.5880d,    7.2400c,\\ 8.0100c,
8.0100c, 7.9300c, 7.2500c,\\ 7.8800c, 7.8800c,  2.3300d,  7.2500c,\\
7.8800c, 7.8500c, 2.3930d, 7.8500c,\\ 2.4580d,  2.5250d,  1.8178e }&
\pbox{5cm}{\vspace{0.1cm}2.7200c,  3.4900c,  3.4900c,  3.8600c,\\
3.4900c,  3.8600c,  3.8600c,  4.2500c,\\ 3.1400c,  3.8600c, 3.8600c,
3.8600c,\\ 3.4900c,  3.8600c,  3.8600c,  5.7800c,\\ 3.1400c,
3.8600c, 3.4900c,  3.4900c,\\ 3.4900c,  3.1400c,  3.1400c,
1.3700c,\\ 3.4900c, 3.1400c,  3.1400c,  1.1600c,\\ 2.8100c, 1.1600c,
9.7000b, 1.2200c,\\ 3.4900c,  3.8600c,  3.8600c,  4.2500c,\\
3.8600c, 4.2500c, 3.8600c, 5.3300c,\\ 3.4900c,  3.8600c,  3.8600c,
5.7800c,\\ 3.8600c, 5.7800c, 5.7800c,  3.4900c,\\ 3.1400c,  3.1400c,
1.3700c, 3.1400c,\\ 1.3700c, 1.1600c,  1.0100c,  2.8100c,\\ 1.1600c,
1.1600c, 1.2200c, 9.7000b,\\ 1.4500c,  1.4500c,  3.9940d }&
\pbox{5cm}{\vspace{0.1cm}2.8900c,  3.6100c,  3.2400c,  4.0000c,\\
3.2400c,  4.0000c,  4.0000c,  4.0000c,\\ 3.2400c,  3.6100c, 3.6100c,
4.0000c,\\ 3.6100c,  4.0000c,  4.0000c,  5.7600c,\\ 3.2400c,
3.6100c, 3.6100c,  3.2400c,\\ 3.6100c,  3.2400c,  3.2400c,
1.4400c,\\ 3.2400c, 2.8900c,  2.8900c,  1.2100c,\\ 2.8900c, 1.0000c,
1.0000c, 1.2100c,\\ 3.2400c,  4.0000c,  4.0000c,  4.4100c,\\
3.6100c, 4.0000c, 4.0000c, 5.2900c,\\ 3.6100c,  4.0000c,  4.0000c,
5.7600c,\\ 4.0000c, 5.7600c, 5.7600c,  3.6100c,\\ 3.2400c,  3.2400c,
1.4400c, 2.8900c,\\ 1.2100c, 1.2100c,  8.1000b,  2.8900c,\\ 1.0000c,
1.0000c, 1.0000c, 8.1000b,\\ 1.2100c,  1.4400c,  3.8440d }
\\ \noalign{\hrule height 2pt}
$\rho_{ggggeg}$ &  1.1466e  &  1.0372e   &  5.0180d  &  4.9000d \\
\hline
Pr. r. s. &
\pbox{5cm}{\vspace{0.1cm}1.3520d,  2.0450d,  2.0250d,  2.8330d,\\
1.9010d,  2.7880d,  2.6010d,  3.1690d,\\ 1.8320d,  2.6450d, 2.4250d,
2.8730d,\\ 2.4650d,  3.6500d,  2.5000d,  1.9700d,\\ 1.9010d,
2.8850d, 2.6450d,  3.2570d,\\ 2.6000d,  3.8420d,  2.8730d,
2.5290d,\\ 2.4650d, 3.6250d,  2.5970d,  2.0570d,\\ 3.3320d, 5.3330d,
1.5850d, 3.2170d,\\ 1.9720d,  2.8850d,  2.6450d,  3.3700d,\\
2.6930d, 3.9650d, 2.9530d, 2.6570d,\\ 2.5540d,  3.7440d,  2.5970d,
2.1690d,\\ 3.4450d, 5.3380d, 1.6810d,  3.3130d,\\ 2.6930d,  4.0690d,
2.9530d, 2.7560d,\\ 3.6250d, 5.5570d,  2.1050d,  4.5140d,\\ 3.4180d,
5.4920d, 1.6810d, 3.5300d,\\ 5.1850d,    2.7450d,  4.2850e }&
\pbox{5cm}{\vspace{0.1cm}5.4900c,  6.8500c,  6.8500c,  7.6500c,\\
6.5000c,  7.2400c,  7.2400c,  7.7200c,\\ 6.1300c,  7.2400c, 7.2400c,
7.7200c,\\ 7.2400c,  8.0100c,  6.9800c,  7.8500c,\\ 6.5000c,
7.2400c, 7.2400c,  7.7200c,\\ 7.2400c,  8.0100c,  7.7200c,
7.8800c,\\ 7.2400c, 8.0100c,  7.2500c,  7.8800c,\\ 8.0100c, 1.4690d,
7.8500c, 2.3930d,\\ 6.5000c,  8.0200c,  7.6500c,  7.7200c,\\
7.2400c, 8.3200c, 7.7200c, 7.8800c,\\ 7.2400c,  8.0100c,  7.2500c,
7.8800c,\\ 8.0100c, 1.4690d, 7.8500c,  2.3930d,\\ 7.2400c,  8.3200c,
7.7200c, 7.9300c,\\ 8.0100c, 1.3250d,  7.8800c,  2.3300d,\\ 8.0100c,
1.3960d, 7.8500c, 2.3300d,\\ 1.5440d,    2.4580d,  1.7620e }&
\pbox{5cm}{\vspace{0.1cm}  2.8100c,  3.4900c,  3.4900c,  3.8600c,\\
3.4900c,  3.8600c,  3.8600c,  3.4900c,\\ 3.1400c,  3.8600c, 3.4900c,
3.1400c,\\ 3.4900c,  3.8600c,  3.1400c,  1.3700c,\\ 3.4900c,
3.8600c, 3.8600c,  3.4900c,\\ 3.8600c,  3.8600c,  3.1400c,
1.6000c,\\ 3.8600c, 3.8600c,  3.1400c,  1.3700c,\\ 3.8600c, 5.3300c,
1.1600c, 1.0100c,\\ 3.4900c,  3.8600c,  3.8600c,  3.4900c,\\
3.8600c, 4.2500c, 3.4900c, 1.6000c,\\ 3.8600c,  3.8600c,  3.1400c,
1.3700c,\\ 3.8600c, 5.3300c, 1.1600c,  1.0100c,\\ 3.8600c,  4.2500c,
3.4900c, 1.6000c,\\ 3.8600c, 4.9000c,  1.3700c,  8.2000b,\\ 3.8600c,
5.3300c, 1.1600c, 8.2000b,\\ 5.3300c,    1.2200c,  3.5060d }&
\pbox{5cm}{\vspace{0.1cm}  2.8900c,  3.6100c,  3.6100c,  4.0000c,\\
3.2400c,  4.0000c,  3.6100c,  3.6100c,\\ 3.2400c,  4.0000c, 3.6100c,
3.2400c,\\ 3.6100c,  4.0000c,  2.8900c,  1.2100c,\\ 3.2400c,
4.0000c, 3.6100c,  3.6100c,\\ 4.0000c,  4.0000c,  3.2400c,
1.4400c,\\ 3.6100c, 4.0000c,  3.2400c,  1.2100c,\\ 3.6100c, 5.2900c,
1.0000c, 8.1000b,\\ 3.6100c,  4.0000c,  3.6100c,  3.6100c,\\
4.0000c, 4.0000c, 3.2400c, 1.4400c,\\ 3.6100c,  4.0000c,  3.2400c,
1.4400c,\\ 4.0000c, 5.2900c, 1.2100c,  8.1000b,\\ 4.0000c,  4.0000c,
3.2400c, 1.6900c,\\ 4.0000c, 4.8400c,  1.2100c,  6.4000b,\\ 4.0000c,
5.2900c, 1.2100c, 8.1000b,\\ 5.2900c,    1.0000c,  3.4810d }
\\ \noalign{\hrule height 2pt}
\end{tabular}}}
\caption{\footnotesize The same as Table (\ref{s9}) but
$\frac{\gamma_{1}}{\lambda}=\frac{4\theta_{6}}{5}$,
$\frac{\gamma_{2}}{\lambda}=\frac{7\theta_{6}}{9}$,
$\frac{\gamma_{3}}{\lambda}=\frac{5\theta_{6}}{8}$,
$\frac{\gamma_{4}}{\lambda}=\frac{3\theta_{6}}{4}$,
$\frac{\gamma_{5}}{\lambda}=\frac{6\theta_{6}}{7}$ and
$\frac{\gamma_{6}}{\lambda}=\frac{9\theta_{6}}{10}$.} \label{s10}
\end{table}
\\ \indent
First, it is easy to see that the sum of all probabilities for all
marked and unmarked states at any time in the absence of dissipation
rates is equal to one, see Fig.\ref{ff2}.
Also, from Figs.\ref{ff3}, Fig.\ref{f1}, Fig.\ref{f2} and the above
tables, one can note that the sum of all probabilities in the
presence of dissipation at different times are gradually decreased
than one, and reaches almost zero at high values of dissipation
rates.
\\ \indent
Second, from Tables (\ref{s1}-\ref{s10}), one observes and
distinguishes between different values of times for the
probabilities of each marked state at  weak values of the
dissipation rates.
Moreover, we observe for times $t_{Np}$, and a weak dissipation,
that the probabilities of each marked states have higher
probabilities than any marked state probabilities through other
times.
Further increasing of $\gamma_{i}$ and for different values of
times, one can not distinguish the probabilities of some marked
states for the different number of qubits, where some of un-marked
states due to the influence of large dissipation  exchange their
values either increase or decrease the probabilities of marked
states.
Moreover, if dissipation rates are taken to be
$\gamma_{i}=\lambda\theta_{N}$, $i=1,2,...,N$, one determines and
observes at different times the probabilities of some marked states
for different values of the number of $N$.
However, when $\gamma_{i}>\lambda\theta_{N}$, we can not distinguish
any marked states  where very small values for the probability of
marked state is shown.
Which means that both marked and un-marked states have the same
chance to contribute.
One concludes that at $\gamma_{i}>\lambda\theta_{N}$, the system is
destroyed completely as well as it can not search for any desired
state whatever the number of qubits.
It is interesting to discuss the influence of large values of
dissipation on the probabilities of marked states at these optional
times in the following scenario:
\\ \indent
Through the optional times
$\frac{0.331\pi\hbar}{2^{N}\lambda\theta_{N}}$ and the number of
qubits up to $N=5$, we can not distinguish and observe the
probability of any marked state  for any number of qubits except the
probabilities of marked states of the kind $q_{2}=\{gg, eg, ge\}$,
$q_{3}=\{ggg, egg, geg, gge\}$, $q_{4}=\{gggg, eggg, gegg, ggeg,
ggge\}$ and $q_{5}=\{ggggg, egggg, geggg, ggegg, gggeg, gggge\}$.
As for the case of $N\geq6$, we can not distinguish the
probabilities of any marked states, where the  probabilities  of
un-marked states  show larger values compared with the probability
of marked state which shows a small values.
\\ \indent
At times $\frac{0.566\pi\hbar}{2^{N}\lambda\theta_{N}}$, one
distinguishes the probability of marked state  for $2-$, $3-$ and
$4-$qubit but for 5-qubit one can not distinguish the probabilities
of any marked states except the marked states which are of the kind
$q_{5}$ one can distinguish the probabilities for it.
While through this time when $N\geq6$, we can not distinguish and
determine the probability of any marked state, where both marked and
un-marked states have the same chance to contribute in the search.
\\ \indent
Also, at times $t_{Np}$, we distinguish the probability of any
marked state for different number of qubits until $N=5$ except the
probabilities of the marked states $|eeee\rangle$ and
$|eeeee\rangle$.
For $N=6$, one can not observe and determine the probability of any
marked state  except the marked states of the kind $q_{6}=\{gggggg,
eggggg, gegggg, ggeggg, gggegg,$ $ggggeg, ggggge\}$, where the value
of any marked state  probability for this kind is distinguished and
shows small value and larger than the probabilities of the remaining
states.
However, we expect through these times for $N>6$ that one can not
distinguish and determine the probability of any marked state.
\\ \indent
Finally, at times $\frac{\pi\hbar}{2^{N}\lambda\theta_{N}}$, it is
shown that the probability of any marked state  for the different
number of qubits until $N=4$ can be distinguished.
As for 5-qubit and 6-qubit, we can not distinguish the probability
of some marked state, see Tables (\ref{s8},\ref{s10}), except the
marked states of the kind $q_{5}$ and $q_{6}$.
After that, through these times, one can not observe the
probabilities of any marked states at $N=7$ and more.
Consequently, through the large dissipation values, most marked
states are distinguished through the optional times $t_{Np}$ for the
different number of qubits.
As a result, one can distinguish and observe, at large values of
dissipation, the probabilities of many different marked states for
multi qubits at different values of times.
These observations give an advantage of using the dynamical quantum
search algorithm with the dissipation are other quantum search
algorithms.

\end{document}